\documentclass[twocolumn,prc,floatfix,showpacs,preprintnumbers,amsmath,amssymb]{revtex4}

\usepackage{color}
\usepackage{graphicx}
\usepackage{dcolumn}    % Align table columns on decimal point
\usepackage{bm}         % bold math
\newcommand{\bsigma}{\mbox{\boldmath$\sigma$}}
\newcommand{\btau}{\mbox{\boldmath$\tau$}}
\begin{document}
% >>>>>>>>>>>>>>>>>>>>>>>>>>>>>>>>>>>>>>>>>>>>>>>>>>>>>>>>>>>>>>>>>>>>
% TITLE AND AUTHORS.
%

\title{Low-lying dipole response: isospin character and collectivity 
in ${}^{68}$Ni, ${}^{132}$Sn and  ${}^{208}$Pb}

\author{X. Roca-Maza\textsuperscript{1}}
\email{xavier.roca.maza@mi.infn.it}
\author{G. Pozzi\textsuperscript{2}}
\email{giac.pozzi@gmail.com}
\author{M. Brenna\textsuperscript{1,2}}
\email{marco.brenna@mi.infn.it}
\author{K. Mizuyama\textsuperscript{1}}
\email{mizukazu147@gmail.com}
\author{G. Col\`o\textsuperscript{1,2}}
\email{gianluca.colo@mi.infn.it}

\affiliation{\textsuperscript{1} INFN, sezione di Milano, via Celoria 16, 
             I-20133 Milano, Italy\\
            \textsuperscript{2} Dipartimento di Fisica, Universit\`a degli 
             Studi di Milano, via Celoria 16, I-20133 Milano, Italy}

\date{\today} 

% >>>>>>>>>>>>>>>>>>>>>>>>>>>>>>>>>>>>>>>>>>>>>>>>>>>>>>>>>>>>>>>>>>>>
% ABSTRACT, PACS.
%
\begin{abstract}
The isospin character, the collective or single-particle nature, and the 
sensitivity to the slope of the nuclear symmetry energy of the low-energy 
isovector dipole response (known as pygmy dipole resonance) are nowadays 
under debate. In the present work we study, within the fully self-consistent 
non-relativistic mean field (MF) approach based on Skyrme Hartree-Fock plus 
Random Phase Approximation (RPA), the measured even-even nuclei ${}^{68}$Ni, 
${}^{132}$Sn and ${}^{208}$Pb. To analyze the model dependence in the predictions 
of the pygmy dipole strength, we employ three different Skyrme parameter sets.
We find that both the isoscalar and the isovector dipole responses of all three 
nuclei show a low-energy peak that increases in magnitude, and is shifted to 
larger excitation energies, with increasing values of the slope of the symmetry 
energy at saturation. We highlight the fact that the collectivity associated 
with the RPA state(s) contributing to this peak is different in the isoscalar 
and isovector case, or in other words it depends on the external probe. While 
the response of these RPA states to an isovector operator does not show a clear 
collective nature, the response to an isoscalar operator is recognizably collective, 
for {\it all} analyzed nuclei and {\it all} studied interactions. 
\end{abstract}

\pacs{21.60.Jz, 24.30.Gd, 21.10.Pc, 21.60.Ev, 21.10.Re, 21.65.Ef, 21.10.Gv, 25.20.-x, 25.60.-t}
 
%  21.10.Gv Nucleon distributions and halo features
%  21.10.Pc Single-particle levels and strength functions
%  21.10.Re Collective levels
%  21.60.Ev Collective models
%  21.60.Jz Nuclear Density Functional Theory and extensions (includes Hartree-Fock and random-phase approximations)
%  21.65.Ef Symmetry energy
%  24.30.Cz Giant resonances 
%  24.30.Gd Other resonances 
%  25.20.-x Photonuclear reactions
%  25.60.-t Reactions induced by unstable nuclei
 
%\keywords{}

\maketitle

% >>>>>>>>>>>>>>>>>>>>>>>>>>>>>>>>>>>>>>>>>>>>>>>>>>>>>>>>>>>>>>>>>>>>
% INTRO.
%
\section{Introduction}
\label{introduction}
Collective phenomena in atomic nuclei have been in the past \cite{bort1998, hara2001} and 
constitute in the present \cite{paar2007} one of the most active and interesting topics of 
research in nuclear physics. Experimental data on giant resonances have allowed us to determine 
fundamental properties  associated with the nuclear interaction in the nuclear medium, such as 
the nuclear incompressibility, the isoscalar effective mass at saturation, and the nuclear symmetry 
energy at some subsaturation density \cite{blai1995, agra2003, colo2004, rute2005,rein1999,trip2008}. 
With the advent of new experimental facilities employing rare isotope beams (RIBs) 
\cite{tanh1995, geis1995, muel2001}, the possibility of studying exotic modes in unstable nuclei is 
nowadays feasible. The low-energy peak present in the isovector dipole response of proton-deficient and 
neutron-rich nuclei, 
the so called Pygmy Dipole Resonance (PDR) \cite{klim2007}, has been experimentally investigated 
in several cases such as ${}^{17-22}$O \cite{leis2001}, ${}^{44,48}$Ca \cite{hart2000,hart2004}, 
${}^{68}$Ni \cite{weil2009}, ${}^{116-120,124,130,132}$Sn \cite{adri2005,fult1969} and ${}^{208}$Pb 
\cite{ryez2002}. In addition, recent extensive theoretical calculations indicate that such a low-energy 
peak is a common property of neutron-rich nuclei \cite{tsun2011}. However, for this PDR, the isoscalar 
character, the collective or single particle nature and the sensitivity to the density dependence of 
the symmetry energy are nowadays under debate 
\cite{klim2007,tsun2011,carb2010,piek2011,rein2010,gamb2011,bara2011,yuks2011}
Hereinafter, we will refer to the low-energy peak present in the isovector dipole response as pygmy 
dipole strength (PDS) instead of the commonly used PDR since the resonant (collective) nature of such a 
peak has not been confirmed yet. 

Such an observable is not only important for nuclear 
structure applications. It also impacts on the determination of reaction rates in the astrophysical 
$r$-process \cite{paar2007}. While some theoretical investigations consider the PDS as a collective 
phenomena \cite{klim2007,tsun2011,carb2010,piek2011}, others ended up with opposite conclusions 
\cite{rein2010}. Within the formers, the PDS is basically understood as a resonant oscillation of the 
neutron skin against the isospin saturated proton-neutron core. In agreement with this picture, 
in Ref.~\cite{carb2010} it has been found that 
a linear relation between the energy weighted sum rule (EWSR) exhausted by the PDS 
and the density dependence of the nuclear symmetry energy exists.
The symmetry energy is a basic property of the nuclear equation of 
state that plays a crucial role in a variety of physical systems. From the very big: the size of a neutron 
star or its composition and structure \cite{latt2001,roca2008}; to the very small: the neutron skin thickness 
of a heavy nucleus \cite{brow2000,cent2009,roca2011}. 
Within the latters, opposite to such a picture, the PDS has been postulated just as a shell 
effect dependent on the particular nucleus under study: in
particular, the authors of Ref. \cite{rein2010} argue that the strong 
fragmentation shown by the strength function within their 
theoretical calculations is indicative of 
such a single particle shell effect \cite{rein2011}. 
As an example of the complexity of the problem, 
in Ref.~\cite{gamb2011} the case of ${}^{48}$Ca have 
been studied in detail. The authors concluded that only 
one of the low-lying excitations within the energy range 5-10 MeV 
can be described as a pygmy resonance.  
Finally, it is also important to mention that within some  
of the works in which the PDS is assumed to be collective, it has been stated that such a low-energy peak 
in the dipole response may, indeed, correspond to a toroidal mode \cite{vret2001, ryez2002, urba2011} or that it 
is weakly dependent on the isovector part of the nuclear effective interaction \cite{bara2011}.

For our study of the PDS on the measured even-even neutron-rich nuclei $^{68}$Ni, $^{132}$Sn and $^{208}$Pb 
that we take as representative of different mass regions, 
we adopt the fully self-consistent non-relativistic mean field (MF) 
approach based on Skyrme Hartree-Fock (HF) plus RPA. 
The MF approach provide a unique framework for the study of 
 all nuclei along the periodic table except the lightest ones. 
Such models typically display a  rather small root mean square (rms) 
deviation on binding energies when compared with  a large set of experimental data
\cite{paar2007,bend2003,vret2005} and are able, through the RPA approach, to predict the main features of 
Giant Resonances \cite{bort1998, hara2001, paar2007}. To assess the sensitivity of our analysis on the nuclear 
model, we employ three Skyrme parameter sets, namely SGII \cite{sgii}, SLy5 \cite{sly5} and SkI3 \cite{ski3}. 
Since the low-energy isovector dipole response of neutron-rich 
nuclei may be related with the density derivative of the symmetry at saturation, the set of chosen models have 
been selected due to the wide range displayed for their predicted values of the $L$ parameter. Such a parameter is 
defined as $L\equiv 3\rho_\infty [\partial c_{\rm sym}(\rho)/\partial \rho]_{\rho_\infty}$ where $c_{\rm sym}(\rho)$ is the 
symmetry energy, $\rho$ is the nucleon density and $\rho_\infty$ is the nuclear saturation density. All the 
studied nuclei are spherical and double-magic. This renders our HF calculations relatively simple and the analysis 
clearer since neither pairing nor deformation should be included. 

First and foremost, we are interested in the 
theoretical study of the main features displayed by the low-energy RPA 
state that give rise to the largest contribution to the PDS or, 
hereinafter, {\it RPA-pygmy} state. In our work, we shall 
investigate the isoscalar or isovector character displayed by the 
transition densities associated to the {\it RPA-pygmy} state, 
and the most relevant particle-hole ($ph$) excitations 
contributing to such a state. In particular, we will 
emphasize that different operators 
will produce a different number of coherent contributions 
from $ph$ amplitudes. This means that in the case
of different experimental probes one will see
the same {\it RPA-pygmy} state with a different associated degree of collectivity.
In the final stage of the preparation of this manuscript we have become aware 
that a similar analysis has been performed in Ref.~\cite{yuks2011}.  

A brief summary of the employed formalism is given in 
Section \ref{formalism} where some  properties of the interactions 
we use are also detailed. In Section \ref{results}, results are presented, analyzed 
and compared with available experimental data. Finally, our conclusions are laid in the last section. 

% >>>>>>>>>>>>>>>>>>>>>>>>>>>>>>>>>>>>>>>>>>>>>>>>>>>>>>>>>>>>>>>>>>>>
% FORMALISM
\section{Formalism}
\label{formalism}
In this section we present the general expression of the Skyrme interaction 
as well as some basic properties of the parametrizations used in our analysis. A brief 
description of the RPA formalism is also presented. We address the reader to 
Refs.~\cite{sly5, vaut1972, bein1975} for further details on the Skyrme interaction. 

\subsection{Skyrme interaction}
The Skyrme interaction is a zero-range, velocity-dependent interaction that describe nucleons 
with space, spin and isospin variables ${\bf r}_i, \bsigma_i$ and $\btau_i$. It is commonly written 
as in Ref.~\cite{sly5},  
\begin{eqnarray}
V({\bf r}_1, {\bf r}_2) &=& t_0(1+x_0P_{\sigma}) \delta({\bf r})\nonumber\\
&+& \frac{1}{2} t_1 (1+x_1P_{\sigma})[{\bf P}'^{2}\delta({\bf r})+
\delta({\bf r}){\bf P}^{2}] \nonumber \\
&+& t_2(1+x_2P_{\sigma}){\bf P}'\cdot\delta({\bf r}){\bf P}\nonumber\\ 
&+&\frac{1}{6} t_3 (1+x_3P_{\sigma})\rho^{\alpha}({\bf R})\delta({\bf r})\nonumber\\
&+& iW_0 ({\bsigma}_1+{\bsigma}_2)\cdot[{\bf P}'\times\delta({\bf r}){\bf P}]~,
\label{potential}
\end{eqnarray}
where ${\bf r} = {\bf r}_1-{\bf r}_2$, ${\bf R} =\frac{1}{2} ({\bf r}_1+{\bf r}_2)$,
${\bf P} = \frac{1}{2i}({\bf \nabla}_1-{\bf \nabla}_2)$, ${\bf P'}$ is the hermitian conjugate
of ${\bf P}$ (acting on the left), $P_{\sigma}=\frac{1}{2}(1+{\bsigma}_1\cdot{\bsigma}_2)$
is the spin-exchange operator. 

As mentioned already in the Introduction, we 
employ three Skyrme interactions: SGII \cite{sgii}, SLy5 \cite{sly5} 
and SkI3 \cite{ski3}; as many others, they
have been accurately calibrated in order to reproduce some bulk 
properties (the binding energies and charge radii) of few selected stable 
nuclei, as well as some empirical nuclear 
matter properties such as the saturation energy and the saturation density 
(and others depending on the specific set).
Throughout this work, we are mainly interested on the sensitivity of the PDS to 
the density derivative of the symmetry energy at saturation \cite{carb2010}. 
Since SGII, SLy5 and SkI3 are characterized by $L$ 
equal to, respectively, 37.63 MeV, 48.27 MeV and 100.52 MeV, 
they span a quite broad range (comparable with the one
spanned by most of the modern and commonly used MF models available in the literature
\cite{cent2009, roca2011}).

\subsection{Random Phase Approximation}
The discrete RPA method is well-known from textbooks \cite{ring1980,rowe1980}. 
In our self-consistent approach, we build the residual interaction ($V_{\rm residual}^{qq^{\prime}}$) for the proton-proton ($qq^{\prime}$=$pp$),
neutron-neutron ($qq^{\prime}$=$nn$) and proton-neutron ($qq^{\prime}$=$pn$) 
channels  from the Skyrme-HF energy density 
functional, namely $V_{\rm residual}^{qq^{\prime}}\equiv 
\delta^2 E_{\rm HF}/\delta\rho_q\delta\rho_{q^\prime}$. Then
we solve fully self-consistently the RPA equations 
by means of the matrix formulation like in Refs. 
\cite{frac2005,tsil2006}.  One should note that the
continuum is discretized by setting the system in a large box.

For any operator $\hat F_{JM}$ the (reduced) transition strength 
or probability is given by
\begin{eqnarray}
B(EJ, \tilde 0 \rightarrow \nu) &=&
\left\vert\langle \nu \vert\vert \hat F_{J}
\vert\vert \tilde{0} \rangle \right\vert^2 \nonumber\\
&=&\left\vert \sum_{ph}
\left( X_{ph}^{(\nu)}
+ Y_{ph}^{(\nu)} \right) \langle p \vert\vert \hat F_{J}
\vert\vert h \rangle \right\vert^2,
\label{rts}
\end{eqnarray}
where $\langle \nu \vert\vert \hat F_{J} \vert\vert \tilde{0} 
\rangle$ is the reduced 
matrix element of $\hat F_{JM}$ (see, e.g., Ref. \cite{bohr1969}). 
The initial 
state in all studied nuclei, $\vert\tilde{0}\rangle$, correspond to the RPA ground-state 
with zero total angular momentum and $\vert\nu\rangle$ stands for a generic RPA excited state. 
The latter equation is also written in an alternative notation that
will turn out to be useful for our present 
purposes. That is, each RPA transition $\vert\tilde{0}\rangle \rightarrow \vert\nu\rangle$ 
excited via $\hat F_{JM}$ is composed by all considered particle-hole ($ph$) pairs that couple 
to a total angular momentum $JM$. The relative contribution of each $ph$ excitation to the reduced 
matrix element $\langle \nu \vert\vert \hat F_{J} \vert\vert \tilde{0} \rangle$ is accounted
by the $X_{ph}^{(\nu)}$ and $Y_{ph}^{(\nu)}$ RPA amplitudes
that specify a given eigenvector of the RPA secular matrix 
\cite{ring1980}. For the analysis of the 
single particle or collective character of a given excitation in the 
response function, it is convenient to write the reduced amplitude as follows:      
\begin{equation}
A_{ph}(EJ,\tilde 0 \rightarrow \nu)=\left( X_{ph}^{(\nu)}
+ Y_{ph}^{(\nu)} \right) \langle p \vert\vert \hat F_{J}
\vert\vert h \rangle .
\label{ampl}
\end{equation}
This is because Eq.~(\ref{ampl}) 
allows one to determine the coherency (relative sign) and magnitude 
($\vert A_{ph}(EJ,\tilde 0 \rightarrow \nu)\vert$) of all the $ph$ 
contributions to the reduced transition probability. 
An RPA state
is claimed to be a resonant excitation if the corresponding reduced amplitude 
is composed by several $ph$ excitations 
similar in magnitude and adding coherently.  

The strength function is defined as usual, 
\begin{equation}
S(E) = \sum_\nu \vert \langle \nu \vert\vert \hat F_{J}
\vert\vert \tilde 0 \rangle \vert^2 \delta(E-E_\nu),
\label{strength}
\end{equation}
where $E_\nu$ is the eigenenergy associated to the RPA state 
$\vert \nu \rangle$.
Its moments can be calculated as 
\begin{equation}
m_k = \int dE\ E^k S(E) = \sum_\nu \vert \langle \nu \vert\vert
\hat F_{J} \vert\vert \tilde 0 \rangle \vert^2 E_\nu^k.
\end{equation}

Another quantity of interest that characterizes the relationship of each excited 
state with the ground state, is the transition density. 
Its integral with a multipole 
operator gives the corresponding transition amplitude 
of that operator. With the 
help of the $X^{(\nu)}$ and $Y^{(\nu)}$ amplitudes of a 
given RPA state $\vert \nu\rangle$,
one can construct the radial part of its transition density defined by
$\delta\rho_\nu( {\bf r})\equiv \langle \nu \vert \hat\rho( {\bf r}) \vert \tilde
0 \rangle = \delta\rho_\nu(r) Y^*_{JM}(\hat r)$ as follows, 
\begin{eqnarray}
\delta\rho_\nu(r) &=& \frac{1}{\sqrt{2J+1}} \sum_{ph} \left( X_{ph}^{(\nu)} + Y_{ph}^{(\nu)} \right)\nonumber\\ 
& &\times \langle p \vert\vert Y_J \vert\vert h \rangle
\frac{u_p(r)u_h(r)}{r^2}~,
\label{td}
\end{eqnarray}
where $u_\alpha(r)$ is the solution to the Skyrme-HF radial equations 
corresponding to the 
single particle state $\alpha$. 
Note that the summations in the expression above can be done 
for neutrons or protons separately. This allows one to calculate the neutron and proton 
transition densities $\delta\rho_{\nu_q}(r)$ ($q=n,p$) and 
define accordingly the isoscalar (IS) and 
isovector (IV) transition densities as
\begin{eqnarray}
\delta\rho_\nu^{\rm (IS)}(r) & \equiv &  \delta\rho_{\nu_n}(r) + \delta\rho_{\nu_p}(r) \quad {\rm and}\\
\delta\rho_\nu^{\rm (IV)}(r) & \equiv &  \delta\rho_{\nu_n}(r) - \delta\rho_{\nu_p}(r).
\label{is-iv-td}
\end{eqnarray}
The interest of the transition densities
relies on the fact that their spatial shape reveal the nature of the 
excitations: volume or surface type, isoscalar or isovector, etc. Moreover, 
they can be used as input in calculations of inelastic scattering cross 
sections.  More details of our implementation of RPA can be
found in Ref. \cite{colo2011}.

As our theoretical study will be devoted to the low energy dipole response in even-even 
neutron-rich nuclei, we define the isoscalar (IS) and isovector
(IV) dipole operators ($J=1$) used for the different calculations:     
\begin{equation}
\hat F_{1M}^{\rm (IS)} = \sum_{i=1}^A r^3_i Y_{1M} (\hat r_i),
\label{is_op}
\end{equation}
\begin{equation}
\hat F_{1M}^{\rm (IV)} = \sum_{i=1}^A r_i Y_{1M} (\hat r_i) \tau_z(i).
\label{iv_op}
\end{equation}
Note that the lowest order term in the IS dipole operator 
coming from the expansion 
of the Bessel functions ($r$) does not reproduce a physical excitation but a translation of the 
whole system. This is the reason why the IS dipole operator 
in Eq.~(\ref{is_op}) is 
proportional to the following term ($r^3$) 
in such an expansion. 

\begin{figure}[h!]
\includegraphics[clip=true,width=0.9\linewidth]{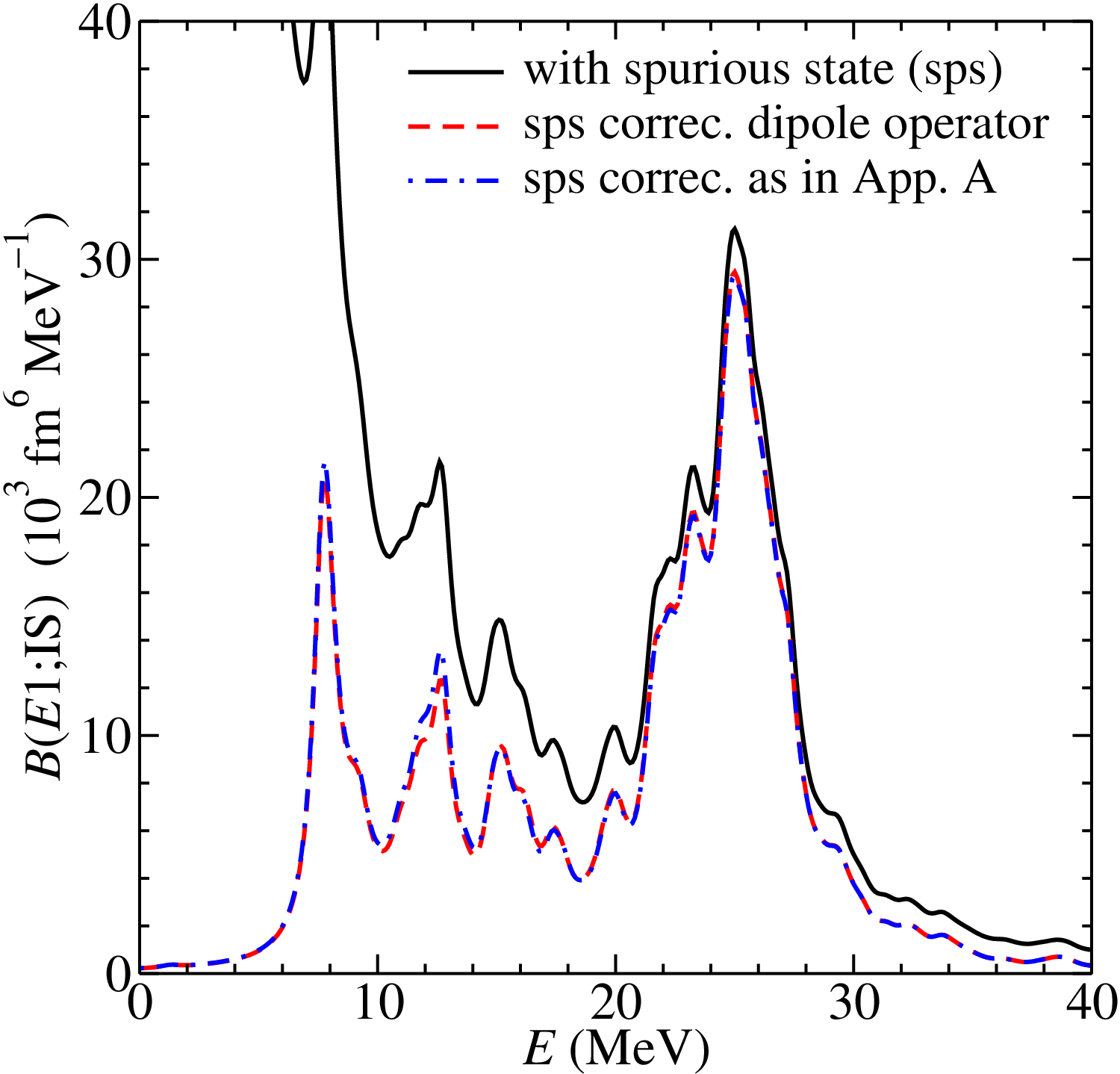}\\
\caption{(Color online) 
Strength function in the case of the SLy5 interaction 
\cite{sly5} for a test nucleus (${}^{208}$Pb) as a function of the excitation energy for three cases: 
(i) the spurious state has not been subtracted (solid line), (ii) the spurious state has been subtracted 
by correcting the isoscalar dipole operator Eq.~(\ref{is_op}) (dashed line), and (iii) the spurious state 
has been subtracted as explained in the Appendix \ref{spurious} (dot-dashed line).} 
\label{fig1}
\end{figure}

The translational mode should in 
principle be decoupled from the physical excitations within the RPA. 
However,  as in any numerical implementation, 
the decoupling is not perfect. Therefore, part of this state, 
known as the spurious state,  overlaps with the 
physical RPA states. 
There are different ways to correct this overlap 
\cite{ring1980, kazu2011}. Our prescription is detailed in 
Appendix \ref{spurious}, 
where we show how one can subtract the spurious state from the neutron and 
proton transition densities. The reliability of our method can be seen in Fig.~\ref{fig1} 
where we compare the strength function ---calculated by convoluting the corresponding reduced transition 
probability of (Eq.~\ref{rts}) with a Lorenzian of 1 MeV width--- for the isoscalar dipole response predicted 
by the SLy5 interaction \cite{sly5} for a test nucleus (${}^{208}$Pb)
in three cases: in one case the spurious state has not 
been subtracted (solid curve), in the second case 
the spurious state has been subtracted 
by correcting the isoscalar dipole 
operator Eq.~(\ref{is_op}) with  the addition of a term 
$-\eta r_iY_{1M} (\hat r_i)$ where 
$\eta = 5\langle r^2 \rangle/3$ like in Ref. \cite{giai1981} 
(dashed line), and finally in the last case 
the spurious state has been subtracted 
as explained in Appendix \ref{spurious} (dot-dashed line). 

From this figure one clearly sees that 
the different prescriptions for correcting the spurious state 
are  completely equivalent. The advantage of our method
relies on the fact that, by construction, we exactly subtract the spurious state from the neutron and proton 
transition densities, and these are among the quantities that we discuss in detail below.  

In the case of the IV dipole operator, we have to subtract the dipole motion associated 
with the displacement of the neutron and proton center-of-mass. It can be done by modifying 
the operator (\ref{iv_op}) as follows, 
\begin{equation}
\hat F_{1M}^{\rm (IV)} = \frac{Z}{A} \sum_{n=1}^N r_n Y_{1M}(\hat r_n) -
\frac{N}{A} \sum_{p=1}^Z r_p Y_{1M}(\hat r_p).
\label{iv_dipo_op}
\end{equation}

Last but not least, calculations beyond RPA may of course
to some extent change the quantitative picture. It is known that
correlations associated with coupling with 2 particle-
2 hole (2p-2h) configurations, or particle-phonon
configurations, tend to shift the strength downward and increase
its fragmentation. However, at least in the cases that have
been studied in Refs.~\cite{sarchi,litvinova}, these changes
do not destroy the qualitative features associated with the 
PDS, namely the isospin character, the overall behavior 
of the transition densities and the collective or single-particle
character of the states. We leave aside in our study the
debated case of Ca isotopes discussed in Refs. 
\cite{gamb2011,tert2007,papa2011}. 
                 
% >>>>>>>>>>>>>>>>>>>>>>>>>>>>>>>>>>>>>>>>>>>>>>>>>>>>>>>>>>>>>>>>>>>>
% RESULTS
%
\section{Results}
\label{results}

In this Section we present a detailed study of the dipole response of 
$^{68}$Ni, $^{132}$Sn and $^{208}$Pb as predicted by the Skyrme interactions 
SGII, SLy5 and SkI3 
within the formalism described in Section \ref{formalism}. The results are organized 
as follows. We present for each nucleus the isoscalar and isovector dipole strength 
functions. 

In order to have a  simple estimate for the collectivity displayed by the different 
dipole responses, we plot also the reduced transition probabilities in single particle 
units (s.p.u., or 
Weisskopf units \cite{bohr1969}). Such unit is based 
on a macroscopic approach.  One evaluates the average 
transition rate of a typical excitation in terms 
of the angular momentum carried by the probe, and the mass radius 
of the nucleus under analysis; in this way, the result
is nucleus-independent. By following 
Ref.~\cite{bohr1969}, one can calculate the isovector dipole response in Weisskopf units 
accounting for the center of mass correction as,
\begin{equation}
B_{\rm W}^{\rm (IV)} (E1) =\frac{3^3 R^2}{4^3\pi}\times
   \left\{ \begin{array}{rl} ~\left(\frac{N}{A}\right)^2 &{\rm for~protons}\\
                                                         &                 \\  
                           \left(-\frac{Z}{A}\right)^2   &{\rm for~neutrons}
           \end{array} \right. 
\label{estim-rtp-iv}
\end{equation} 
where the radius $R$ is taken to be $r_0 A^{1/3}$ 
and $r_0$ is the  radius of the average sphere that one nucleon 
occupies, at the standard 
saturation density of 0.16 fm${}^{-3}$ (that is, $r_0=1.14$ fm). 
For the isoscalar dipole case, once the spurious state
has been subtracted as explained in the previous section, one finds
\begin{equation}
B_{\rm W}^{\rm (IV)} (E1) =\frac{3}{4\pi}\left(\frac{1}{2}R^3-\eta\frac{3}{4}R\right)^2
=\frac{3R^6}{4^3\pi}
\label{estim-rtp-is}
\end{equation}
with $\eta=\frac{5}{3}\langle r^2 \rangle$ 
and $\langle r^2 \rangle=\frac{3}{5}R^2$. The sum over all 
nucleons in Eqs.~(\ref{estim-rtp-iv}) and 
(\ref{estim-rtp-is}) coincides in good approximation 
(around 10\%$-$20\%) with the corresponding total RPA strength, although 
there exist some mass dependence: heavier nuclei are better reproduced by the 
estimate provided by the Weisskopf units. Such a unit allows us to account {\it qualitatively} for the nature of different 
excitations since a given RPA state will contribute with several single particle units if it is collective. 
Moreover, it also enables the comparison between the results obtained 
for different nuclei. 

Then, we focus on the low-energy region in order to 
investigate the {\it RPA-pygmy} state leading to the PDS. We show, first, the neutron 
and proton transition densities associated with such a state for each interaction and 
nucleus. The analysis of the transition densities may be very illustrative since 
they allow one to distinguish some spatial details related to the dynamics of every 
excitation. For example,  one could understand 
if either nucleons from the surface or from the interior of the nucleus 
are contributing more to the excitation, and this is
crucial to estimate which reaction is more efficient
in exciting this mode. Besides this, the comparison 
between the neutron and proton transition densities informs us about the relative motion 
of neutrons with respect to protons, or in other words, on the isoscalar or isovector character 
of each RPA state. 

 For this aim, we also 
use a local criterion to study quantitatively the isoscalar 
and isovector splitting of the {\it RPA-pygmy} state \cite{paar2009} based on the following. 
At each radial distance $r_{i}$, where $i=1 \ldots N$ at which the neutron and 
proton transition densities are calculated, we define that a 
certain RPA state is 70\% isoscalar if at least the 70\% of the calculated points fulfill the condition 
$\vert \delta\rho_\nu^{\rm (IS)}(r)\vert > \vert \delta\rho_\nu^{\rm (IV)}(r)\vert$. Moreover, we will exploit 
the possibility of analyzing the isoscalar or isovector nature of the {\it RPA-pygmy} state in 
different regions of the nucleus. Specifically, we impose the above defined criteria of 
isoscalarity in two additional regions: one in 
the internal part of the nucleus, {\it i.e.} from 0 fm 
to R/2 and the other in external part of the nucleus, namely from R/2 to R. 

Finally, we analyze the most relevant particle-hole ($ph$) excitations contributing 
to the {\it RPA-pygmy} state. To this end, we calculate the magnitude and sign (coherency) 
with which each $ph$ excitation contribute to the isovector (IV) and isoscalar (IS) dipole 
reduced transition probability ($B(E1;\xi)$ where $\xi=$IV or IS). For that, we have used the 
isoscalar $A_{ph}^q(E1;\xi=IS)$ and isovector $A_{ph}^q(E1;\xi=IV)$ reduced amplitudes defined in 
Eq.~(\ref{ampl}). 

%
% >>>>>>>>>>>>>>>>>>>>>>>>>>>>>>>>>>>>>>>>>>>>>>>>>>>>>>>>>>>>>>>>>>>>
%
\subsection{Strength functions in $^{68}$Ni, $^{132}$Sn and $^{208}$Pb}

In this subsection we analyze the main features displayed by the strength function 
Eq.~(\ref{strength}) associated to the isovector and isoscalar dipole response of 
$^{68}$Ni, $^{132}$Sn and $^{208}$Pb. It has been calculated by convoluting the 
corresponding reduced transition probability, Eq.~(\ref{rts}), with a Lorenzian of 
1 MeV width. The low-energy dipole response of all studied nuclei 
has been measured \cite{ryez2002, adri2005, weil2009}. 

\begin{figure}[h!]
\includegraphics[clip=true,width=0.9\linewidth]{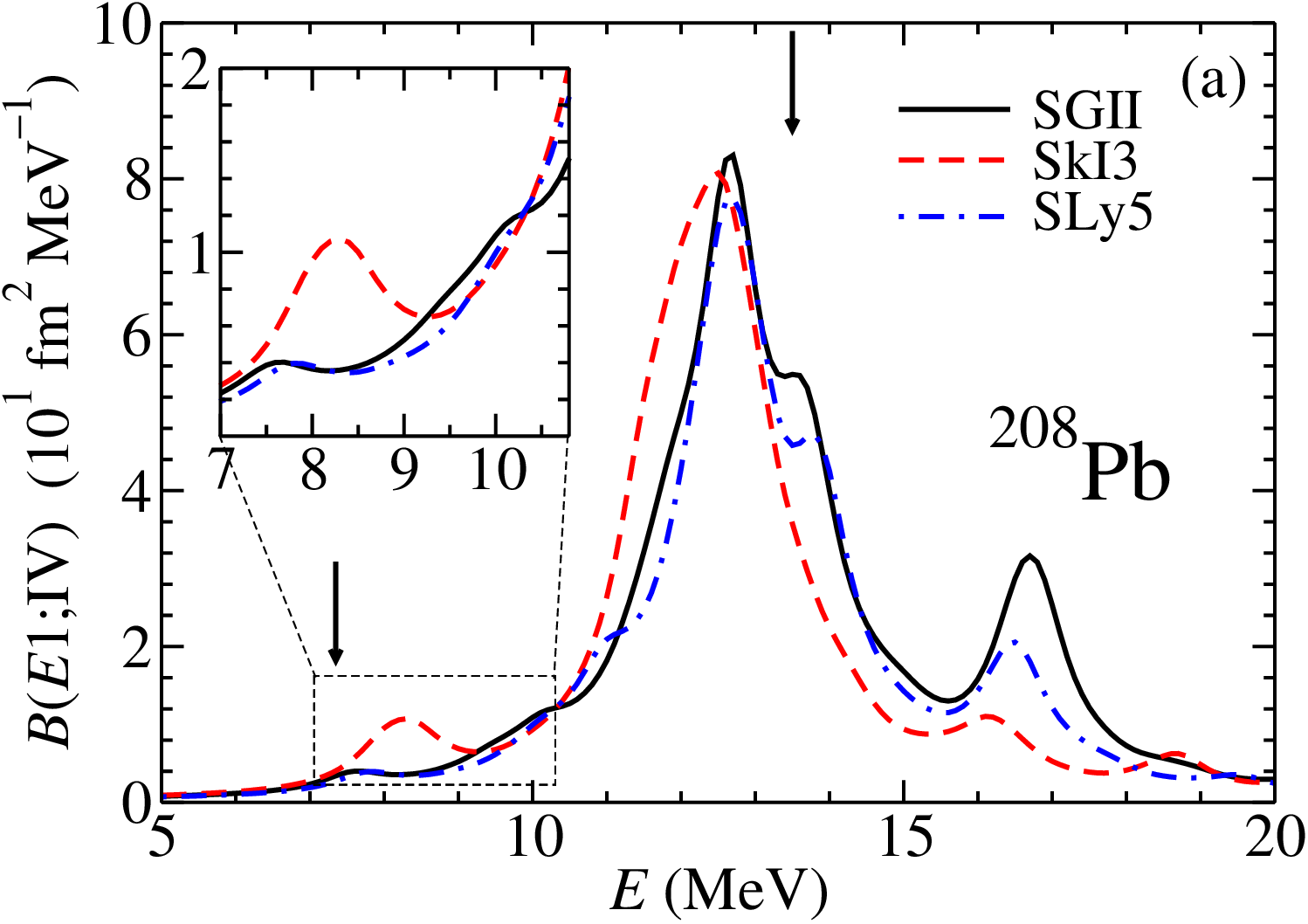}\\
\includegraphics[clip=true,width=0.9\linewidth]{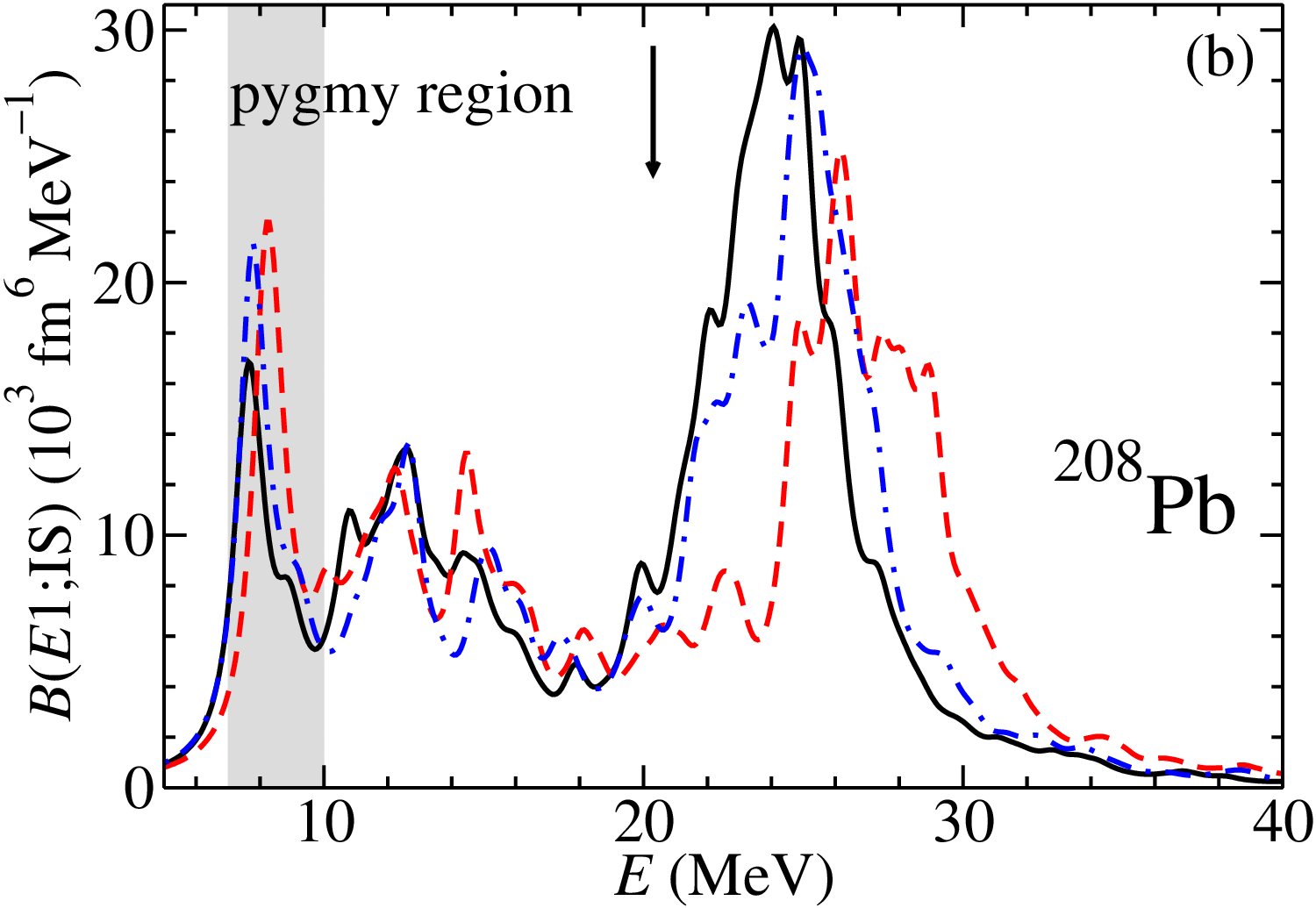}
\caption{(Color online) Strength function corresponding to the isovector (a) and isoscalar (b) 
dipole response of ${}^{208}$Pb as a function of the excitation energy. The inset in 
(a) displays in a larger scale the pygmy region. In both figures the predictions of 
SGII, SLy5 and SkI3 are depicted. Black arrows indicate the experimental 
centroid energies for the PDS ($E=7.37$ MeV within a window of $6-8$ MeV) 
\cite{ryez2002}, for the ISGDR ($E=20.3\pm 2$ MeV \cite{garg1999}) and the 
energy peak for the IVGDR ($E=13.43$ MeV and a total width of $2.42$ 
MeV \cite{berm1975}).} 
\label{fig2}
\end{figure}
\begin{table}[h!]
\begin{center}
\caption{Excitation energy $E$ 
and isoscalar ($\xi=IS$) and isovector ($\xi=IV$) reduced transition 
probabilities $B(E1;\xi)$ 
corresponding to the {\it RPA-pygmy} states of ${}^{68}$Ni, ${}^{132}$Sn and 
${}^{208}$Pb as predicted by SGII, SLy5 and SkI3 interactions.}
\label{rpapygmy}
\begin{tabular}{rcccc}
\hline\hline
           &force& $E$ &$B(E1;IS)$ & $B(E1;IV)$ \\
           &     &[MeV]& [fm${}^6$]& [fm${}^2$] \\
\hline
           & & & & \\
           & & & & \\
${}^{68}$Ni & SGII & ~9.77 & 1.9$\times 10^{3}$ & 1.4 \\
           & SLy5 & ~9.30 & 1.7$\times 10^{3}$ & 0.8 \\
           & SkI3 &10.45 & 3.0$\times 10^{3}$ & 3.6 \\
\hline
           & & & & \\
${}^{132}$Sn&SGII & ~8.52 & 3.3$\times 10^{3}$ & 1.2 \\
           &SLy5 & ~8.64 & 1.0$\times 10^{4}$ & 1.6 \\
           &SkI3 & ~9.23 & 1.1$\times 10^{4}$ & 7.4 \\
\hline
${}^{208}$Pb&SGII & ~7.61 & 1.7$\times 10^{4}$ & 2.9 \\
           &SLy5 & ~7.74 & 2.8$\times 10^{4}$ & 2.8 \\
           &SkI3 & ~8.01 & 1.9$\times 10^{4}$ & 6.6 \\
\hline\hline
\end{tabular}
\end{center}
\end{table}

We start analyzing the results for ${}^{208}$Pb. In Fig.~\ref{fig2}a we show the strength function 
corresponding to the isovector dipole response as a function of the excitation energy. The inset 
displays in a larger scale the pygmy region. In Fig.~\ref{fig2}b, the same quantities are shown 
but this time for the isoscalar dipole response as a function of the excitation energy. In both 
figures, the predictions of the three selected interactions are shown. 
The centroid energies of the PDS and the Isoscalar Giant Dipole Resonance (ISGDR) 
as well as the energy peak in the Isovector Giant Dipole Resonance (IVGDR)   
for ${}^{208}$Pb as predicted by the employed interactions ($E=7.6-8.0$ MeV, $E=20-21$ MeV 
and $E=12-13$ MeV, respectively) fairly agree with the experimental data 
($E=7.37$ MeV within a window of $6-8$ MeV \cite{ryez2002}, $E=20.1-20.5$ MeV \cite{garg1999} 
and $E=13.43$ MeV and a total width of $2.42$ 
MeV \cite{berm1975}, respectively). Consequently, 
the RPA predictions of SGII, SLy5 and SkI3 may allow us to elucidate the microscopical 
structure and properties of the PDS. In Table \ref{rpapygmy} the excitation energy and 
isoscalar and isovector reduced transition probabilities of the {\it RPA-pygmy} state 
---{\it i.e.} the RPA state which give rise to the largest peak in the PDS region--- are 
detailed for all the sudied nuclei as predicted by SGII, SLy5 and SkI3. In the case of 
${}^{208}$Pb we find an excitation energy of $E = 7.61$ MeV for SGII, $E = 7.74$ MeV for 
SLy5 and $E = 8.01$ MeV for SkI3. We qualitatively observe 
that the low-energy peak found in the IV and IS dipole responses of ${}^{208}$Pb shows an 
increasing and outward trend with the excitation energy as the value of the parameter $L$ 
increases. This behavior is in agreement with Ref.~\cite{carb2010} where the energy weighted 
sum rule or $m_1$ for the PDS was found to be linearly correlated with $L$ in mean-field models. 

\begin{figure}[t]
\includegraphics[clip=true,width=0.9\linewidth]{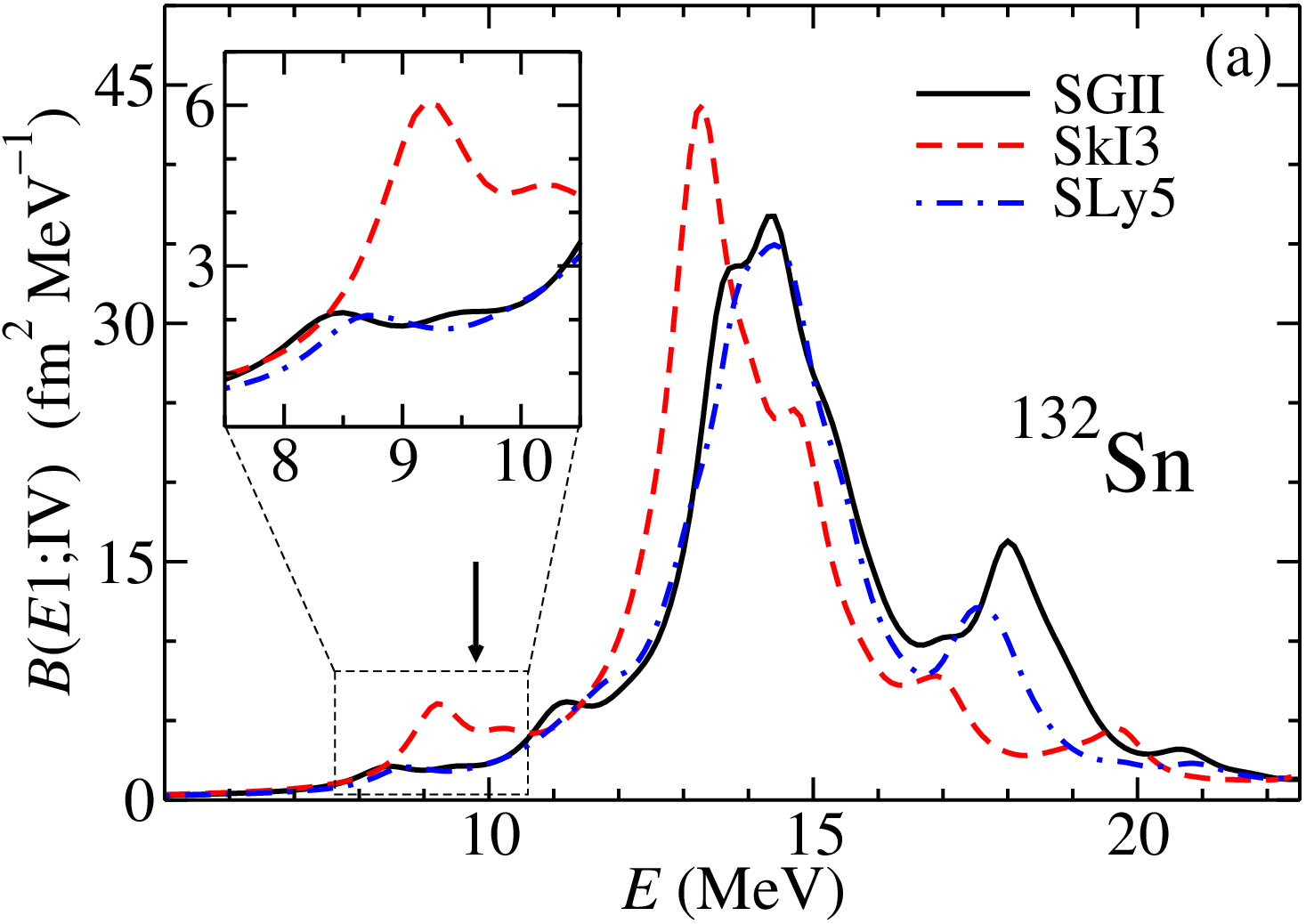}\\
\includegraphics[clip=true,width=0.9\linewidth]{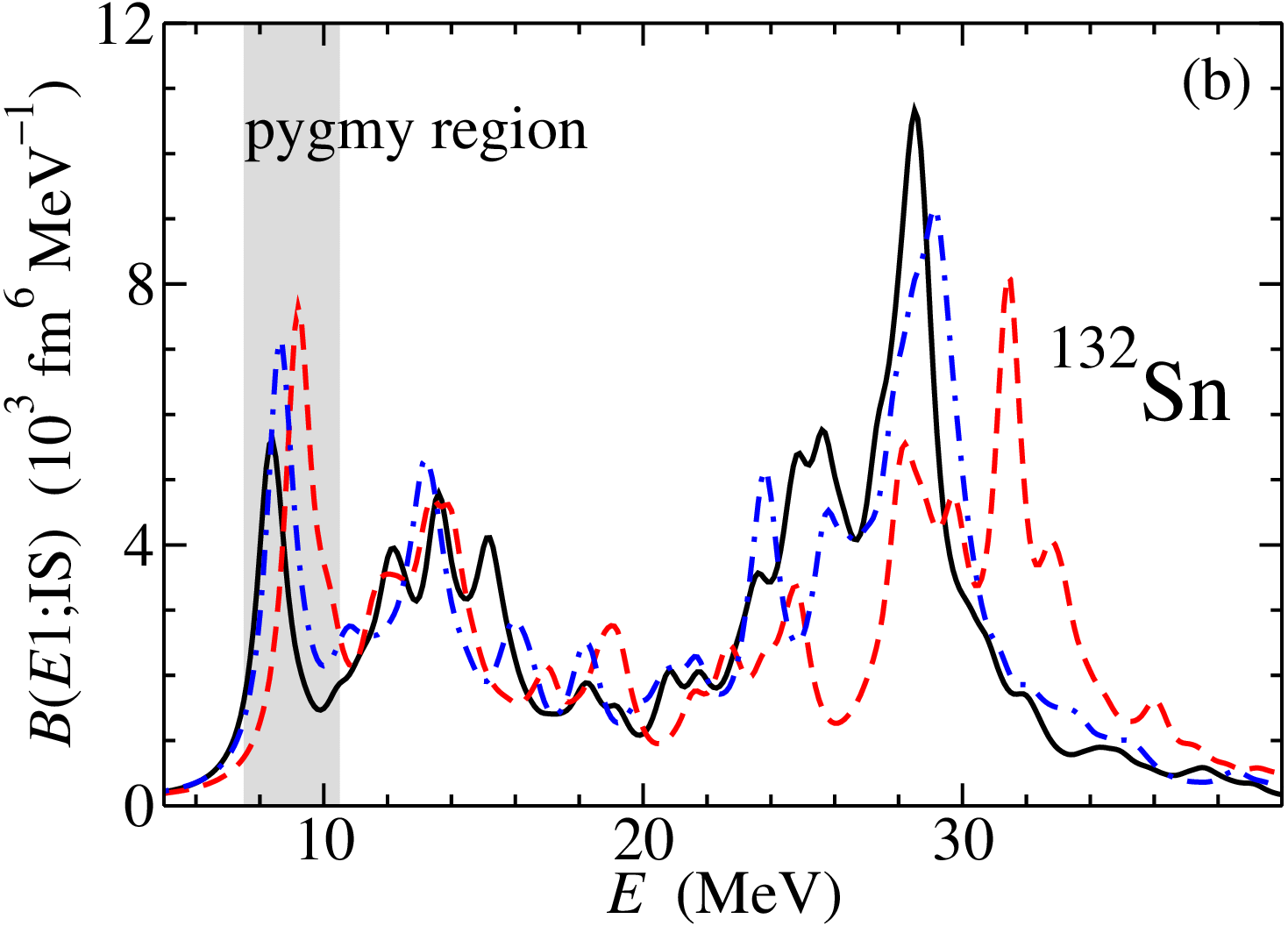}
\caption{(Color online) Same as Fig.~\ref{fig2} for ${}^{132}$Sn. The experimental 
value for the peak energy of the PDS ($E=9.8\pm 0.7$ MeV) is indicated by a 
black arrow \cite{adri2005}.} 
\label{fig3}
\end{figure}
\begin{figure}[t]
\includegraphics[clip=true,width=0.9\linewidth]{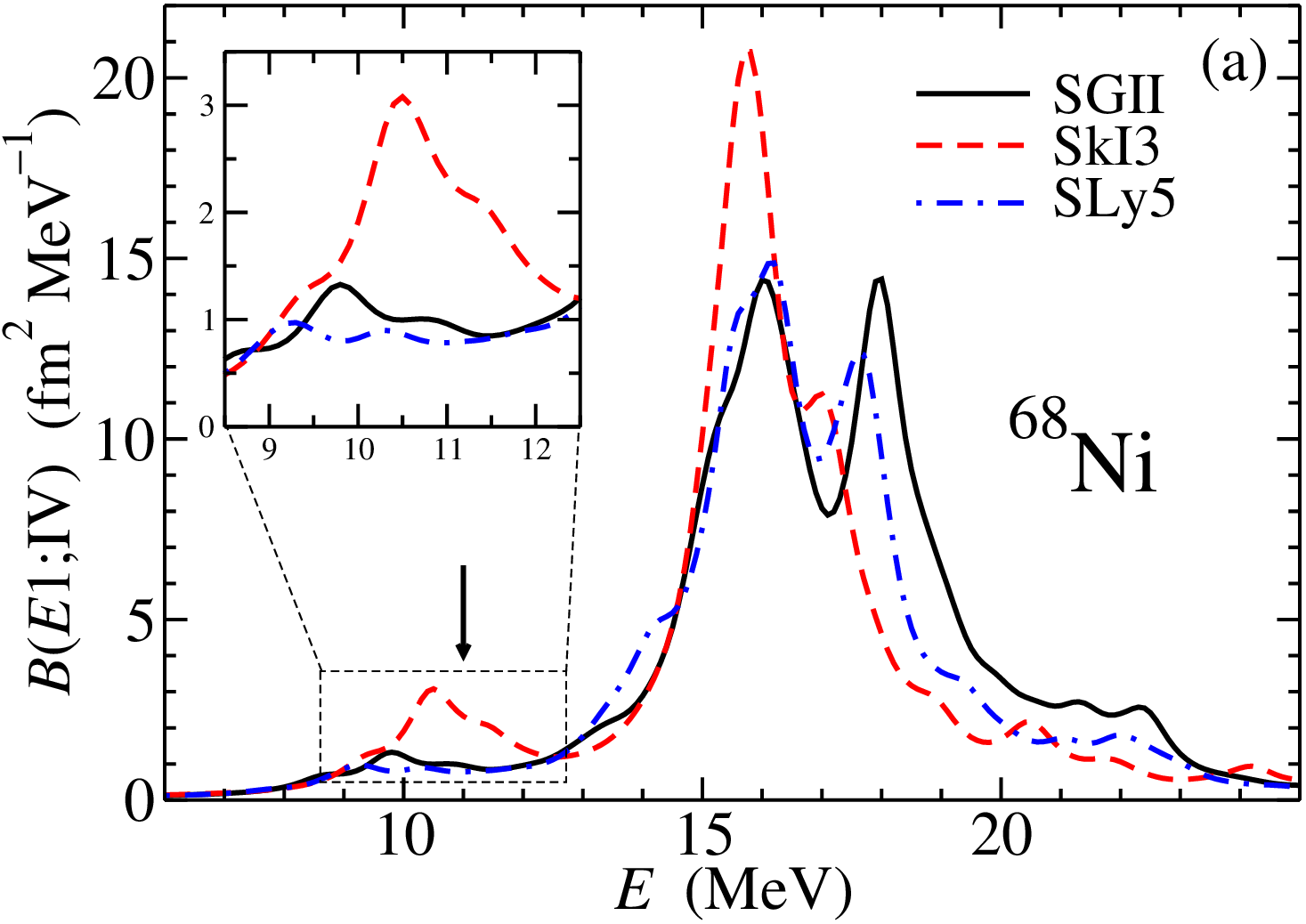}\\
\includegraphics[clip=true,width=0.9\linewidth]{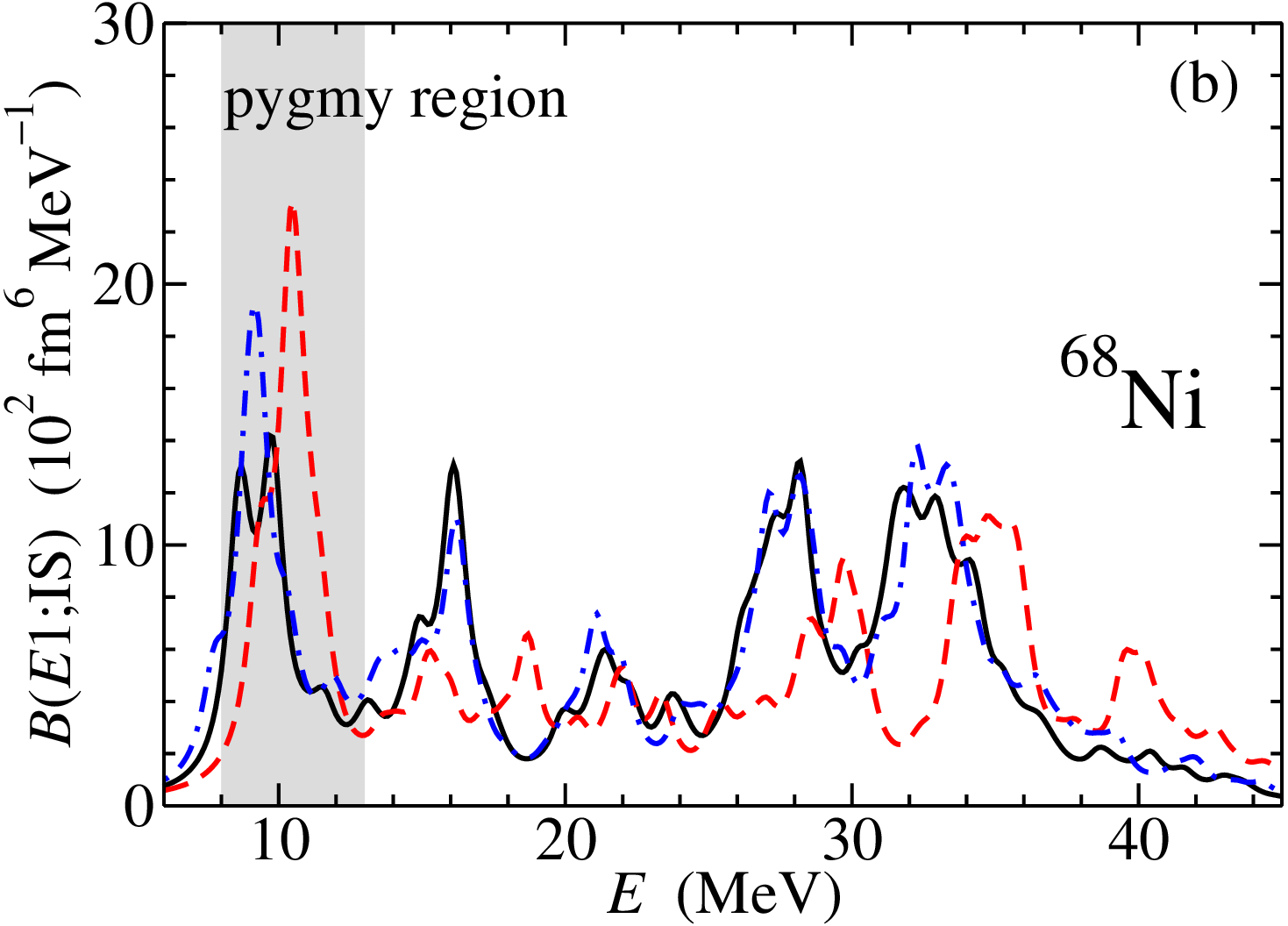}
\caption{(Color online) Same as Fig.~\ref{fig2} for ${}^{68}$Ni. The experimental 
value for the peak energy of the PDS ($E=11$ MeV and an energy width estimated to 
be less than 1 MeV) is indicated by a black arrow \cite{weil2009}.} 
\label{fig4}
\end{figure}

In the case of ${}^{132}$Sn and ${}^{68}$Ni, the strength functions for the  
dipole response are depicted in Figs.~\ref{fig3}a and \ref{fig4}a (IV) and 
Figs.~\ref{fig3}b and \ref{fig4}b (IS), respectively. 
Again, the predictions of SGII, SLy5 and SkI3 ($E=8.5-9.2$ MeV for ${}^{132}$Sn and $E=9.3-10.4$ 
MeV for ${}^{68}$Ni) are in rather good agreement with the 
measured data ($E=9.1-10.5$ MeV for ${}^{132}$Sn \cite{adri2005} and $E=11$ MeV and an energy width estimated to 
be less than 1 MeV for ${}^{68}$Ni \cite{weil2009}). In the case of ${}^{132}$Sn, the {\it RPA-pygmy} state predicted by SGII 
correspond to a state with an excitation energy of $E = 8.52$ MeV while SLy5 predicts 
$E = 8.64$ MeV and SkI3 $E = 9.23$ MeV. And for ${}^{68}$Ni, the values predicted by SGII, 
SLy5 and SkI3 for the excitation energy of the {\it RPA-pygmy} state are respectively: 
$E = 9.77$ MeV, $E = 9.30$ MeV and $E = 10.45$ MeV. Qualitatively in both nuclei, it 
seems again that the larger the value of $L$, the higher 
the values predicted for the 
excitation energy and the larger the different peaks arising in the low-energy region 
(see Figs.~\ref{fig3} and \ref{fig4}). In addition, we observe for all nuclei that the PDS 
is an order of magnitude smaller than the IVGDR and that its isoscalar counterpart is of 
the same order of magnitude than the corresponding ISGDR.    
 
%
% >>>>>>>>>>>>>>>>>>>>>>>>>>>>>>>>>>>>>>>>>>>>>>>>>>>>>>>>>>>>>>>>>>>>
%

%
\begin{figure*}[t]
\includegraphics[clip=true,width=0.34\linewidth]{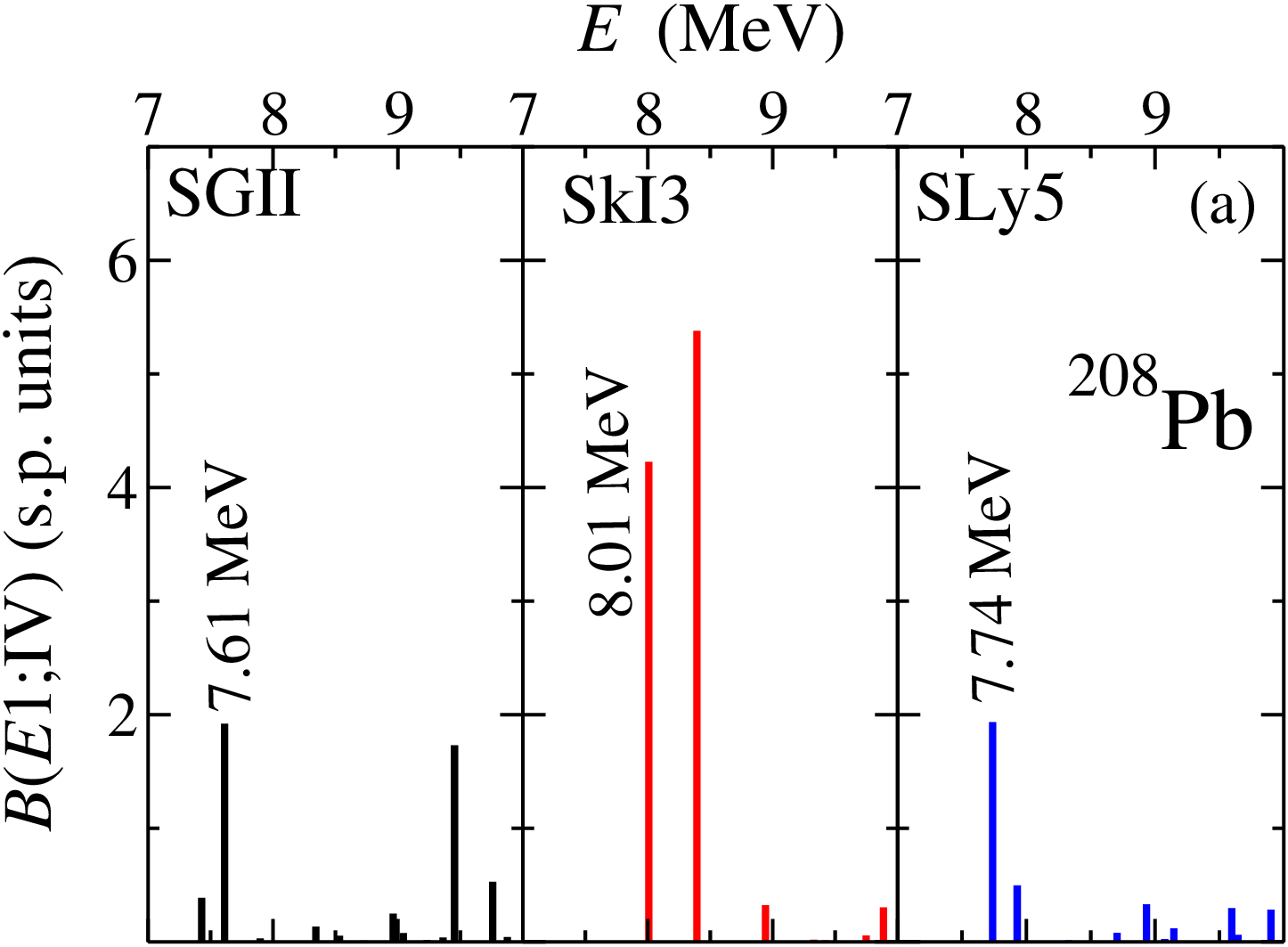}
\includegraphics[clip=true,width=0.30\linewidth]{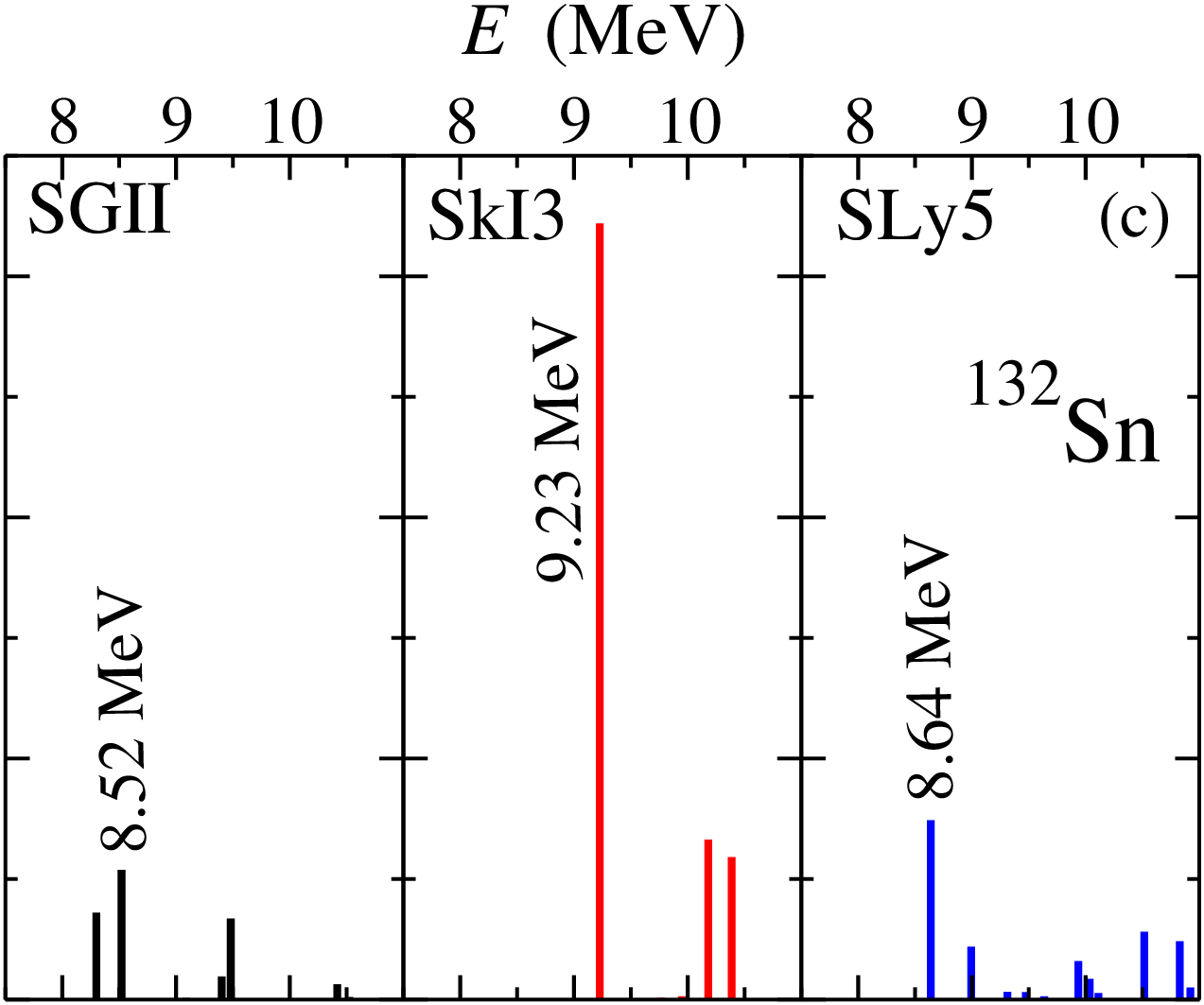}
\includegraphics[clip=true,width=0.34\linewidth]{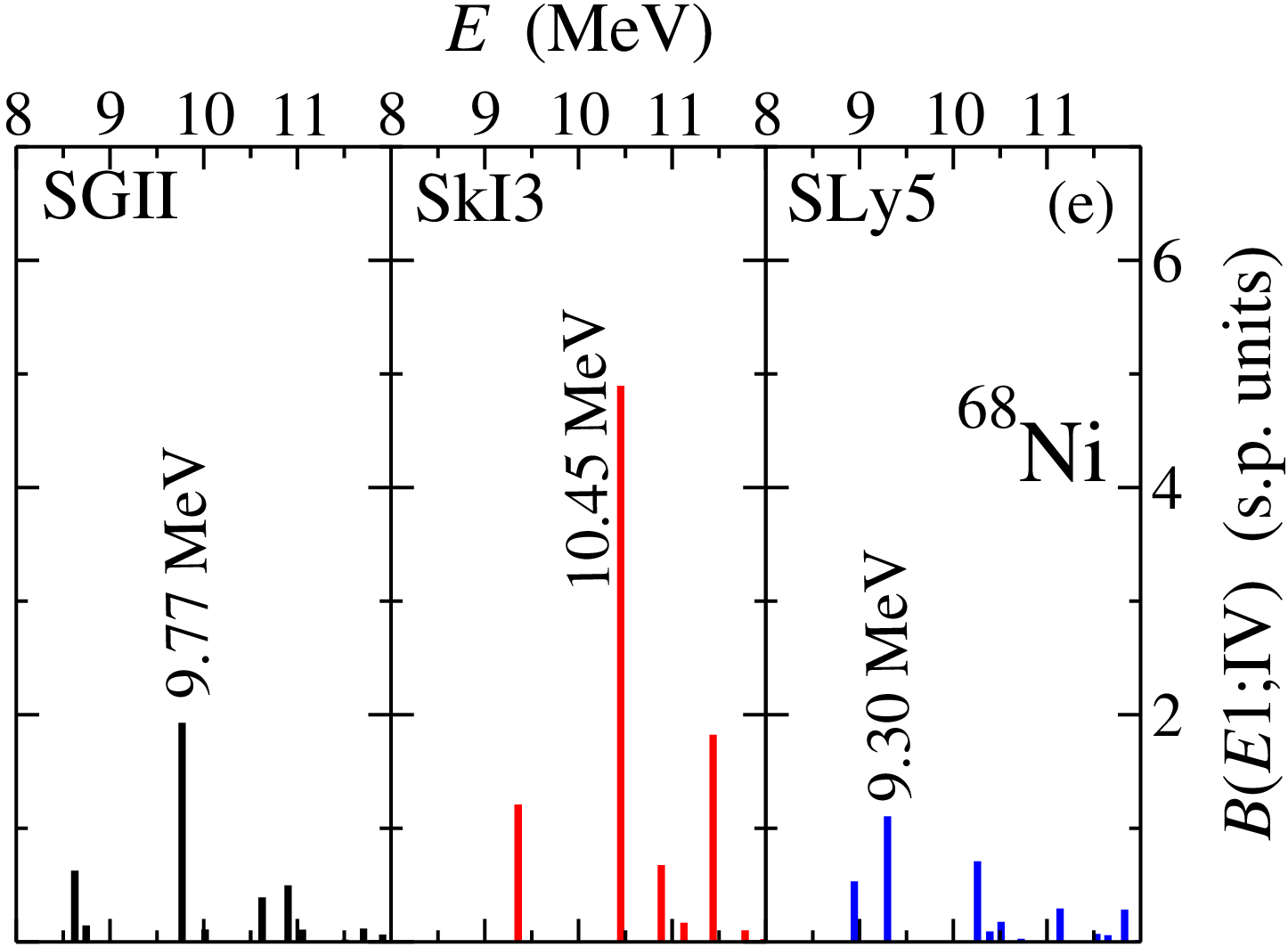}\\
\includegraphics[clip=true,width=0.34\linewidth]{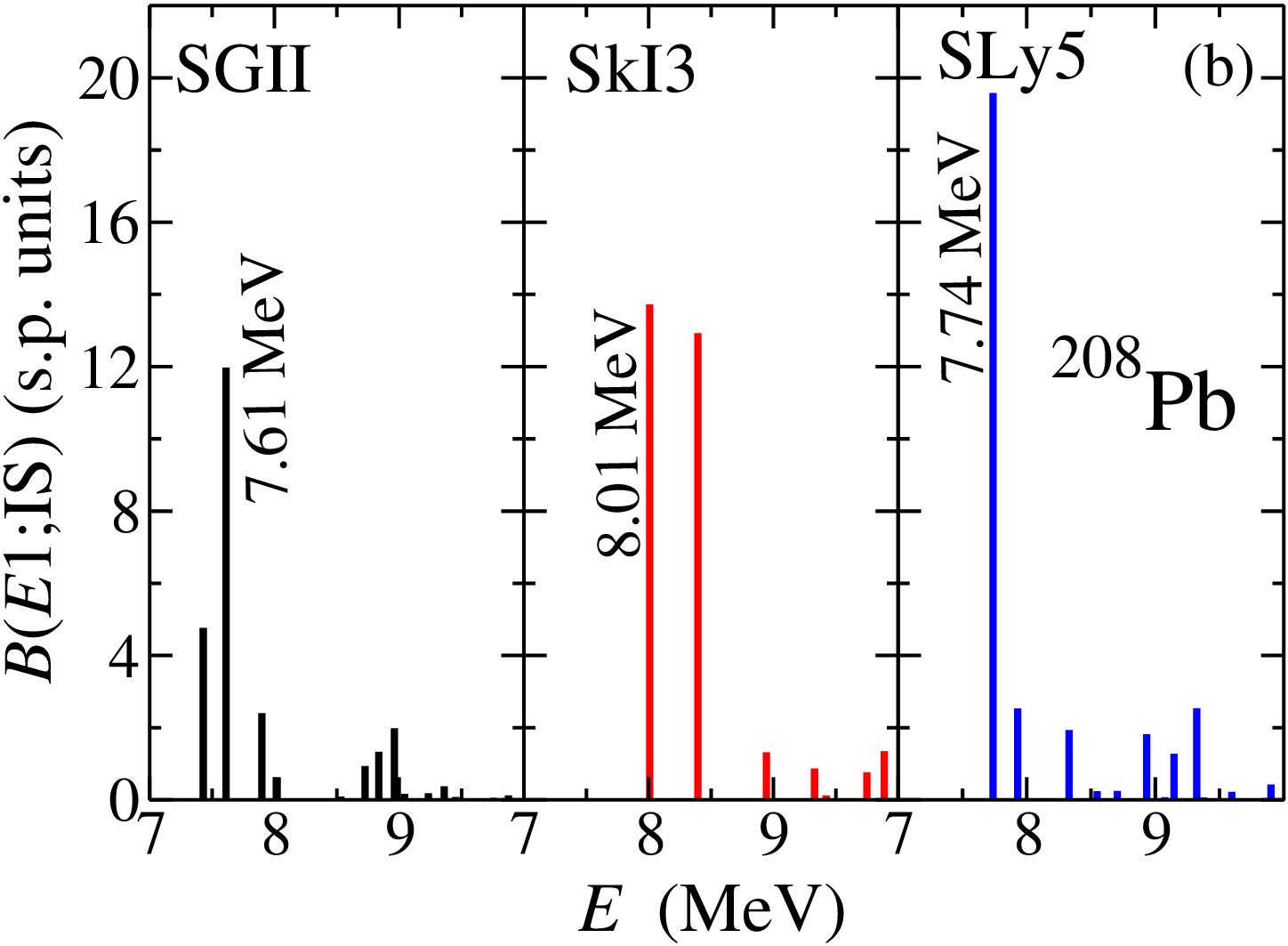}
\includegraphics[clip=true,width=0.30\linewidth]{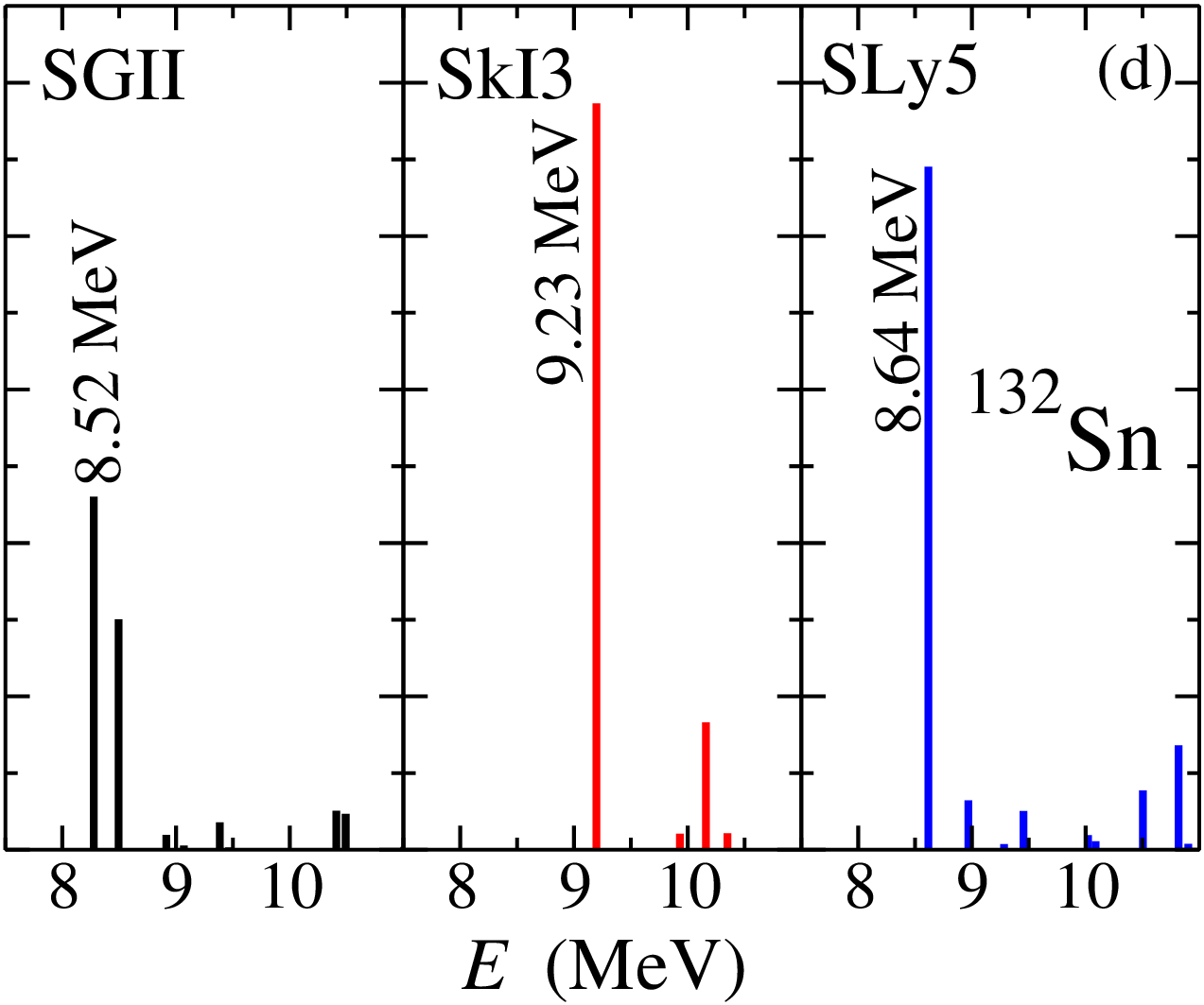}
\includegraphics[clip=true,width=0.34\linewidth]{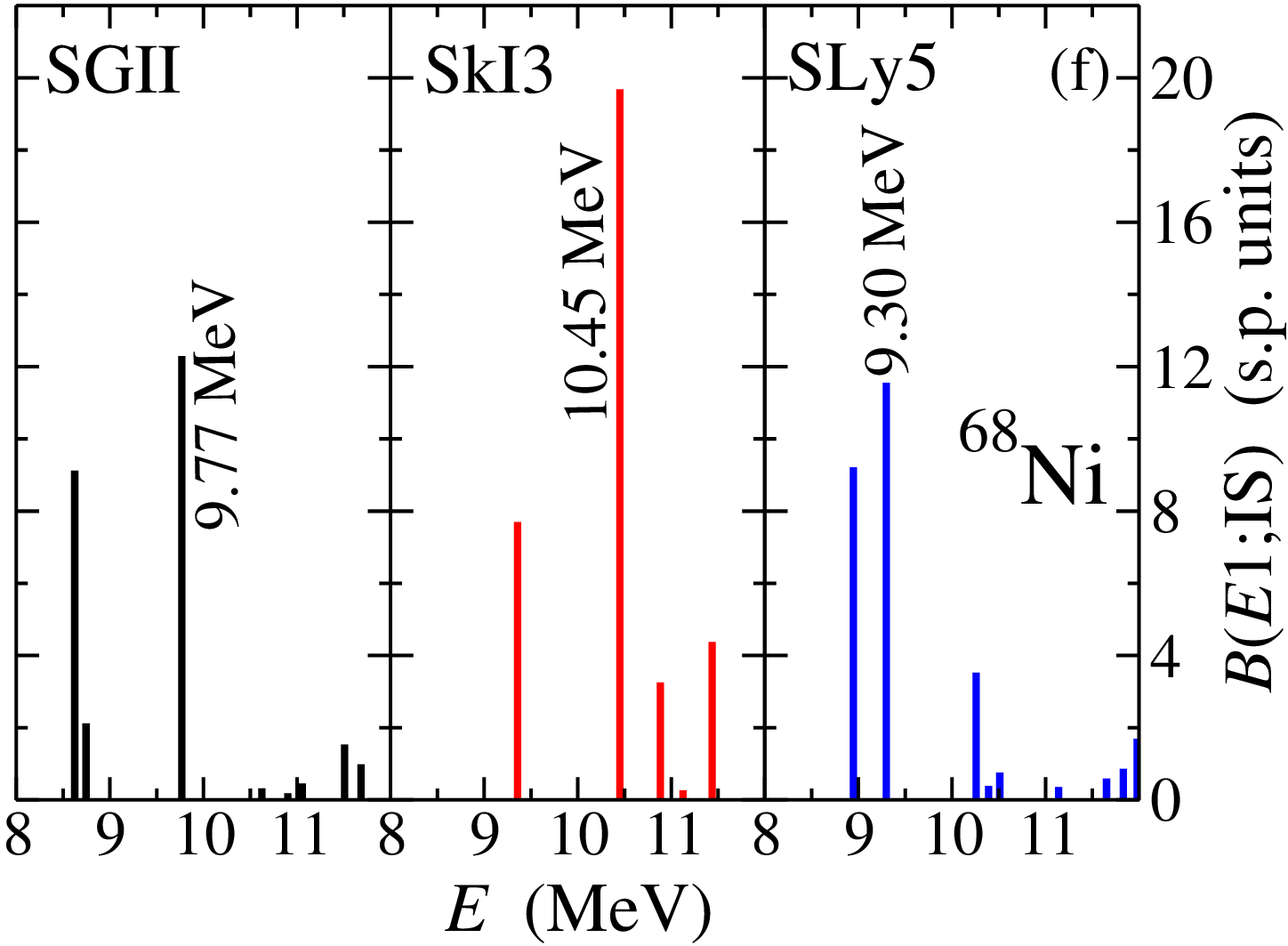}
\caption{(Color online) Reduced transition probabilities for the isovector dipole response 
[(a), (c) and (e)] 
and isoscalar dipole response [(b), (d) and (f)], 
in the case of ${}^{208}$Pb [(a) and (b)], ${}^{132}$Sn [(c) and (d)] 
and ${}^{68}$Ni [(e) and (f)] in s.p. units, as a function of the excitation 
energy and as predicted by the selected MF interactions. Note that we only show the energy region 
relevant for our study of the {\it RPA-pygmy} state.} 
\label{fig5}
\end{figure*}

\subsection{Reduced transition probability in single particle units 
as an indicator of collectivity}
\label{rtpspu}

In Fig.~\ref{fig5}, we focus on the relevant region for the study of PDS and 
show the reduced transition probabilities in single particle units 
[see Eqs.~(\ref{estim-rtp-iv}) 
and (\ref{estim-rtp-is}) and the related discussion]. 
The excitation energies of the {\it RPA-pygmy} state are 
also depicted. We display again both the 
isovector (Figs.~\ref{fig5}a, \ref{fig5}c and \ref{fig5}e) and isoscalar (Figs.~\ref{fig5}b, \ref{fig5}d and 
\ref{fig5}f) dipole responses. 

Firstly, we focus on the isovector dipole response of ${}^{208}$Pb 
(see Fig.~\ref{fig5}a). 
 Our calculations predict an {\it RPA-pygmy} state 
characterized by $\approx$ 2-4 single particle units: 
this result does not pin down clearly the nature of the state. 
As a reference, the RPA state leading to 
the largest values of the reduced transition strength in the 
IVGDR contribute with about 30 single particle 
units if the strength is fragmented, and with 
more than 60 if the strength is concentrated in one single peak. 
This is a clear indication of the 
collective nature of the IVGDR. From Fig.~\ref{fig5}b, where the isoscalar or compression dipole 
response of the same nucleus is depicted for the used Skyrme interactions, the RPA state 
leading to the pygmy peak is contributing with 15-20 single particle units, 
very 
similarly in magnitude to those displayed by the largest peak in the same isoscalar response 
at larger excitation energies and that can be seen in Fig.~\ref{fig2}b. These large values indicate 
the collective character 
of the {\it RPA-pygmy} state when it is excited by an isoscalar probe. 

In Figs.~\ref{fig5}c and \ref{fig5}e (IV) and Figs.~\ref{fig5}d and \ref{fig5}f (IS), 
the reduced transition probabilities in single particle units for the case 
of ${}^{132}$Sn and 
${}^{68}$Ni are depicted, respectively. Note that we show only the low energy region. As 
mentioned, the single particle units normalize the absolute value of $B(E1;\xi)$ for all nuclei. 
This statement is qualitatively fullfiled by all the interactions: 
$B(E1,IV)$ corresponding to the {\it RPA-pygmy} state in ${}^{68}$Ni, ${}^{132}$Sn and 
${}^{208}$Pb as calculated with SGII lead to 1.9, 1.1 and 1.9 s.p. units, respectively. For the case 
of SLy5, we find 1.1, 1.5 and 1.9 s.p. units, respectively. And, finally, for the 
case of SkI3, we find 4.9, 6.4 and 4.2 s.p. units, respectively. For the isoscalar response,
the $B(E1,IS)$ of the same excited states in ${}^{68}$Ni, ${}^{132}$Sn and 
${}^{208}$Pb as calculated 
with SGII lead to 12,  9 and  12 s.p. units, respectively. For the 
case of SLy5, we find 11, 18 and 19 s.p. units, respectively. And, finally, for the case 
of SkI3, we find 20, 19 and 14 s.p. units, respectively.
   
Despite the fact that the reduced transition probabilities in s.p. units can 
be only used as a qualitative estimator of the collectivity displayed by a given 
RPA excitation, we can conclude that while the isoscalar dipole response of all studied 
nuclei and within all employed interactions seems to indicate that the {\it RPA-pygmy} 
state develope a certain amount of collectivity, 
the isovector response of the same excited 
state does not provide a clear trend: the collectivity displayed is very weak and depends on 
the used model.    

%
% >>>>>>>>>>>>>>>>>>>>>>>>>>>>>>>>>>>>>>>>>>>>>>>>>>>>>>>>>>>>>>>>>>>>
%

%
\begin{figure*}[t]
\includegraphics[clip=true,width=0.343\linewidth]{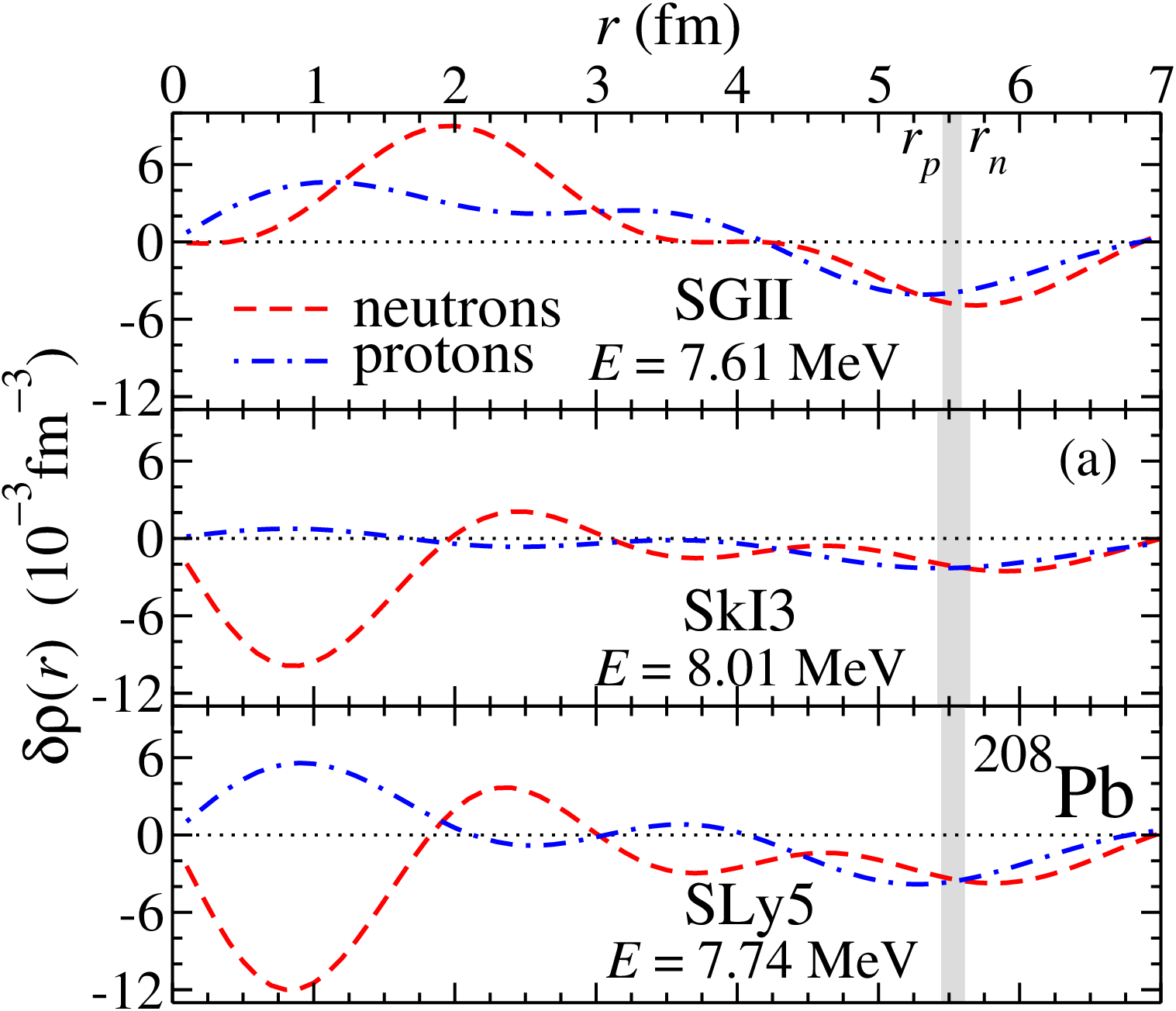}
\includegraphics[clip=true,width=0.316\linewidth]{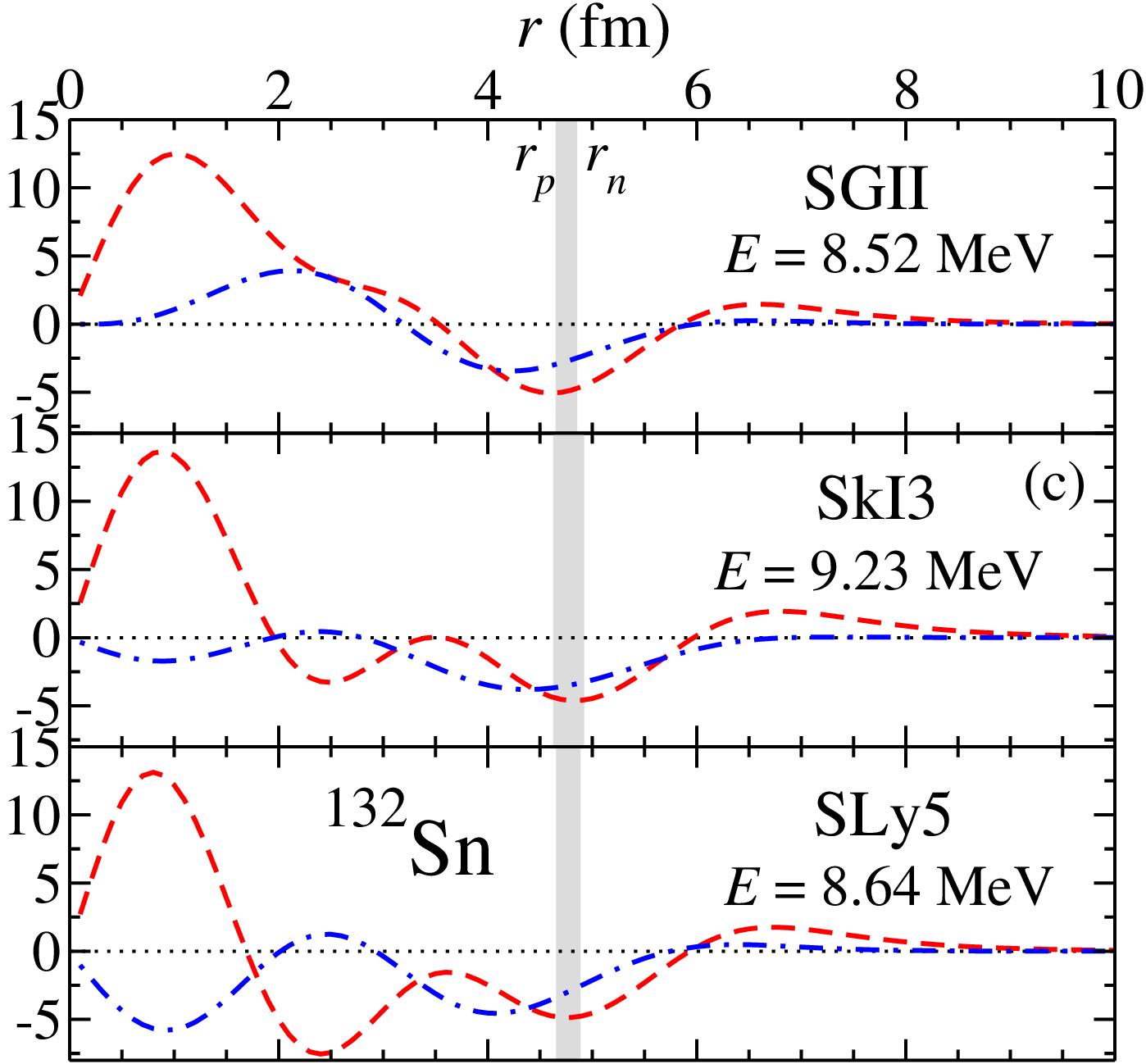}
\includegraphics[clip=true,width=0.316\linewidth]{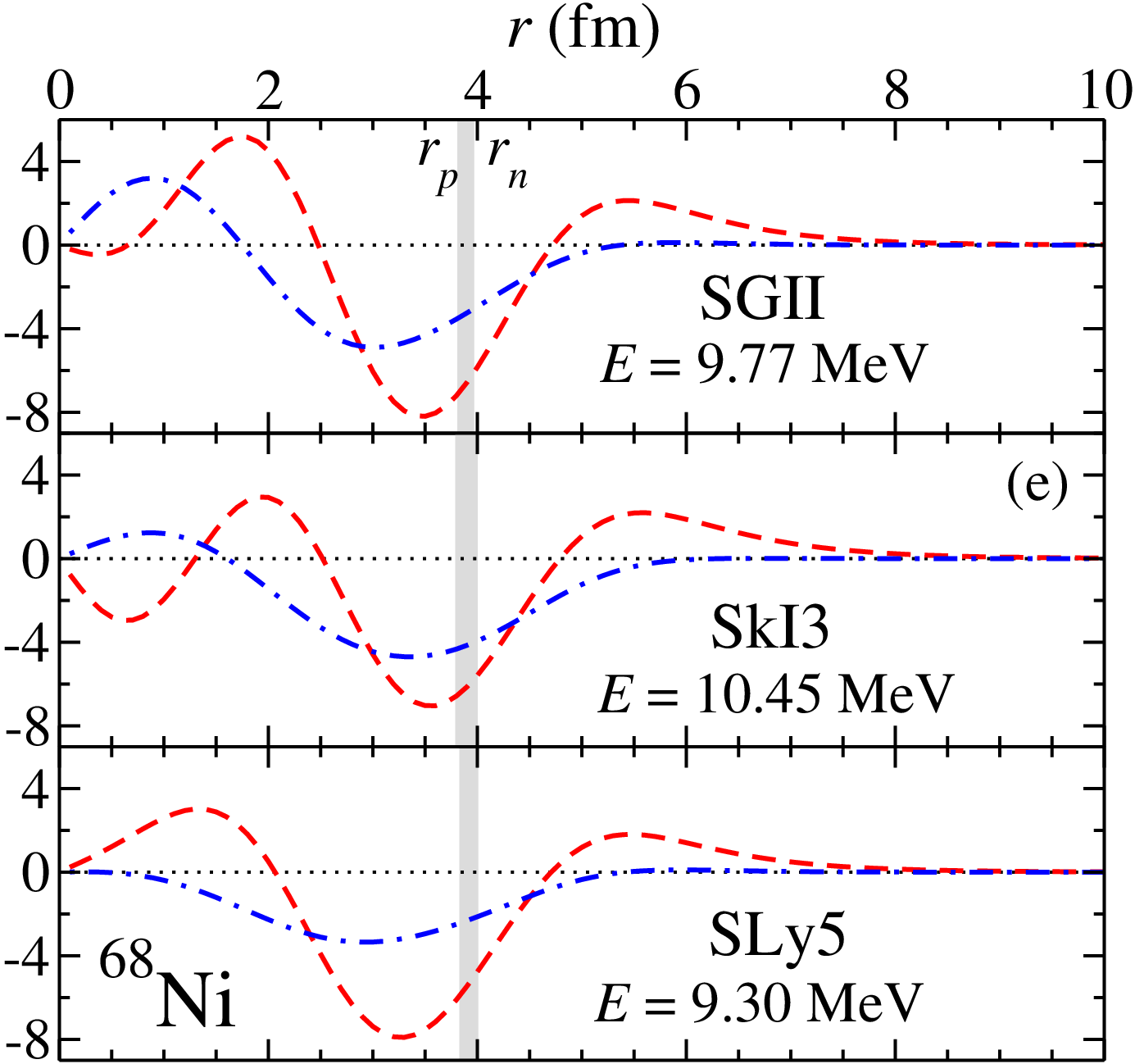}\\
\includegraphics[clip=true,width=0.343\linewidth]{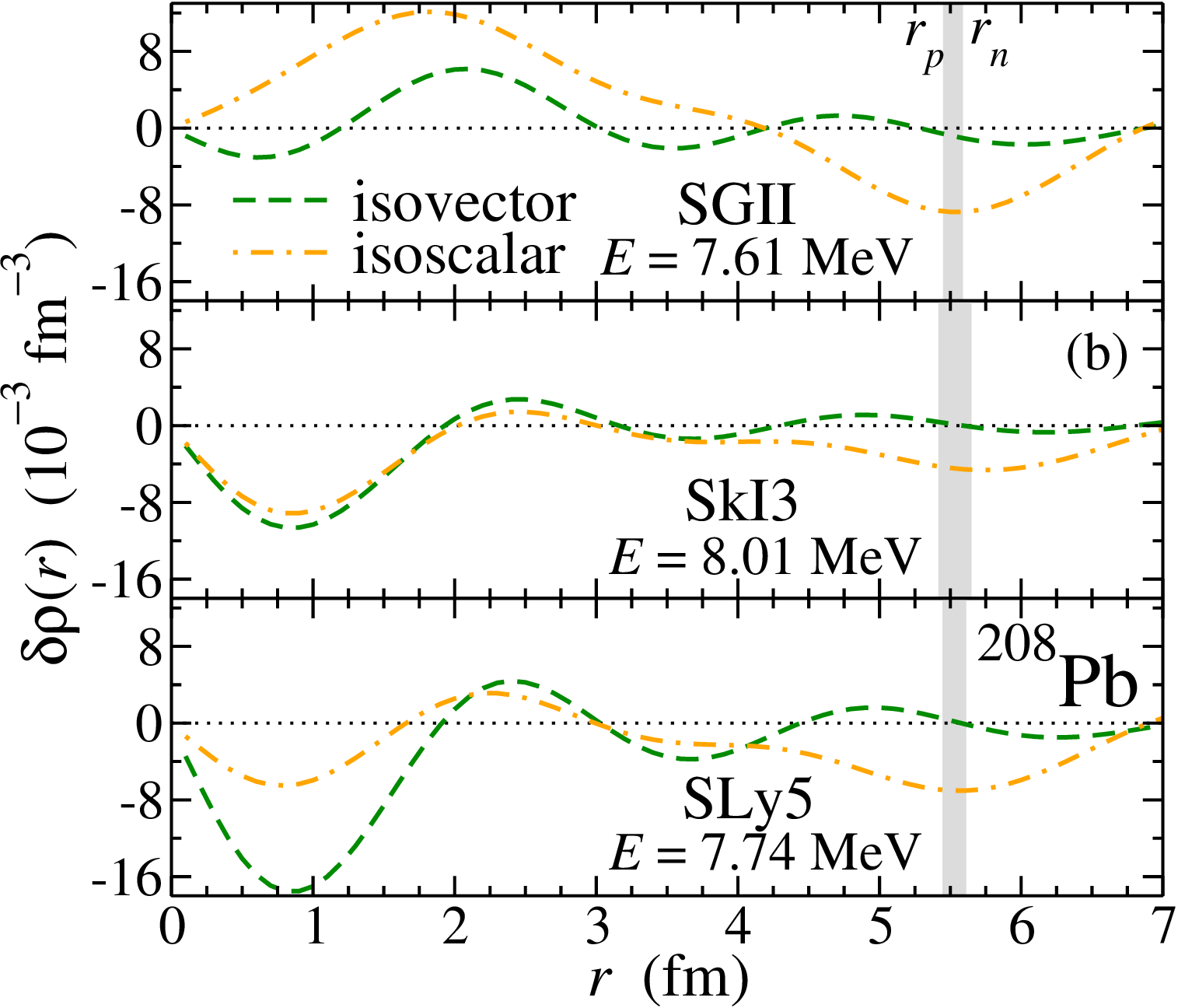}
\includegraphics[clip=true,width=0.316\linewidth]{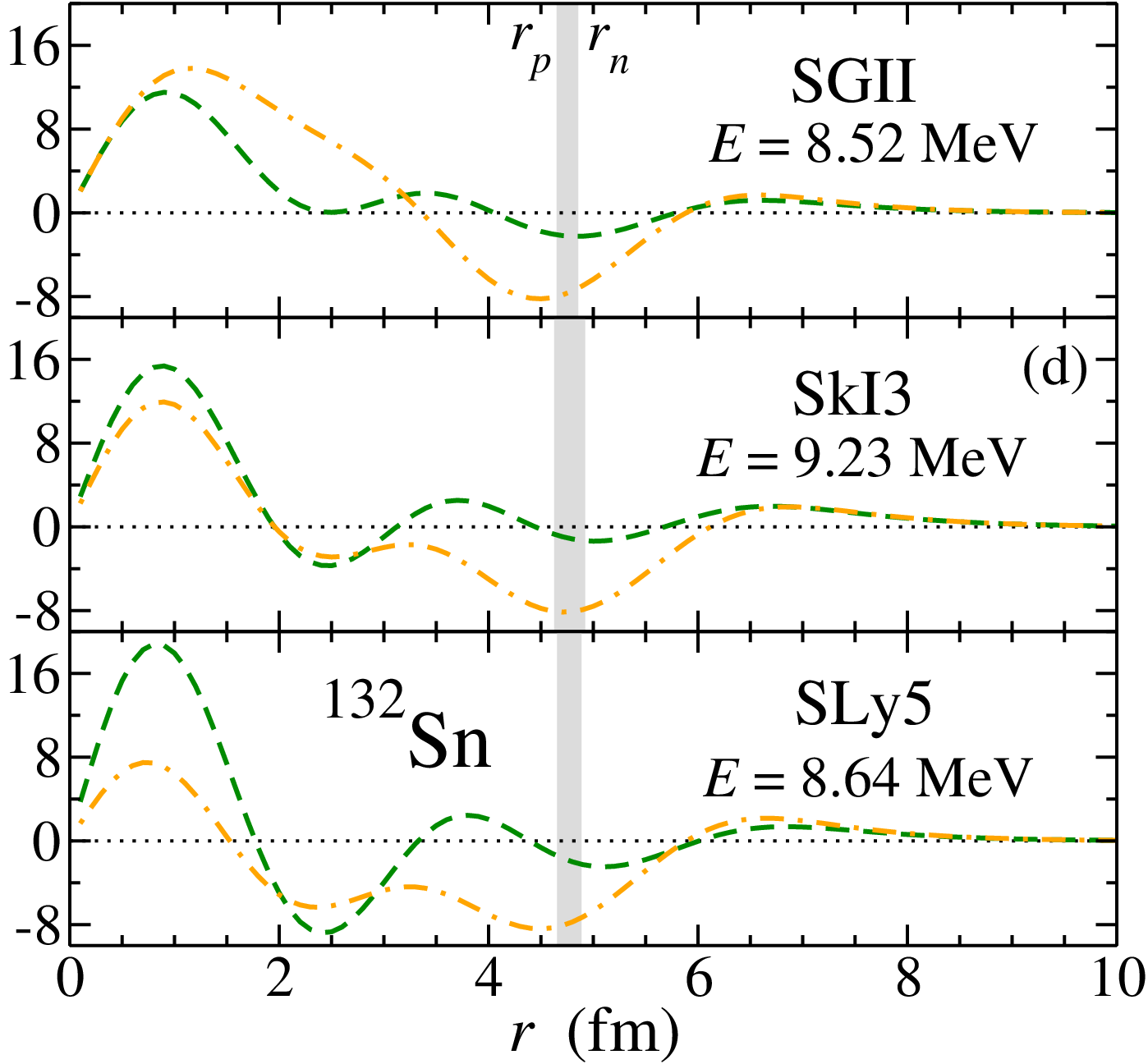}
\includegraphics[clip=true,width=0.316\linewidth]{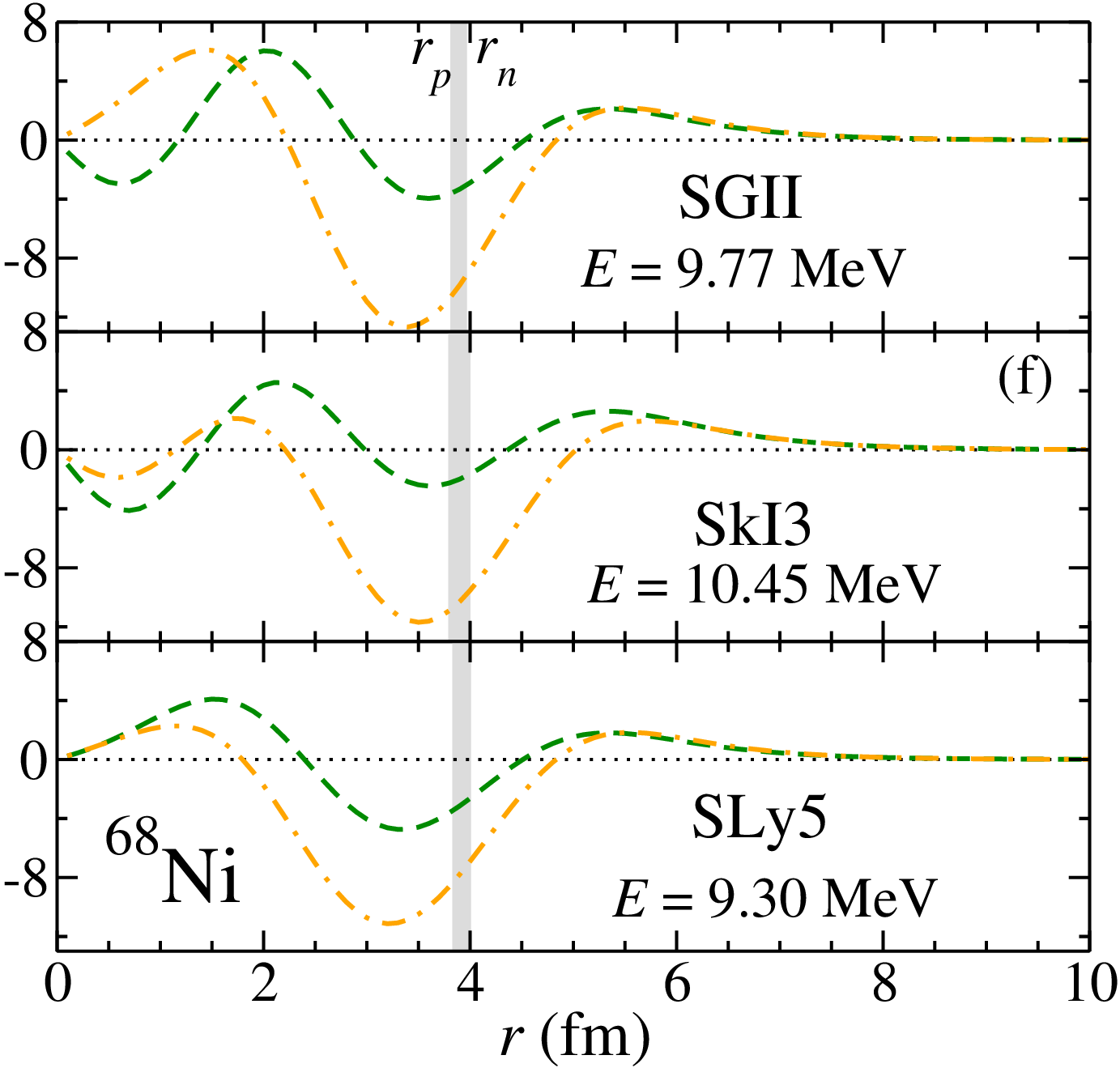}
\caption{(Color online) Neutron and proton [(a), (c) and (e)] 
and isoscalar and isovector [(b), (d) and (f)]
transition densities of the {\it RPA-pygmy} state, 
as a function of the radial distance, 
for ${}^{208}$Pb [(a) and (b)], ${}^{132}$Sn [(c) and (d)] 
and ${}^{68}$Ni [(e) and (f)]. 
The predictions of SGII, SLy5 and SkI3 are shown. Proton ($r_p$) and neutron ($r_n$) 
rms radii are indicated for each interaction by the edges of the grey region.} 
\label{fig6}
\end{figure*}

\subsection{The low-energy RPA states: isoscalar or isovector character?}

A collective excited state  can be said to be {\it purely} 
isovector if the transition 
densities of protons and neutrons scale as Z and N, respectively, 
and have opposite phase. 
On the other side, a collective excited state can be 
defined as {\it purely} isoscalar 
if the neutron 
and proton transition densities scale in absolute value in the
same way, but have the same sign. These 
two cases are extreme. As isospin is not a good quantum 
number in finite nuclei, and is more and more broken 
as the neutron excess increases, 
the most common 
situation corresponds to a mixture of a certain degree of isoscalarity and 
isovectoriality, that can be better seen by looking at 
the neutron and proton transition densities. 

The isoscalar or isovector nature of the low-energy RPA states has 
been already studied in 
Ref. \cite{paar2009} in 
the case of ${}^{140}$Ce. In this work, it was found that the
low-lying dipole states 
of ${}^{140}$Ce are split into two groups depending on their isospin 
structure. More recently, similar conclusions were 
found in a study of the pygmy dipole 
strength in ${}^{124}$Sn \cite{endr2010}, where it has 
been stated that the theoretical calculations 
were dominated by a low-lying isoscalar component basically due to oscillations of the neutron 
skin thickness of the nucleus under study. It is important to note that both investigations 
were reported to be in qualitative agreement with the available experimental data.   

On the basis of the above mentioned works and the definitions given 
in Eqs.~(\ref{is-iv-td}) and 
in the text around, we present a more systematic study of the isospin
structure of the low-energy RPA states as predicted by the 
forces SGII, SLy5 and SkI3 for the studied 
${}^{68}$Ni, ${}^{132}$Sn and ${}^{208}$Pb nuclei. 
First of all, we plot the neutron and proton, as well as the  
isoscalar and isovector transition densities 
corresponding to the {\it RPA-pygmy} state 
as predicted by each interaction in order to illustrate how these low-energy transition 
densities behave. 

We show the neutron and proton transition densities 
in Figs.~\ref{fig6}a, \ref{fig6}c and \ref{fig6}e, and 
the isoscalar and 
isovector transition densities in 
Figs.~\ref{fig6}b, \ref{fig6}d and \ref{fig6}f, respectively. 
All of them, correspond to the {\it RPA-pygmy} state. The position of the 
proton ($r_p$) and neutron ($r_n$) rms radii 
corresponds to the edges of the grey 
region that defines in this way the neutron skin thickness 
predicted by each interaction. 

For the case of ${}^{208}$Pb, it can be seen from 
Fig.~\ref{fig2}a that neutrons and 
protons oscillate differently depending on the interaction 
 but in all cases the surface has a dominant isoscalar 
character. On the contrary, 
the interior or bulk region is not dominated by the isoscalar 
or isovector component but 
it is a mixture of them. The isoscalar or isovector 
dominance is better seen in Fig.~\ref{fig6}b. 
At the surface of the nucleus the isovector transition 
density of the {\it RPA-pygmy} state 
is very close to zero, while the isoscalar one is not.        

In Figs.~\ref{fig6}c (protons and neutrons) 
and \ref{fig6}d (IS and IV), we display 
the transition densities for the case of ${}^{132}$Sn. 
 It is interesting to note that
the situation is very similar to the one 
found in ${}^{208}$Pb. 

The neutron and proton transition densities 
corresponding to the {\it RPA-pygmy} state 
in ${}^{68}$Ni are depicted in Fig.~\ref{fig6}e, 
and the corresponding isoscalar and isovector ones are 
displayed in Fig.~\ref{fig6}f.    
The behavior of the different transition densities 
is predicted to be very similar within the studied models. 
This did not hold for ${}^{132}$Sn and ${}^{208}$Pb where 
some qualitative differences arose. Therefore, it is  even
more clear in this case that the interior of ${}^{68}$Ni is 
not dominated by isoscalar or isovector 
components. At the surface of the nucleus, 
the isoscalar part dominates but the isovector part is not very 
small as it happened for ${}^{132}$Sn or ${}^{208}$Pb. 

\begin{figure}[t!]
\includegraphics[clip=true,width=0.9\linewidth]{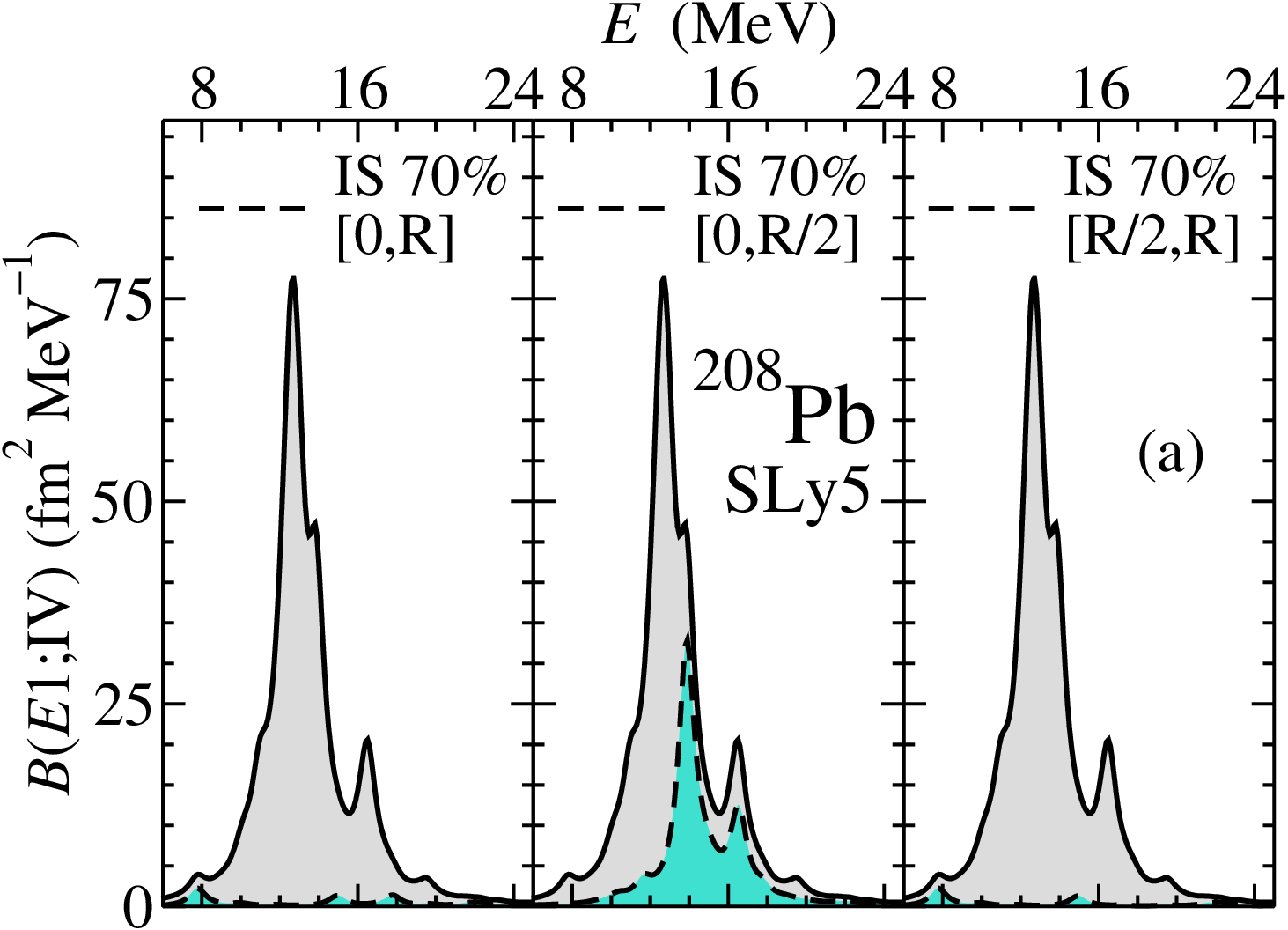}\\
\includegraphics[clip=true,width=0.9\linewidth]{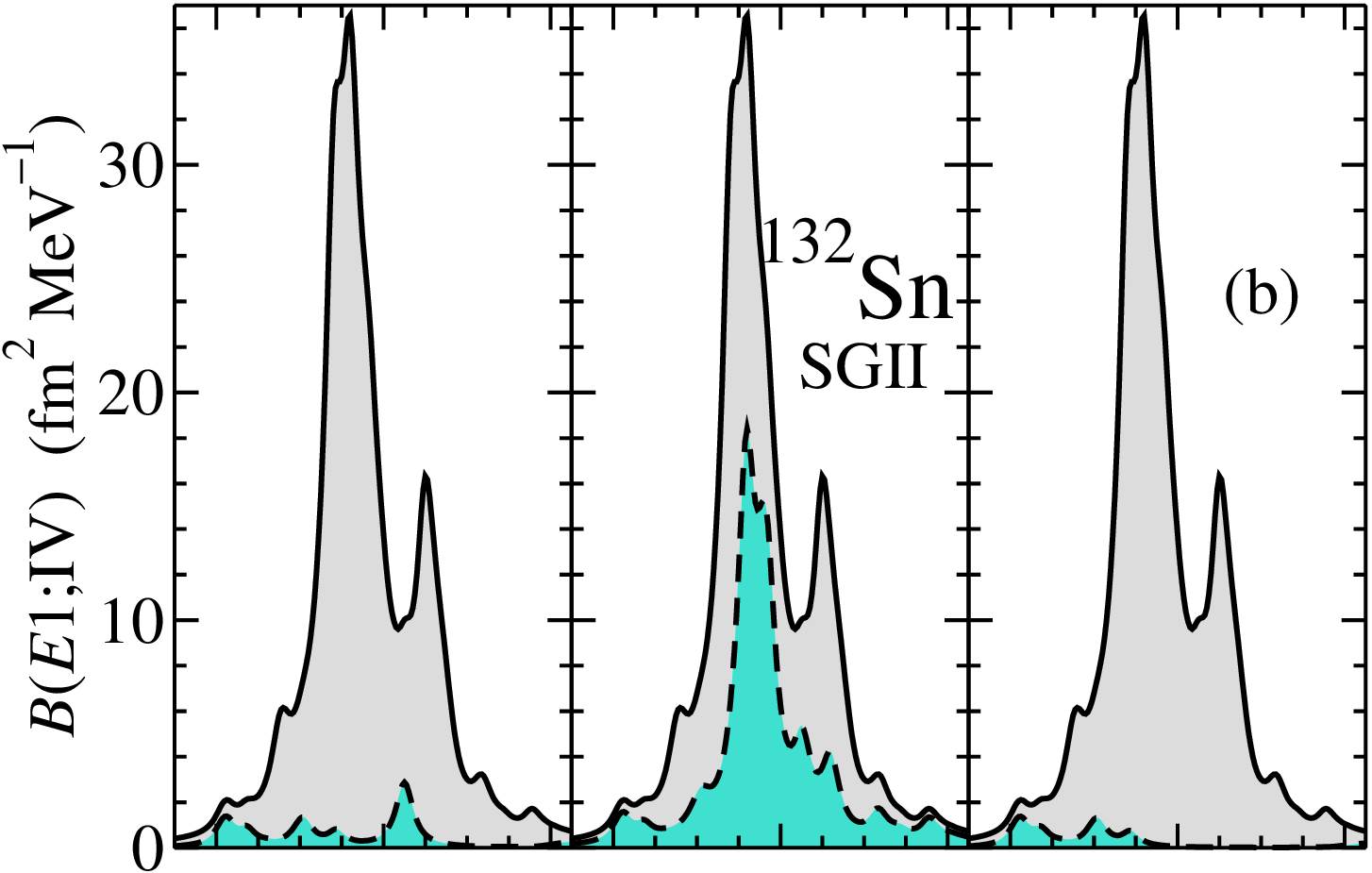}\\
\includegraphics[clip=true,width=0.9\linewidth]{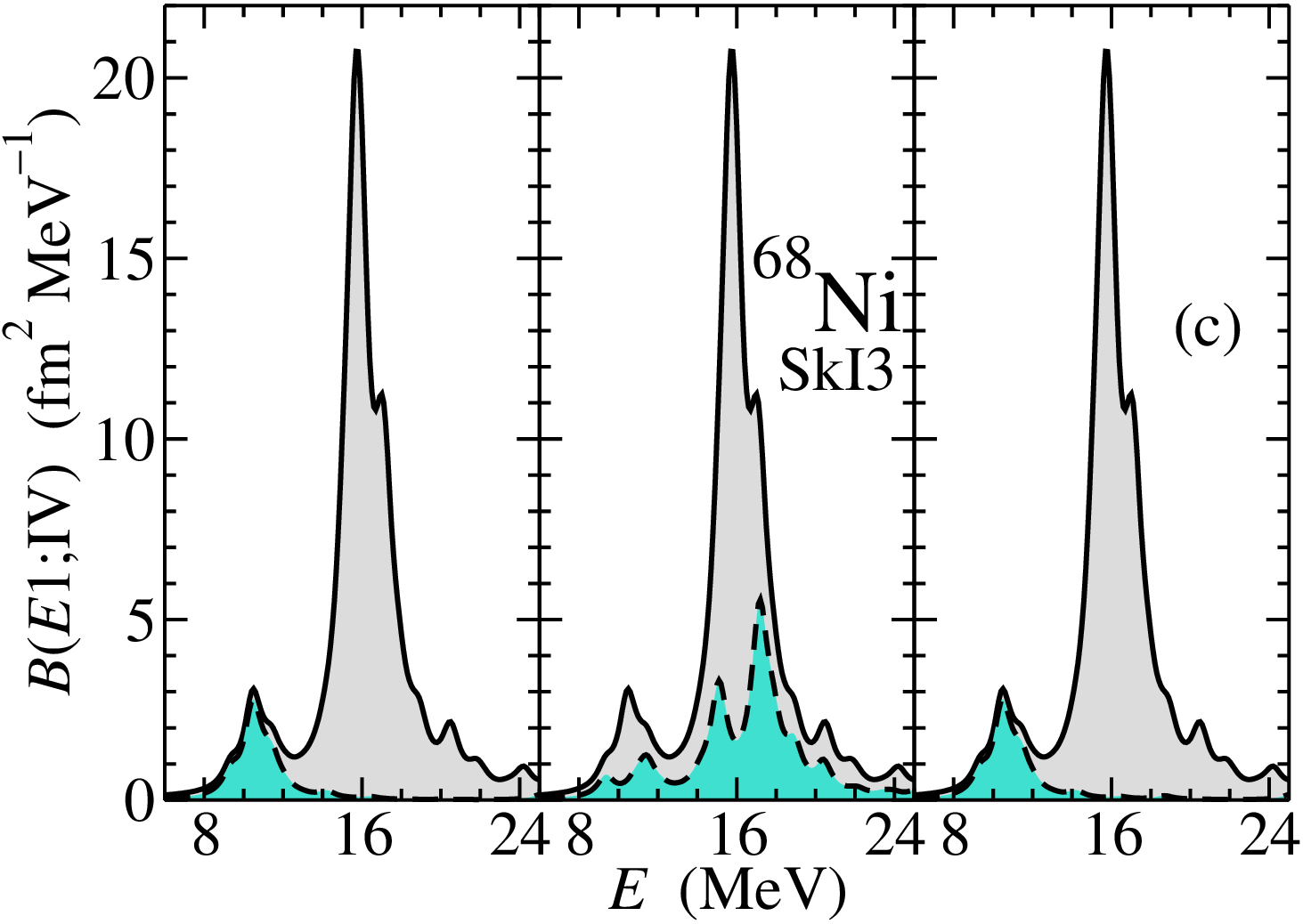}
\caption{(Color online) Strength function corresponding to the isovector 
dipole response of ${}^{208}$Pb (a), 
${}^{132}$Sn (b) and ${}^{68}$Ni (c) 
as a function of the excitation energy (solid lines), 
and partial contribution due to 
those states which are at least 70\% isoscalar 
(dashed line): for each of the panels, the RPA states that are at 
least 70\% isoscalar between 0 and $R$ are considered 
in the panel at the left hand side, 
those which have this feature between 0 and $R/2$ are 
considered in the central panel, and those which are
like that between $R/$2 and $R$ are considered 
in the panel at the right hand side. See text for further 
explanations.} 
\label{fig7}
\end{figure}

Then, we apply our criteria for defining a 70\% 
isoscalar RPA state [see text after 
Eq.~(\ref{is-iv-td})] to all calculated excited states 
and plot their contribution to the isovector dipole 
strength function. We calculate the same quantity 
for different regions. First, we apply the criteria to those states that are 70\% isoscalar in 
the region between 0 and $R$, where $R=r_0A^{1/3}$ (left panels in 
Fig.~\ref{fig7}), then to those which are 70\% isoscalar in the 
internal part of the nucleus, namely between 0 
and $R/2$  (central panels in the same figure), and 
finally to those which are 70\% isoscalar 
in the external part of the nucleus between $R/2$ and $R$  
(right panels of the same figure). 
Specifically, in Fig.~\ref{fig7}a, 
we show the above mentioned calculations for ${}^{208}$Pb 
as predicted by SLy5 (dashed line). As a guidance, we also 
show the total isovector dipole stregth function (solid line). The results predicted by the other 
interactions in the case of ${}^{208}$Pb are very similar and we are not showing them. From  
such a figure, it is evident that the RPA states which are mostly isoscalar in the whole region [0,R] 
(left panel in Fig.~\ref{fig7}a) and in the external part of the nucleus [R/2,R] (right panel in Fig.~\ref{fig7}a) 
are essentially the same ones since both give rise to almost the same contributions: most of the PDS and a small 
contribution to the rest of the strength function. In the central panel,
where 
we represent those RPA states that are 
70\% isoscalar in the internal region of the nucleus [0,R/2], 
we confirm that 
there is no essential contribution from them 
to the state giving rise to the PDS.

In Figs.~\ref{fig7}b and \ref{fig7}c, we show the same 
quantities but for ${}^{132}$Sn as predicted by SGII and for ${}^{68}$Ni as predicted 
by SkI3, respectively. Qualitatively, the 
same results found for ${}^{208}$Pb are now found in 
these figures. Again, we do 
not show the results for the other studied interactions since they are very similar and the same conclusions can be drawn.   

Our results 
indicate that one is allowed to qualitatively distinguish 
the PDS from the IVGDR, and state that while 
the latter strength is basically isovector and involves 
the motion of  mainly internal nucleons, the former is more  
isoscalar than isovector and involves the motion of 
external nucleons, that are mainly neutrons in a neutron rich nucleus.    

% 

%
% >>>>>>>>>>>>>>>>>>>>>>>>>>>>>>>>>>>>>>>>>>>>>>>>>>>>>>>>>>>>>>>>>>>>
%
\subsection{The most relevant particle-hole excitations in the low-energy region}

The RPA states are build as a superposition of particle-hole 
($ph$) excitations.
A given RPA state shows a collective character under 
the action of an external 
operator if there are many $ph$ excitations 
providing non negligible 
contributions (each associated with a transition
amplitude and the sum 
of $X_{ph}^{(\nu)}$ and $Y_{ph}^{(\nu)}$) 
that add coherently in Eq.~(\ref{rts}). 
Therefore, within our approach, the 
collective nature of a peak in the response function of a nucleus is always hidden in the 
RPA states. 

For these reasons, our purpose in this subsection is to analyze the contributions of the 
different $ph$ excitations to the {\it RPA-pygmy} states depending on the operator used 
to excite the nucleus. In particular, we study the isoscalar and isovector dipole responses 
of the {\it RPA-pygmy} states since we are most interested in them. 
Such a study will allow 
us to quantify the collectivity displayed 
by each RPA state depending on the operator, nucleus 
and interaction used for the theoretical calculations, 
and detect if there are common trends 
in the microscopic structure.
    
\begin{figure}[t!]
\includegraphics[clip=true,width=0.9\linewidth]{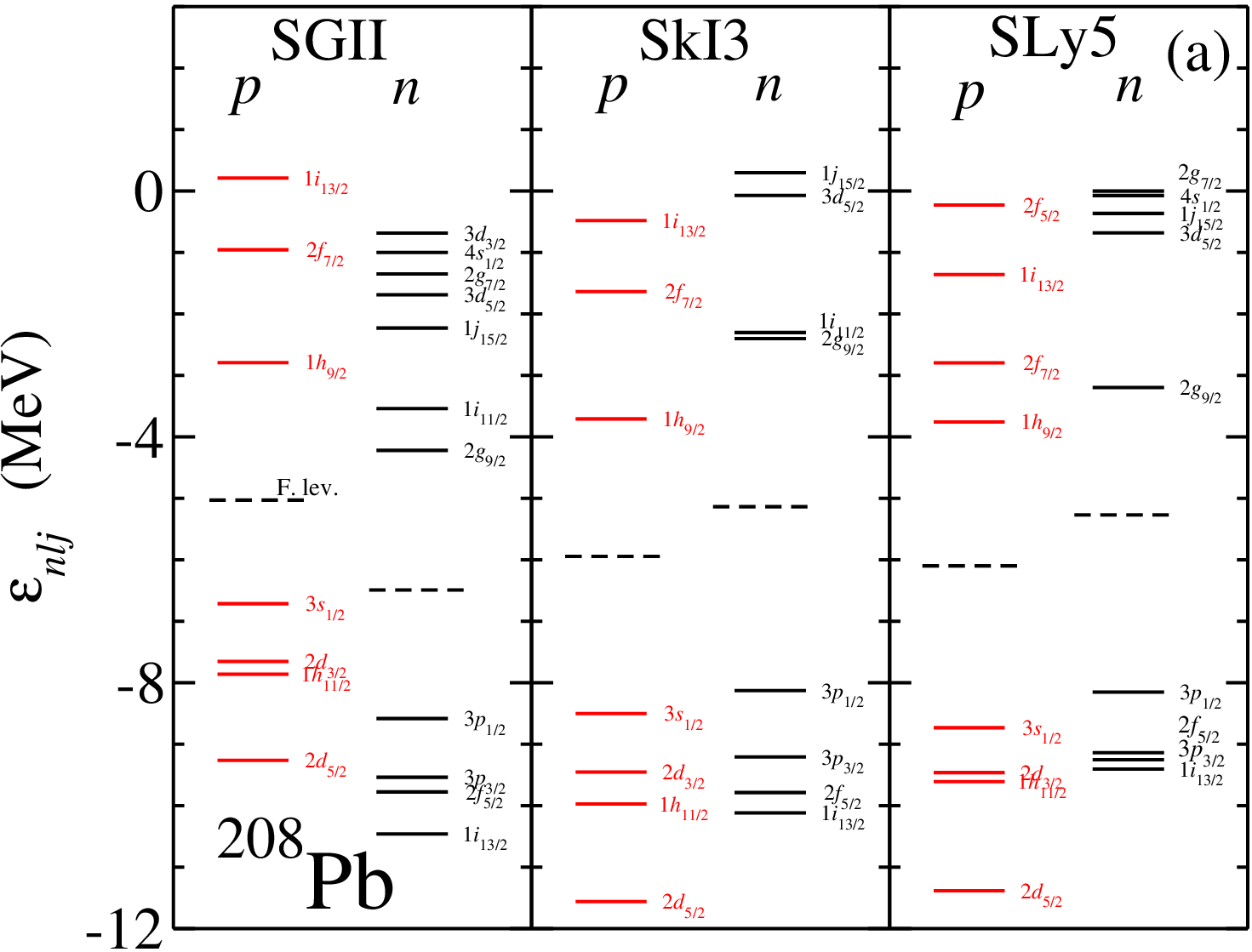}\\
\includegraphics[clip=true,width=0.9\linewidth]{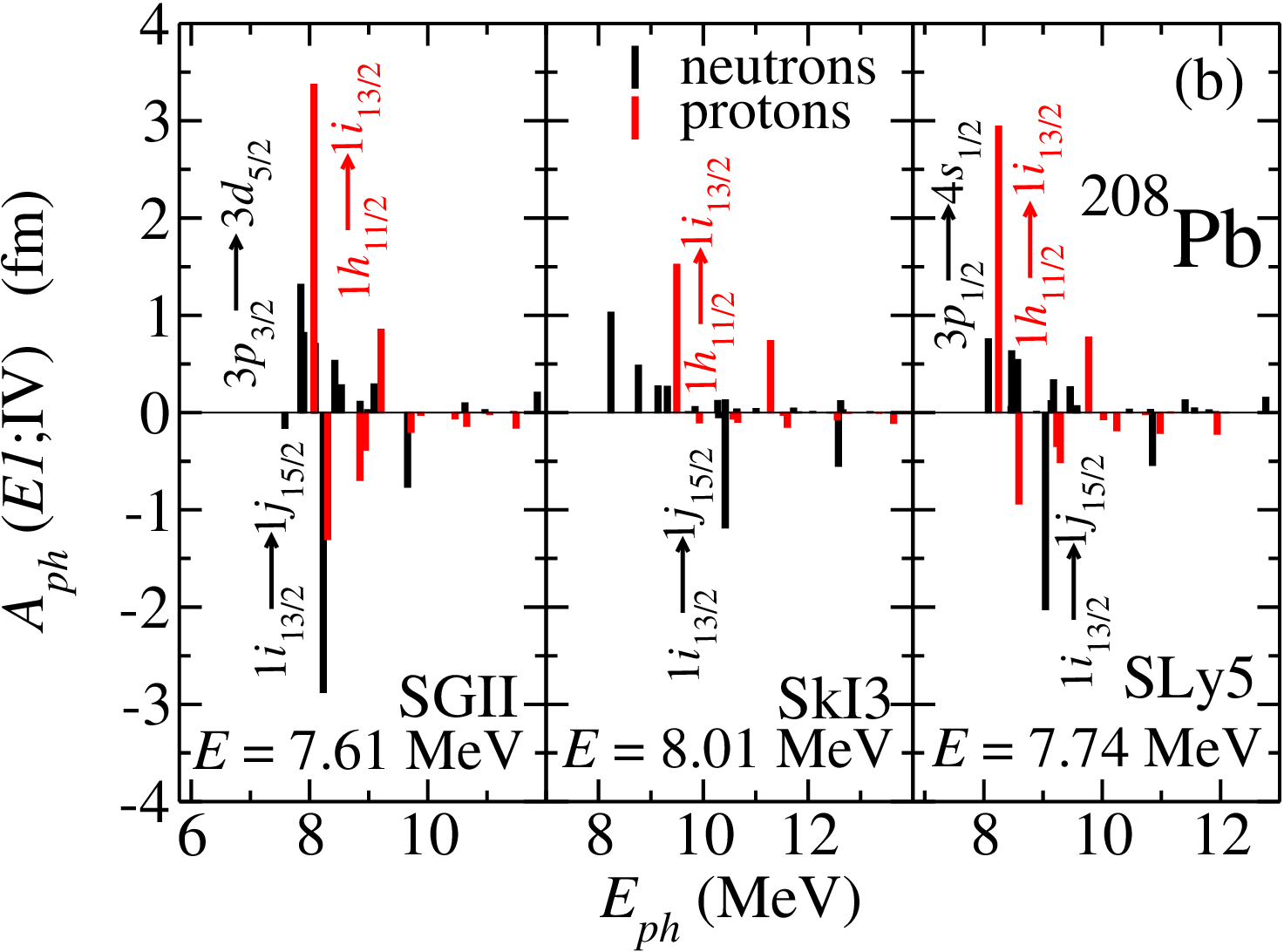}\\
\includegraphics[clip=true,width=0.9\linewidth]{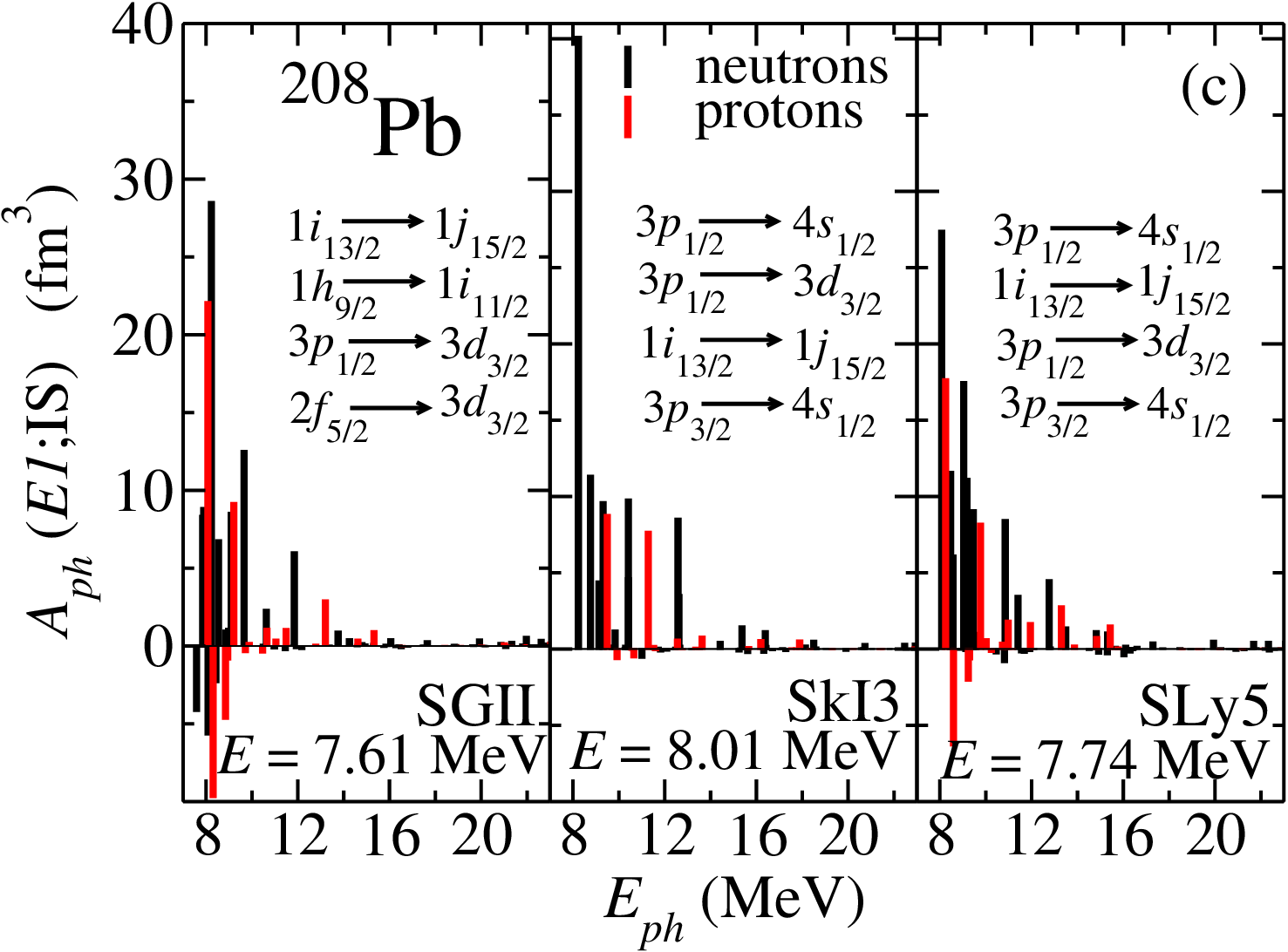}
\caption{(Color online) Proton and neutron single particle levels of ${}^{208}$Pb as 
predicted by the different MF models (a). The Fermi level 
is indicated by a dashed 
black line. All $ph$ contributions 
to the isovector reduced amplitude corresponding 
to the ${}^{208}$Pb {\it RPA-pygmy} state as 
a function of the $ph$ excitation energy (c). 
All studied models are shown. Largest neutron $ph$ contributions are also listed in decreasing 
order from top to bottom. Same as (b) but for the isoscalar reduced amplitude (c).} 
\label{fig8}
\end{figure}

First of all, we show in Fig.~\ref{fig8}a the 
neutron (black) and proton (red) 
single-particle MF levels close to 
the  Fermi energy 
for ${}^{208}$Pb as predicted by SGII (left panel), 
SLy5 (right panel) and SkI3 (middle panel). 
The proton levels show a rather similar ordering and spacing 
for all the interactions. On the contrary, the neutron 
levels differ  more 
when different interactions are compared. 

In Figs.~\ref{fig8}b and \ref{fig8}c, 
also for the case of ${}^{208}$Pb, we show all the neutron (black) 
and proton (red) $ph$ contributions to 
the reduced amplitude $A_{ph}^q(E1;\xi)$ [see Eq.~(\ref{ampl})]  
as a function of their excitation energy 
for the isovector and isoscalar dipole responses, respectively. 
Both reduced amplitudes have 
been calculated for the case of the {\it RPA-pygmy} 
state predicted by the different 
MF models. Notice that not all contributions 
can be seen from these figures since most of them are very small. 

It is evident from Fig.~\ref{fig8}b that the contributions 
of the most relevant $ph$ excitations to the isovector 
reduced amplitude are only a few in number 
and  there is some amount of
destructive interference. Accordingly, 
we have seen that their total contribution 
to the isovector reduced transition strength 
in s.p. units do not clearly exceed one. 
Opposite to that, it is also evident 
from Fig.~\ref{fig8}c that the contributions 
of the most relevant $ph$ excitations to the 
isoscalar reduced transition amplitude
are basically coming from neutron transitions, 
and that most of them add coherently. This is, actually, one of the 
basic features to assert that a given RPA state is collective. 

In Table \ref{aph-table-pb} in Appendix \ref{table-aph}, 
we show the numerical details of the ten 
neutron and proton $ph$ excitations providing 
the largest contributions to the 
isovector and isoscalar $A_{ph}^q(E1;\xi)$. 
We also indicate in the figures and table the 
involved single particle orbitals 
corresponding to the largest contributions 
to the reduced amplitudes. In particular, we 
generally find within all the employed models 
that the dynamics of the  
low-energy isoscalar dipole response 
of ${}^{208}$Pb seems to be governed by the 
excitations of the outermost neutrons, namely those 
that 
form the neutron skin thickness of this nucleus. 
From the analysis of the $ph$ contributions, 
we conclude that while the 
low-energy isoscalar dipole response of ${}^{208}$Pb 
arising form the {\it RPA-pygmy} state can 
be considered as a collective 
mode in all studied models, the PDS can not.  

\begin{figure}[t!]
\includegraphics[clip=true,width=0.9\linewidth]{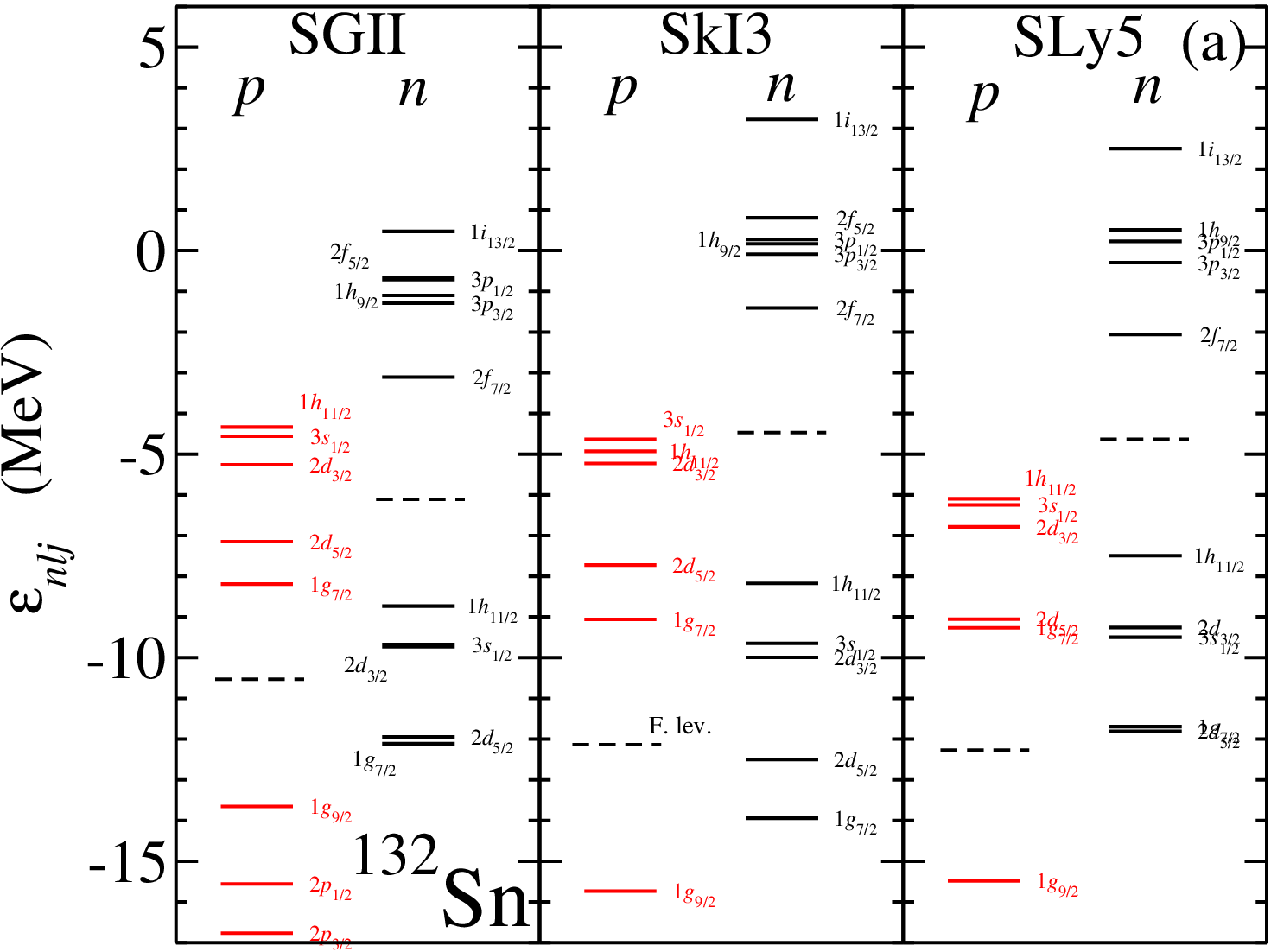}\\
\includegraphics[clip=true,width=0.9\linewidth]{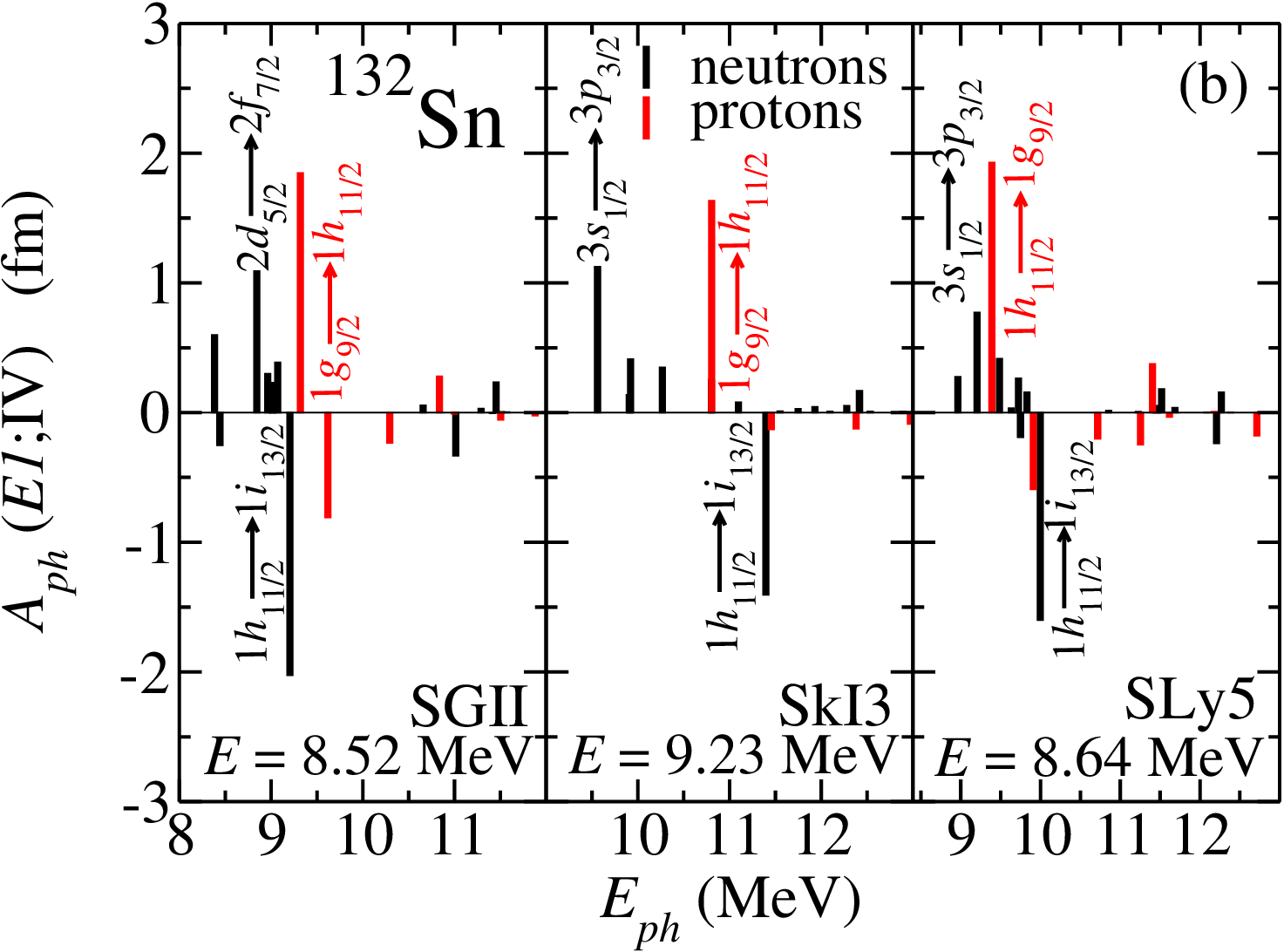}\\
\includegraphics[clip=true,width=0.9\linewidth]{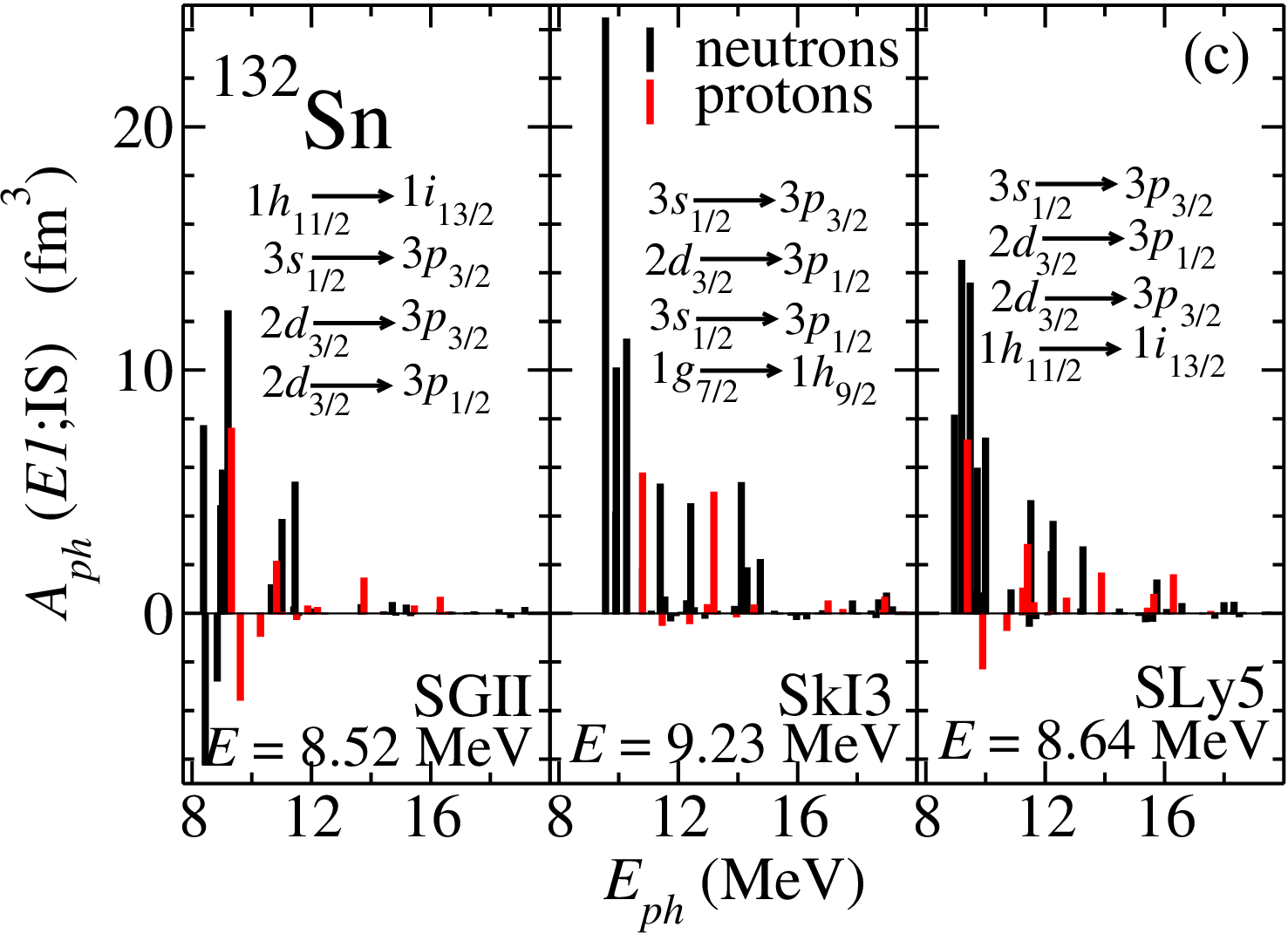}
\caption{(Color online) Same as Fig.~\ref{fig8} for the case of 
${}^{132}$Sn.} 
\label{fig9}
\end{figure}
\begin{figure}[t!]
\includegraphics[clip=true,width=0.9\linewidth]{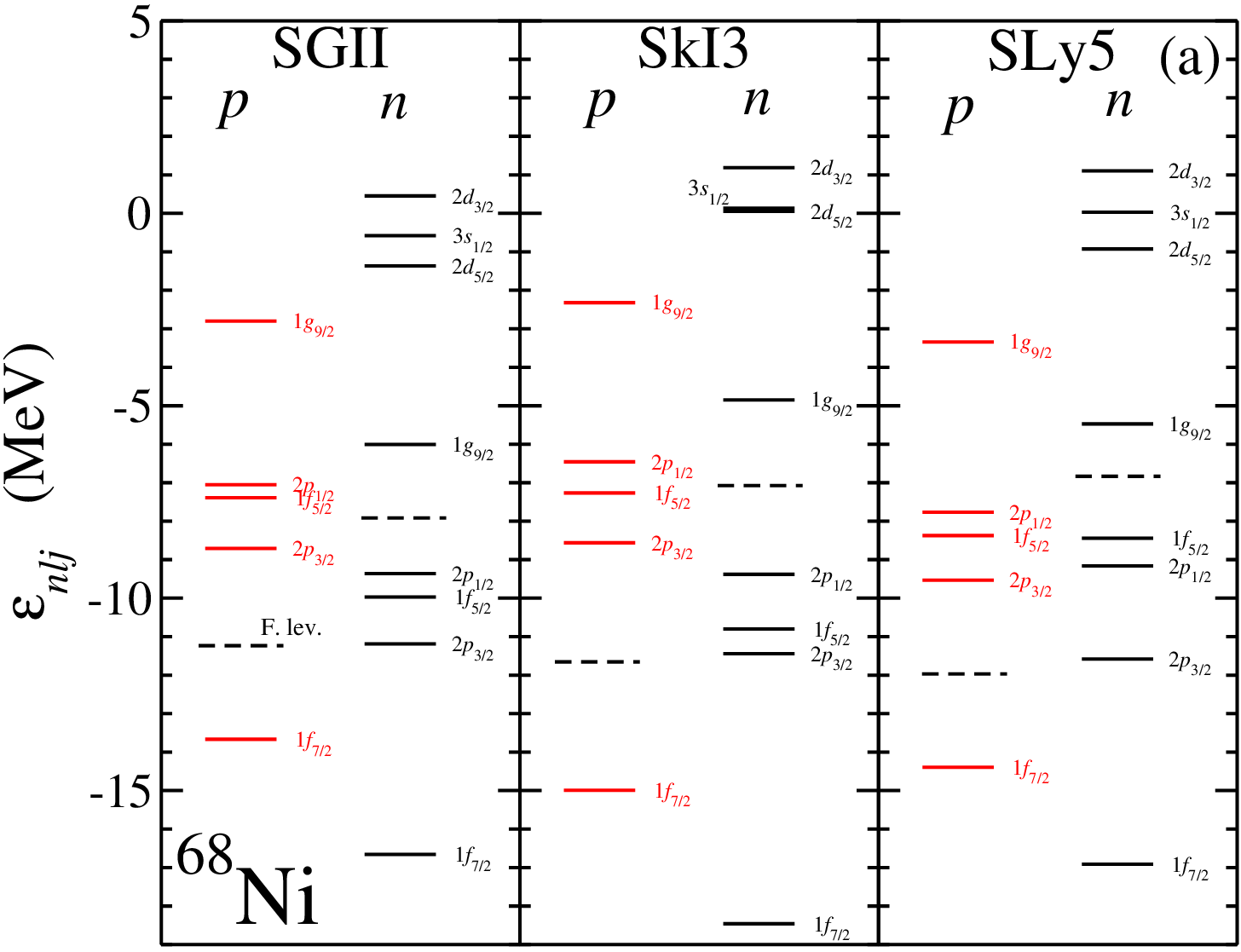}\\
\includegraphics[clip=true,width=0.9\linewidth]{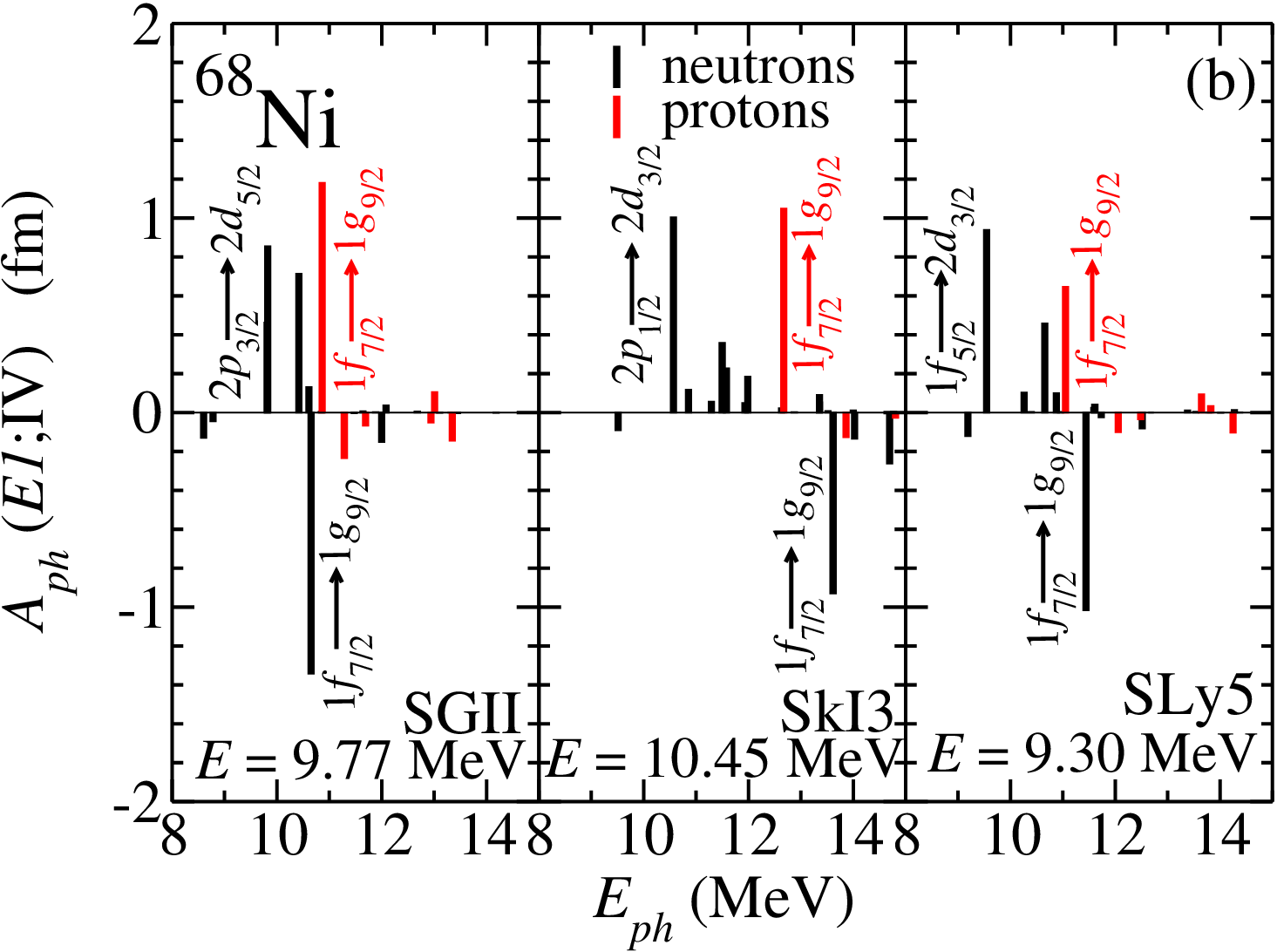}\\
\includegraphics[clip=true,width=0.9\linewidth]{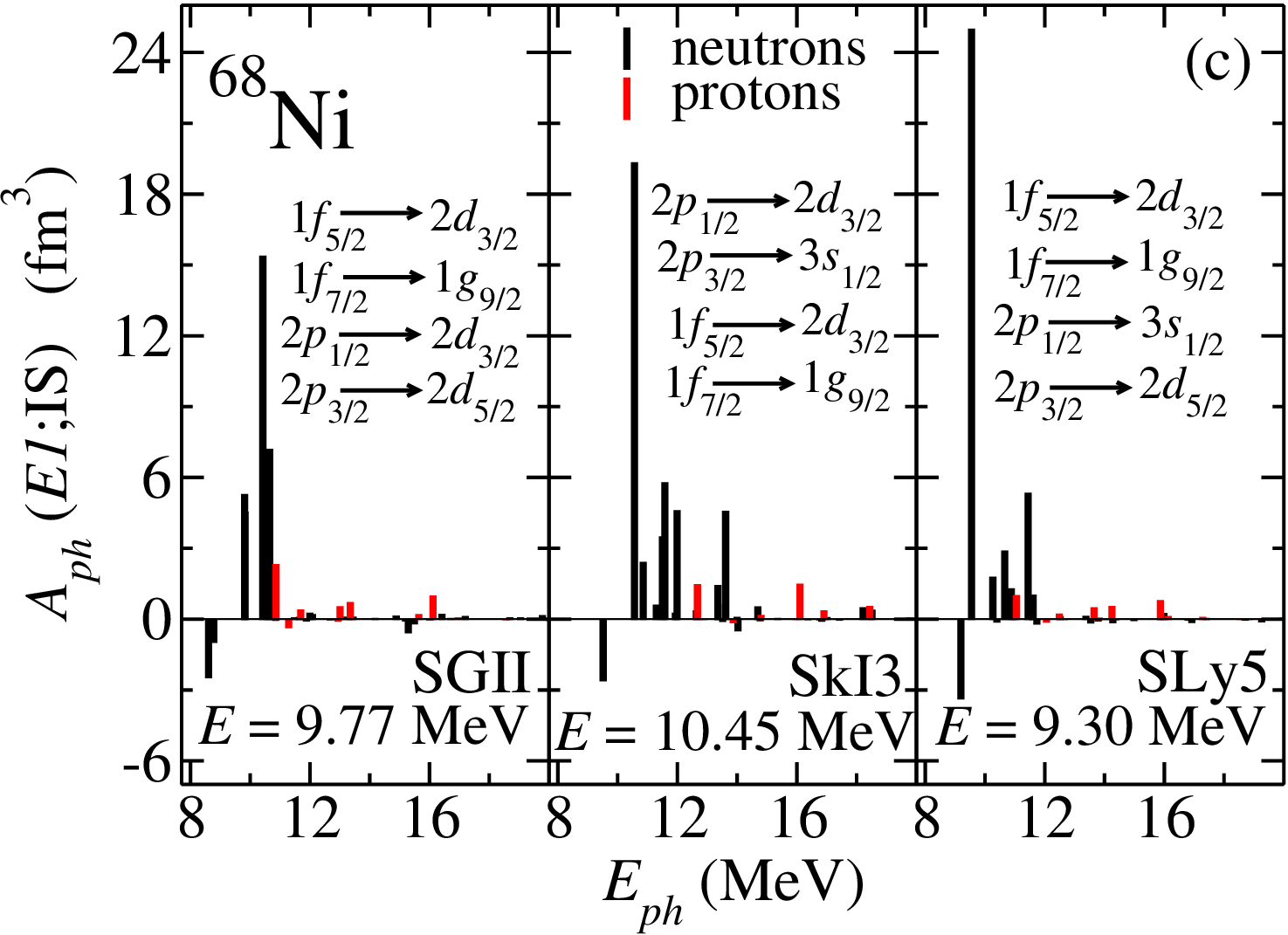}
\caption{(Color online) Same in Fig.~\ref{fig8} for the case of 
${}^{68}$Ni.} 
\label{fig10}
\end{figure}

Given the relevance of the analysis of the different 
reduced amplitudes $A_{ph}^q(E1;\xi)$,
we show the same figures also for ${}^{132}$Sn 
and ${}^{68}$Ni where the same features found in ${}^{208}$Pb 
can be observed. The neutron and proton 
single particle levels of these nuclei are displayed 
in Figs. \ref{fig9}a and \ref{fig10}a, respectively. 
For the case of ${}^{132}$Sn, the proton single 
particle levels display a similar spacing and ordering for 
the different models while the neutron ones 
do not. On the contrary, the ordering and spacing 
of the neutron and proton single particle levels in 
${}^{68}$Ni are very similar within all studied forces. 
This difference do not affect the qualitative 
structure of the isovector and isoscalar 
reduced amplitudes for ${}^{132}$Sn (Figs.~\ref{fig9}b 
and \ref{fig9}c, 
respectively) and ${}^{68}$Ni (Figs.~\ref{fig10}b 
and \ref{fig10}c, respectively). 
As in the case of ${}^{208}$Pb, while the isovector dipole 
responses of these nuclei do not show in the corresponding
reduced amplitudes more than a few relevant neutron 
and proton $ph$ excitations (with some amount of
destructive interference),
the isoscalar reduced transition amplitudes display 
a more clear collective behavior 
of the neutron $ph$ excitations coming from the 
transitions of the outermost levels (see Tables \ref{aph-table-sn} 
and \ref{aph-table-ni} of the Appendix \ref{table-aph} for some numerical details).

% >>>>>>>>>>>>>>>>>>>>>>>>>>>>>>>>>>>>>>>>>>>>>>>>>>>>>>>>>>>>>>>>>>>>
% CONCLUSIONS
\section{Conclusions}
\label{conclusions}
The collective or single-particle character of the low-energy dipole 
response has been carefully 
studied  in three nuclei, representative of different 
mass regions and neutron excess. To that end, we have analyzed within the fully 
self-consistent non-relativistic 
MF Skyrme Hartree-Fock plus RPA approach, the measured even-even nuclei ${}^{68}$Ni, 
${}^{132}$Sn and ${}^{208}$Pb. 
In order to investigate the sensitivity of the low-energy 
isovector and isoscalar dipole responses of the studied nuclei on the slope of the 
symmetry energy, we employed three Skyrme interactions: 
SGII, SLy5 and SkI3. They 
have been selected because they cover a wide range 
for the predicted values of the $L$ parameter. We 
have qualitatively found that the isoscalar as well as the isovector dipole responses for 
all studied nuclei show a low-energy peak in the strength function 
that increases in magnitude and 
is shifted to larger energies with increasing values of $L$. The behavior of the isovector 
dipole response of these nuclei is in agreement with recent findings for which a much larger 
set of relativistic and non-relativistic MF interactions were used \cite{carb2010}. 

From our analysis of ${}^{68}$Ni, ${}^{132}$Sn and ${}^{208}$Pb, we have also seen that 
the collectivity associated with the RPA states giving rise to the PDS show up 
depending on the nature of the probe used for exciting the nucleus: 
 in particular, there is systematically more collectivity 
in the isoscalar than in the isovector transitions.

To detail this conclusion more, the 
studied {\it RPA-pygmy} states consistently display a strong 
isoscalar character, although a non negligible isovector component is always observed. 
Our results do not support a clear collective nature of the
isovector response. This is opposite to what 
happens for all studied interactions 
when the same nuclei are excited by an isoscalar dipole operator. 
In this, the 
low-energy peak appearing in the strength function 
is recognizably collective and basically 
due to the outermost neutrons ---in other words, to the neutrons 
forming the neutron skin
known to be present in ${}^{208}$Pb \cite{prex2011,roca2011} 
and likely to developed also in 
${}^{68}$Ni and ${}^{132}$Sn.

Therefore, we can conclude that the isoscalar dipole 
oscillations of neutron rich nuclei arising 
from the {\it RPA-pygmy} states can be understood within the Skyrme-HF plus RPA approach as a collective 
motion of the outermost neutrons in neutron rich nuclei. Such a statement should be confirmed 
by further experimental investigations. 

% >>>>>>>>>>>>>>>>>>>>>>>>>>>>>>>>>>>>>>>>>>>>>>>>>>>>>>>>>>>>>>>>>>>>
% ACKNOWLEDGMENTS
%\acknowledgments 

% >>>>>>>>>>>>>>>>>>>>>>>>>>>>>>>>>>>>>>>>>>>>>>>>>>>>>>>>>>>>>>>>>>>>
% Appendix 
\appendix
\section{Spurious state}
\label{spurious}
We subtract the spurious state from the proton ($q=p$) and neutron ($q=n$) transition 
densities $\delta\rho_{\nu}^q(r)$ [where $\nu$ characterizes 
the RPA state, cf. Eq.~(\ref{td})] 
by imposing, first, that the translational operator which 
is proportional to the radial coordinate $r$ 
does not give any finite transition amplitude. This means  
\begin{equation}
\int dr r^2 r \left(\delta\rho_{\tilde{\nu}}^n + \delta\rho_{\tilde{\nu}}^p\right) = 0,  
\label{spurious-is-constraint}
\end{equation}
where $\tilde{\nu}$ denotes the corrected RPA state 
without spurious state contamination. As a second condition, 
we impose on these new transition densities 
that the strength of the IVGDR is not 
modified. That is, we write
\begin{equation}
\int dr r^2 r \left(\frac{Z}{A}\delta\rho_{\tilde{\nu}}^n 
- \frac{N}{A}\delta\rho_{\tilde{\nu}}^p\right) =   \int dr r^2 r \left(\frac{Z}{A}\delta\rho_{\nu}^n 
- \frac{N}{A}\delta\rho_{\nu}^p\right).  
\label{spurious-iv-constraint}
\end{equation}
By writing 
\begin{equation}
\delta\rho_{\tilde{\nu}}^q = \delta\rho_{\nu}^q - \alpha^q\frac{d\rho_{\rm HF}^q(r)}{dr}, 
\label{spurious-td}
\end{equation}
where $\rho_{\rm HF}^q(r)$ is the proton ($q=p$) or neutron ($q=n$) 
density obtained from the self-consistent HF calculations, we find the following 
solution, 
\begin{eqnarray}
\alpha^n &=& \frac{N}{A}\frac{\int dr r^2\ r\delta\rho_\nu}{\int dr r^2\ r \frac{d\rho_{\rm HF}^n}{dr}}\\
\alpha^p &=& \frac{Z}{A}\frac{\int dr r^2\ r\delta\rho_\nu}{\int dr r^2\ r \frac{d\rho_{\rm HF}^p}{dr}}
\label{spurious-alpha}
\end{eqnarray}
where $\delta\rho_\nu=\delta\rho_\nu^n + \delta\rho_\nu^p$. 

\section{Most relevant $ph$ contributions} 
\label{table-aph}
In this Appendix we give the numerical details of the most relevant $ph$ 
contributions to the isoscalar and isovector dipole response of ${}^{68}$Ni, 
${}^{132}$Sn and ${}^{208}$Pb  as predicted by SGII, SLy5 and SkI3 studied in 
Sec.~\ref{results}. Specifically, we show for all nuclei and interactions the 
ten neutron and proton $ph$ excitations with larger contributions to the reduced amplitude 
$A(E1;\xi)$ [see Eq.~(\ref{ampl})].    

\begin{table*}[t!]
\begin{center}
\caption{The ten largest contributions of the proton ($q=p$) and neutron ($q=n$) $ph$ excitations to the 
isovector and isoscalar reduced amplitudes, $A_{ph}^q(E1;\xi)$ where $\xi=$ IV and IS, respectively, 
as predicted by SGII, SLy5 and SkI3 models for ${}^{208}$Pb. The single particle levels involved in the 
transitions corresponding to each $ph$ excitation are also indicated.}
\label{aph-table-pb}
\begin{tabular}{llrrlrrlr}
\hline\hline
   & ~~~~~~~~~~~~SGII & & &~~~~~~~~~~~~SkI3 & & &~~~~~~~~~~~~SLy5 &  \\ 
   & & & & & & & & \\
   & & & &~~~~~~~~~Isovector& & & & \\
   & & & & & & & & \\
   &$E=7.61$MeV &$A_{ph}$ &\hspace{5mm} & $E=8.01$MeV&$A_{ph}$ &\hspace{5mm} & $E=7.74$MeV& $A_{ph}$\\
   &  & [fm] & & & [fm] & & & [fm]\\
\hline
$n$&$1i_{13/2}\rightarrow 1j_{15/2}$&$ -2.87$&&$1i_{13/2}\rightarrow 1j_{15/2}$&$ -1.18$&&$1i_{13/2}\rightarrow 1j_{15/2}$&$ -2.02$\\
   &$3p_{ 3/2}\rightarrow 3d_{ 5/2}$&$  1.31$&&$3p_{ 1/2}\rightarrow 4s_{ 1/2}$&$  1.02$&&$3p_{ 1/2}\rightarrow 4s_{ 1/2}$&$  0.75$\\
   &$3p_{ 1/2}\rightarrow 3d_{ 3/2}$&$  0.81$&&$1h_{ 9/2}\rightarrow 1i_{11/2}$&$ -0.54$&&$3p_{ 1/2}\rightarrow 3d_{ 3/2}$&$  0.63$\\
   &$1h_{ 9/2}\rightarrow 1i_{11/2}$&$ -0.76$&&$3p_{ 1/2}\rightarrow 3d_{ 3/2}$&$  0.48$&&$3p_{ 3/2}\rightarrow 3d_{ 5/2}$&$  0.54$\\
   &$2f_{ 7/2}\rightarrow 2g_{ 9/2}$&$  0.70$&&$3p_{ 3/2}\rightarrow 3d_{ 5/2}$&$  0.27$&&$1h_{ 9/2}\rightarrow 1i_{11/2}$&$ -0.53$\\
   &$2f_{ 5/2}\rightarrow 2g_{ 7/2}$&$  0.53$&&$3p_{ 3/2}\rightarrow 4s_{ 1/2}$&$  0.26$&&$3p_{ 3/2}\rightarrow 4s_{ 1/2}$&$  0.33$\\
   &$2f_{ 5/2}\rightarrow 3d_{ 3/2}$&$  0.28$&&$2f_{ 5/2}\rightarrow 3d_{ 3/2}$&$  0.12$&&$2f_{ 5/2}\rightarrow 3d_{ 3/2}$&$  0.26$\\
   &$3p_{ 3/2}\rightarrow 4s_{ 1/2}$&$  0.28$&&$2f_{ 5/2}\rightarrow 2g_{ 7/2}$&$  0.12$&&$1h_{ 9/2}\rightarrow 2g_{ 7/2}$&$  0.15$\\
   &$1h_{ 9/2}\rightarrow 2g_{ 7/2}$&$  0.20$&&$2f_{ 7/2}\rightarrow 3d_{ 5/2}$&$  0.11$&&$2f_{ 7/2}\rightarrow 3d_{ 5/2}$&$  0.12$\\
   &$3p_{ 1/2}\rightarrow 4s_{ 1/2}$&$ -0.15$&&$3p_{ 3/2}\rightarrow 3d_{ 3/2}$&$  0.05$&&$2f_{ 5/2}\rightarrow 2g_{ 7/2}$&$  0.11$\\
\hline
$p$&$1h_{11/2}\rightarrow 1i_{13/2}$&$  3.37$&&$1h_{11/2}\rightarrow 1i_{13/2}$&$  1.52$&&$1h_{11/2}\rightarrow 1i_{13/2}$&$  2.94$\\
   &$2d_{ 5/2}\rightarrow 2f_{ 7/2}$&$ -1.30$&&$1g_{ 7/2}\rightarrow 1h_{ 9/2}$&$  0.73$&&$2d_{ 5/2}\rightarrow 2f_{ 7/2}$&$ -0.93$\\
   &$1g_{ 7/2}\rightarrow 1h_{ 9/2}$&$  0.85$&&$3s_{ 1/2}\rightarrow 3p_{ 1/2}$&$ -0.14$&&$1g_{ 7/2}\rightarrow 1h_{ 9/2}$&$  0.77$\\
   &$2d_{ 3/2}\rightarrow 2f_{ 5/2}$&$ -0.69$&&$2d_{ 5/2}\rightarrow 3p_{ 3/2}$&$ -0.10$&&$3s_{ 1/2}\rightarrow 3p_{ 3/2}$&$ -0.50$\\
   &$3s_{ 1/2}\rightarrow 3p_{ 3/2}$&$ -0.38$&&$2d_{ 5/2}\rightarrow 2f_{ 7/2}$&$ -0.10$&&$2d_{ 3/2}\rightarrow 2f_{ 5/2}$&$ -0.34$\\
   &$3s_{ 1/2}\rightarrow 3p_{ 1/2}$&$ -0.19$&&$2d_{ 3/2}\rightarrow 2f_{ 5/2}$&$ -0.09$&&$2d_{ 5/2}\rightarrow 3p_{ 3/2}$&$ -0.21$\\
   &$1g_{ 7/2}\rightarrow 2f_{ 5/2}$&$ -0.19$&&$2d_{ 3/2}\rightarrow 3p_{ 1/2}$&$ -0.07$&&$2d_{ 3/2}\rightarrow 3p_{ 1/2}$&$ -0.21$\\
   &$2d_{ 5/2}\rightarrow 3p_{ 3/2}$&$ -0.15$&&$3s_{ 1/2}\rightarrow 3p_{ 3/2}$&$ -0.06$&&$1g_{ 7/2}\rightarrow 2f_{ 5/2}$&$ -0.18$\\
   &$2d_{ 3/2}\rightarrow 3p_{ 1/2}$&$ -0.13$&&$1g_{ 7/2}\rightarrow 2f_{ 5/2}$&$ -0.04$&&$3s_{ 1/2}\rightarrow 3p_{ 1/2}$&$ -0.18$\\
   &$2d_{ 5/2}\rightarrow 2f_{ 5/2}$&$ -0.06$&&$1h_{11/2}\rightarrow 2g_{ 9/2}$&$ -0.02$&&$2d_{ 3/2}\rightarrow 3p_{ 3/2}$&$ -0.06$\\
   & & & & & & & & \\
   & & & &~~~~~~~~~Isoscalar& & & & \\
 & &[fm${}^{3}$] & & &[fm${}^{3}$] & & &[fm${}^{3}$] \\
\hline
$n$&$1i_{13/2}\rightarrow 1j_{15/2}$&$ 28.53$&&$3p_{ 1/2}\rightarrow 4s_{ 1/2}$&$ 40.13$&&$3p_{ 1/2}\rightarrow 4s_{ 1/2}$&$ 27.42$\\
   &$1h_{ 9/2}\rightarrow 1i_{11/2}$&$ 12.54$&&$3p_{ 1/2}\rightarrow 3d_{ 3/2}$&$ 11.37$&&$1i_{13/2}\rightarrow 1j_{15/2}$&$ 17.50$\\
   &$3p_{ 1/2}\rightarrow 3d_{ 3/2}$&$  8.86$&&$1i_{13/2}\rightarrow 1j_{15/2}$&$  9.80$&&$3p_{ 1/2}\rightarrow 3d_{ 3/2}$&$ 11.62$\\
   &$2f_{ 5/2}\rightarrow 3d_{ 3/2}$&$  8.57$&&$3p_{ 3/2}\rightarrow 4s_{ 1/2}$&$  9.64$&&$3p_{ 3/2}\rightarrow 4s_{ 1/2}$&$ 11.16$\\
   &$3p_{ 3/2}\rightarrow 3d_{ 5/2}$&$  8.36$&&$1h_{ 9/2}\rightarrow 1i_{11/2}$&$  8.55$&&$2f_{ 5/2}\rightarrow 3d_{ 3/2}$&$  9.10$\\
   &$3p_{ 3/2}\rightarrow 4s_{ 1/2}$&$  6.79$&&$2f_{ 5/2}\rightarrow 3d_{ 3/2}$&$  4.64$&&$1h_{ 9/2}\rightarrow 1i_{11/2}$&$  8.46$\\
   &$1h_{ 9/2}\rightarrow 2g_{ 7/2}$&$  6.03$&&$3p_{ 3/2}\rightarrow 3d_{ 5/2}$&$  4.44$&&$3p_{ 3/2}\rightarrow 3d_{ 5/2}$&$  6.14$\\
   &$2f_{ 7/2}\rightarrow 2g_{ 9/2}$&$ -5.69$&&$2f_{ 7/2}\rightarrow 3d_{ 5/2}$&$  3.54$&&$1h_{ 9/2}\rightarrow 2g_{ 7/2}$&$  4.51$\\
   &$3p_{ 1/2}\rightarrow 4s_{ 1/2}$&$ -4.18$&&$1h_{ 9/2}\rightarrow 2g_{ 7/2}$&$  1.49$&&$2f_{ 7/2}\rightarrow 3d_{ 5/2}$&$  3.49$\\
   &$2f_{ 5/2}\rightarrow 2g_{ 7/2}$&$ -2.33$&&$3p_{ 3/2}\rightarrow 3d_{ 3/2}$&$  1.22$&&$2f_{ 5/2}\rightarrow 3d_{ 5/2}$&$  1.87$\\
\hline
$p$&$1h_{11/2}\rightarrow 1i_{13/2}$&$ 22.10$&&$1h_{11/2}\rightarrow 1i_{13/2}$&$  8.79$&&$1h_{11/2}\rightarrow 1i_{13/2}$&$ 17.69$\\
   &$2d_{ 5/2}\rightarrow 2f_{ 7/2}$&$ -9.71$&&$1g_{ 7/2}\rightarrow 1h_{ 9/2}$&$  7.69$&&$1g_{ 7/2}\rightarrow 1h_{ 9/2}$&$  8.22$\\
   &$1g_{ 7/2}\rightarrow 1h_{ 9/2}$&$  9.20$&&$2d_{ 5/2}\rightarrow 3p_{ 3/2}$&$  0.82$&&$2d_{ 5/2}\rightarrow 2f_{ 7/2}$&$ -6.31$\\
   &$2d_{ 3/2}\rightarrow 2f_{ 5/2}$&$ -4.69$&&$2d_{ 5/2}\rightarrow 2f_{ 7/2}$&$ -0.65$&&$1g_{ 7/2}\rightarrow 2f_{ 5/2}$&$  2.81$\\
   &$1g_{ 7/2}\rightarrow 2f_{ 5/2}$&$  2.93$&&$2d_{ 3/2}\rightarrow 3p_{ 1/2}$&$  0.62$&&$2d_{ 3/2}\rightarrow 2f_{ 5/2}$&$ -2.07$\\
   &$2d_{ 3/2}\rightarrow 3p_{ 1/2}$&$  1.11$&&$1h_{11/2}\rightarrow 4i_{13/2}$&$  0.60$&&$2d_{ 3/2}\rightarrow 3p_{ 1/2}$&$  1.86$\\
   &$2d_{ 5/2}\rightarrow 3p_{ 3/2}$&$  1.10$&&$1g_{ 7/2}\rightarrow 2f_{ 5/2}$&$  0.59$&&$2d_{ 5/2}\rightarrow 3p_{ 3/2}$&$  1.70$\\
   &$1h_{11/2}\rightarrow 4i_{13/2}$&$  0.96$&&$1h_{11/2}\rightarrow 2g_{ 9/2}$&$  0.56$&&$1h_{11/2}\rightarrow 2g_{ 9/2}$&$  1.55$\\
   &$1h_{11/2}\rightarrow 2g_{ 9/2}$&$  0.95$&&$2d_{ 3/2}\rightarrow 2f_{ 5/2}$&$ -0.54$&&$1g_{ 7/2}\rightarrow 3h_{ 9/2}$&$  1.19$\\
   &$1f_{ 5/2}\rightarrow 2g_{ 7/2}$&$  0.92$&&$1g_{ 7/2}\rightarrow 4h_{ 9/2}$&$  0.46$&&$1h_{11/2}\rightarrow 4i_{13/2}$&$  1.07$\\
\hline\hline
\end{tabular}
\end{center}
\end{table*}
\begin{table*}[t!]
\begin{center}
\caption{Same as in Table \ref{aph-table-pb} for ${}^{132}$Sn.}
\label{aph-table-sn}
\begin{tabular}{llrrlrrlr}
\hline\hline
   & ~~~~~~~~~~~~SGII & & &~~~~~~~~~~~~SkI3 & & &~~~~~~~~~~~~SLy5 &  \\ 
   & & & & & & & & \\
   & & & &~~~~~~~~~Isovector& & & & \\
   & & & & & & & & \\
 &$E=8.52$MeV & $A_{ph}$&\hspace{5mm} & $E=9.23$MeV& $A_{ph}$&\hspace{5mm} & $E=8.64$MeV& $A_{ph}$\\
 & & [fm] & & & [fm]& & & [fm]\\
\hline
$n$&$1h_{11/2}\rightarrow 1i_{13/2}$&$ -2.02$&&$1h_{11/2}\rightarrow 1i_{13/2}$&$ -1.40$&&$1h_{11/2}\rightarrow 1i_{13/2}$&$ -1.59$\\
   &$2d_{ 5/2}\rightarrow 2f_{ 7/2}$&$  1.08$&&$3s_{ 1/2}\rightarrow 3p_{ 3/2}$&$  1.12$&&$3s_{ 1/2}\rightarrow 3p_{ 3/2}$&$  0.77$\\
   &$3s_{ 1/2}\rightarrow 3p_{ 3/2}$&$  0.59$&&$1g_{ 7/2}\rightarrow 1h_{ 9/2}$&$ -0.50$&&$2d_{ 3/2}\rightarrow 3p_{ 1/2}$&$  0.41$\\
   &$2d_{ 3/2}\rightarrow 2f_{ 5/2}$&$  0.38$&&$3s_{ 1/2}\rightarrow 3p_{ 1/2}$&$  0.41$&&$2d_{ 3/2}\rightarrow 3p_{ 3/2}$&$  0.27$\\
   &$1g_{ 7/2}\rightarrow 1h_{ 9/2}$&$ -0.33$&&$2d_{ 3/2}\rightarrow 3p_{ 1/2}$&$  0.34$&&$3s_{ 1/2}\rightarrow 3p_{ 1/2}$&$  0.26$\\
   &$3s_{ 1/2}\rightarrow 3p_{ 1/2}$&$  0.29$&&$2d_{ 3/2}\rightarrow 2f_{ 5/2}$&$  0.25$&&$1g_{ 7/2}\rightarrow 1h_{ 9/2}$&$ -0.23$\\
   &$2d_{ 3/2}\rightarrow 3p_{ 3/2}$&$ -0.25$&&$2d_{ 5/2}\rightarrow 3p_{ 3/2}$&$  0.16$&&$2d_{ 5/2}\rightarrow 2f_{ 7/2}$&$ -0.18$\\
   &$1g_{ 7/2}\rightarrow 2f_{ 5/2}$&$  0.23$&&$2d_{ 3/2}\rightarrow 3p_{ 3/2}$&$  0.13$&&$2d_{ 5/2}\rightarrow 3p_{ 3/2}$&$  0.18$\\
   &$2d_{ 3/2}\rightarrow 3p_{ 1/2}$&$  0.22$&&$1g_{ 7/2}\rightarrow 2f_{ 5/2}$&$  0.09$&&$1g_{ 7/2}\rightarrow 2f_{ 5/2}$&$  0.15$\\
   &$1g_{ 7/2}\rightarrow 2f_{ 7/2}$&$  0.07$&&$2d_{ 5/2}\rightarrow 2f_{ 7/2}$&$  0.07$&&$2d_{ 3/2}\rightarrow 2f_{ 5/2}$&$  0.15$\\
\hline
$p$&$1g_{ 9/2}\rightarrow 1h_{11/2}$&$  1.84$&&$1g_{ 9/2}\rightarrow 1h_{11/2}$&$  1.63$&&$1g_{ 9/2}\rightarrow 1h_{11/2}$&$  1.92$\\
   &$2p_{ 3/2}\rightarrow 2d_{ 5/2}$&$ -0.80$&&$1f_{ 5/2}\rightarrow 1g_{ 7/2}$&$  0.66$&&$2p_{ 3/2}\rightarrow 2d_{ 5/2}$&$ -0.59$\\
   &$1f_{ 5/2}\rightarrow 1g_{ 7/2}$&$  0.27$&&$2p_{ 3/2}\rightarrow 2d_{ 5/2}$&$ -0.12$&&$1f_{ 5/2}\rightarrow 1g_{ 7/2}$&$  0.37$\\
   &$2p_{ 1/2}\rightarrow 2d_{ 3/2}$&$ -0.23$&&$2p_{ 1/2}\rightarrow 2d_{ 3/2}$&$ -0.12$&&$2p_{ 1/2}\rightarrow 3s_{ 1/2}$&$ -0.24$\\
   &$1f_{ 5/2}\rightarrow 2d_{ 3/2}$&$ -0.15$&&$2p_{ 3/2}\rightarrow 3s_{ 1/2}$&$ -0.10$&&$2p_{ 1/2}\rightarrow 2d_{ 3/2}$&$ -0.20$\\
   &$2p_{ 3/2}\rightarrow 3s_{ 1/2}$&$ -0.08$&&$2p_{ 1/2}\rightarrow 3s_{ 1/2}$&$ -0.08$&&$1f_{ 5/2}\rightarrow 2d_{ 3/2}$&$ -0.18$\\
   &$2p_{ 3/2}\rightarrow 2d_{ 3/2}$&$ -0.05$&&$1f_{ 5/2}\rightarrow 2d_{ 3/2}$&$ -0.05$&&$2p_{ 3/2}\rightarrow 3s_{ 1/2}$&$ -0.17$\\
   &$1g_{ 9/2}\rightarrow 2f_{ 7/2}$&$ -0.04$&&$1g_{ 9/2}\rightarrow 2f_{ 7/2}$&$ -0.04$&&$1g_{ 9/2}\rightarrow 2f_{ 7/2}$&$ -0.09$\\
   &$1f_{ 7/2}\rightarrow 2d_{ 5/2}$&$ -0.03$&&$2p_{ 3/2}\rightarrow 2d_{ 3/2}$&$ -0.03$&&$1f_{ 7/2}\rightarrow 2d_{ 5/2}$&$ -0.08$\\
   &$1f_{ 5/2}\rightarrow 2d_{ 5/2}$&$ -0.02$&&$1f_{ 7/2}\rightarrow 1g_{ 7/2}$&$  0.02$&&$1f_{ 5/2}\rightarrow 2d_{ 5/2}$&$ -0.03$\\
 & & & & & & & & \\
 & & & &~~~~~~~~~Isoscalar& & & & \\
 & &[fm${}^{3}$] & & &[fm${}^{3}$] & & &[fm${}^{3}$] \\
\hline
$n$&$1h_{11/2}\rightarrow 1i_{13/2}$&$ 12.42$&&$3s_{ 1/2}\rightarrow 3p_{ 3/2}$&$ 24.45$&&$3s_{ 1/2}\rightarrow 3p_{ 3/2}$&$ 14.49$\\
   &$3s_{ 1/2}\rightarrow 3p_{ 3/2}$&$  7.69$&&$2d_{ 3/2}\rightarrow 3p_{ 1/2}$&$ 11.25$&&$2d_{ 3/2}\rightarrow 3p_{ 1/2}$&$ 13.55$\\
   &$2d_{ 3/2}\rightarrow 3p_{ 3/2}$&$ -6.21$&&$3s_{ 1/2}\rightarrow 3p_{ 1/2}$&$ 10.07$&&$2d_{ 3/2}\rightarrow 3p_{ 3/2}$&$  8.13$\\
   &$2d_{ 3/2}\rightarrow 3p_{ 1/2}$&$  5.87$&&$1g_{ 7/2}\rightarrow 1h_{ 9/2}$&$  5.35$&&$1h_{11/2}\rightarrow 1i_{13/2}$&$  7.18$\\
   &$1g_{ 7/2}\rightarrow 2f_{ 5/2}$&$  5.37$&&$1h_{11/2}\rightarrow 1i_{13/2}$&$  5.29$&&$3s_{ 1/2}\rightarrow 3p_{ 1/2}$&$  5.95$\\
   &$3s_{ 1/2}\rightarrow 3p_{ 1/2}$&$  4.40$&&$2d_{ 5/2}\rightarrow 3p_{ 3/2}$&$  4.49$&&$2d_{ 5/2}\rightarrow 3p_{ 3/2}$&$  4.61$\\
   &$1g_{ 7/2}\rightarrow 1h_{ 9/2}$&$  3.84$&&$2d_{ 3/2}\rightarrow 3p_{ 3/2}$&$  4.14$&&$1g_{ 7/2}\rightarrow 2f_{ 5/2}$&$  3.76$\\
   &$2d_{ 5/2}\rightarrow 2f_{ 7/2}$&$ -2.76$&&$1g_{ 7/2}\rightarrow 2f_{ 5/2}$&$  2.19$&&$1h_{11/2}\rightarrow 3g_{ 9/2}$&$  2.71$\\
   &$1g_{ 7/2}\rightarrow 2f_{ 7/2}$&$  2.24$&&$1h_{11/2}\rightarrow 3g_{ 9/2}$&$  1.84$&&$1g_{ 7/2}\rightarrow 1h_{ 9/2}$&$  2.51$\\
   &$2d_{ 5/2}\rightarrow 3p_{ 3/2}$&$  1.15$&&$2d_{ 3/2}\rightarrow 2f_{ 5/2}$&$  1.81$&&$1g_{ 9/2}\rightarrow 2f_{ 7/2}$&$  1.35$\\
\hline
$p$&$1g_{ 9/2}\rightarrow 1h_{11/2}$&$  7.59$&&$1g_{ 9/2}\rightarrow 1h_{11/2}$&$  5.75$&&$1g_{ 9/2}\rightarrow 1h_{11/2}$&$  7.10$\\
   &$2p_{ 3/2}\rightarrow 2d_{ 5/2}$&$ -3.55$&&$1f_{ 5/2}\rightarrow 1g_{ 7/2}$&$  4.96$&&$1f_{ 5/2}\rightarrow 1g_{ 7/2}$&$  2.81$\\
   &$1f_{ 5/2}\rightarrow 1g_{ 7/2}$&$  2.12$&&$1g_{ 9/2}\rightarrow 2f_{ 7/2}$&$  0.64$&&$2p_{ 3/2}\rightarrow 2d_{ 5/2}$&$ -2.26$\\
   &$1f_{ 5/2}\rightarrow 2d_{ 3/2}$&$  1.43$&&$1f_{ 5/2}\rightarrow 2d_{ 3/2}$&$  0.48$&&$1f_{ 5/2}\rightarrow 2d_{ 3/2}$&$  1.63$\\
   &$2p_{ 1/2}\rightarrow 2d_{ 3/2}$&$ -0.92$&&$2p_{ 3/2}\rightarrow 2d_{ 5/2}$&$ -0.47$&&$1g_{ 9/2}\rightarrow 2f_{ 7/2}$&$  1.56$\\
   &$1g_{ 9/2}\rightarrow 2f_{ 7/2}$&$  0.64$&&$2p_{ 1/2}\rightarrow 2d_{ 3/2}$&$ -0.41$&&$2p_{ 1/2}\rightarrow 3s_{ 1/2}$&$  1.02$\\
   &$1g_{ 9/2}\rightarrow 4h_{11/2}$&$  0.41$&&$1g_{ 9/2}\rightarrow 4h_{11/2}$&$  0.40$&&$1f_{ 7/2}\rightarrow 3g_{ 9/2}$&$  0.80$\\
   &$1f_{ 5/2}\rightarrow 4g_{ 7/2}$&$  0.34$&&$1f_{ 5/2}\rightarrow 4g_{ 7/2}$&$  0.38$&&$1g_{ 9/2}\rightarrow 4h_{11/2}$&$  0.78$\\
   &$1d_{ 3/2}\rightarrow 2f_{ 5/2}$&$  0.33$&&$1g_{ 9/2}\rightarrow 5h_{11/2}$&$  0.35$&&$1f_{ 7/2}\rightarrow 2d_{ 5/2}$&$  0.76$\\
   &$1f_{ 7/2}\rightarrow 3g_{ 9/2}$&$  0.28$&&$2p_{ 1/2}\rightarrow 3s_{ 1/2}$&$  0.33$&&$1d_{ 5/2}\rightarrow 2f_{ 7/2}$&$  0.69$\\
\hline\hline
\end{tabular}
\end{center}
\end{table*}
\begin{table*}[t!]
\begin{center}
\caption{Same as in Table \ref{aph-table-pb} for ${}^{68}$Ni.}
\label{aph-table-ni}
\begin{tabular}{llrrlrrlr}
\hline\hline
   & ~~~~~~~~~~~~SGII & & &~~~~~~~~~~~~SkI3 & & &~~~~~~~~~~~~SLy5 &  \\ 
   & & & & & & & & \\
   & & & &~~~~~~~~~Isovector& & & & \\
   & & & & & & & & \\
 &$E=9.77$MeV & $A_{ph}$&\hspace{5mm} & $E=10.45$MeV& $A_{ph}$&\hspace{5mm} & $E=9.30$MeV& $A_{ph}$\\
 & &[fm] & & &[fm] & & &[fm] \\
$n$&$1f_{ 7/2}\rightarrow 1g_{ 9/2}$&$ -1.34$&&$2p_{ 1/2}\rightarrow 2d_{ 3/2}$&$  1.00$&&$1f_{ 7/2}\rightarrow 1g_{ 9/2}$&$ -1.01$\\
   &$2p_{ 3/2}\rightarrow 2d_{ 5/2}$&$  0.85$&&$1f_{ 7/2}\rightarrow 1g_{ 9/2}$&$ -0.92$&&$1f_{ 5/2}\rightarrow 2d_{ 3/2}$&$  0.93$\\
   &$1f_{ 5/2}\rightarrow 2d_{ 3/2}$&$  0.71$&&$2p_{ 3/2}\rightarrow 2d_{ 5/2}$&$  0.35$&&$2p_{ 3/2}\rightarrow 2d_{ 5/2}$&$  0.45$\\
   &$2p_{ 1/2}\rightarrow 2d_{ 3/2}$&$  0.46$&&$1f_{ 5/2}\rightarrow 2g_{ 7/2}$&$ -0.26$&&$2p_{ 1/2}\rightarrow 3s_{ 1/2}$&$ -0.12$\\
   &$1f_{ 5/2}\rightarrow 1g_{ 7/2}$&$ -0.15$&&$2p_{ 3/2}\rightarrow 3s_{ 1/2}$&$  0.22$&&$2p_{ 1/2}\rightarrow 2d_{ 3/2}$&$  0.10$\\
   &$2p_{ 3/2}\rightarrow 3s_{ 1/2}$&$  0.13$&&$1f_{ 5/2}\rightarrow 2d_{ 3/2}$&$  0.18$&&$1f_{ 5/2}\rightarrow 3d_{ 3/2}$&$  0.09$\\
   &$1f_{ 5/2}\rightarrow 2d_{ 5/2}$&$ -0.12$&&$1f_{ 5/2}\rightarrow 1g_{ 7/2}$&$ -0.13$&&$1f_{ 5/2}\rightarrow 2g_{ 7/2}$&$ -0.08$\\
   &$2p_{ 1/2}\rightarrow 3s_{ 1/2}$&$ -0.04$&&$1f_{ 5/2}\rightarrow 2d_{ 5/2}$&$  0.11$&&$1f_{ 5/2}\rightarrow 2d_{ 5/2}$&$ -0.05$\\
   &$1f_{ 5/2}\rightarrow 3d_{ 3/2}$&$  0.03$&&$2p_{ 1/2}\rightarrow 3s_{ 1/2}$&$ -0.09$&&$2p_{ 3/2}\rightarrow 3s_{ 1/2}$&$  0.04$\\
   &$1f_{ 7/2}\rightarrow 2d_{ 5/2}$&$ -0.03$&&$1f_{ 5/2}\rightarrow 3d_{ 3/2}$&$  0.09$&&$1f_{ 5/2}\rightarrow 1g_{ 7/2}$&$ -0.02$\\
\hline
$p$&$1f_{ 7/2}\rightarrow 1g_{ 9/2}$&$  1.18$&&$1f_{ 7/2}\rightarrow 1g_{ 9/2}$&$  1.04$&&$1f_{ 7/2}\rightarrow 1g_{ 9/2}$&$  0.64$\\
   &$2s_{ 1/2}\rightarrow 2p_{ 3/2}$&$ -0.23$&&$1d_{ 3/2}\rightarrow 1f_{ 5/2}$&$  0.31$&&$1d_{ 3/2}\rightarrow 2p_{ 1/2}$&$ -0.10$\\
   &$1d_{ 3/2}\rightarrow 2p_{ 1/2}$&$ -0.14$&&$2s_{ 1/2}\rightarrow 2p_{ 3/2}$&$ -0.12$&&$2s_{ 1/2}\rightarrow 2p_{ 3/2}$&$ -0.10$\\
   &$1d_{ 3/2}\rightarrow 1f_{ 5/2}$&$  0.10$&&$1d_{ 3/2}\rightarrow 2p_{ 1/2}$&$ -0.07$&&$1d_{ 3/2}\rightarrow 1f_{ 5/2}$&$  0.09$\\
   &$1f_{ 7/2}\rightarrow 2d_{ 5/2}$&$ -0.07$&&$1f_{ 7/2}\rightarrow 2d_{ 5/2}$&$ -0.03$&&$1f_{ 7/2}\rightarrow 2d_{ 5/2}$&$ -0.06$\\
   &$1d_{ 3/2}\rightarrow 2p_{ 3/2}$&$ -0.06$&&$1d_{ 3/2}\rightarrow 2p_{ 3/2}$&$ -0.02$&&$1d_{ 3/2}\rightarrow 2p_{ 3/2}$&$ -0.03$\\
   &$2s_{ 1/2}\rightarrow 2p_{ 1/2}$&$ -0.05$&&$1d_{ 5/2}\rightarrow 1f_{ 5/2}$&$  0.01$&&$2s_{ 1/2}\rightarrow 2p_{ 1/2}$&$  0.03$\\
   &$1d_{ 5/2}\rightarrow 2p_{ 3/2}$&$ -0.03$&&$1f_{ 7/2}\rightarrow 2g_{ 7/2}$&$  0.01$&&$1d_{ 5/2}\rightarrow 2p_{ 3/2}$&$ -0.02$\\
   &$1d_{ 5/2}\rightarrow 1f_{ 5/2}$&$  0.01$&&$1d_{ 5/2}\rightarrow 2p_{ 3/2}$&$ -0.01$&&$1d_{ 5/2}\rightarrow 1f_{ 5/2}$&$  0.01$\\
   &$1f_{ 7/2}\rightarrow 1g_{ 7/2}$&$  0.01$&&$2s_{ 1/2}\rightarrow 2p_{ 1/2}$&$  0.01$&&$1f_{ 7/2}\rightarrow 1g_{ 7/2}$&$  0.01$\\
 & & & & & & & & \\
 & & & &~~~~~~~~~Isoscalar& & & & \\
 & &[fm${}^{3}$] & & &[fm${}^{3}$] & & &[fm${}^{3}$] \\
\hline
$n$&$1f_{ 5/2}\rightarrow 2d_{ 3/2}$&$ 15.35$&&$2p_{ 1/2}\rightarrow 2d_{ 3/2}$&$ 19.31$&&$1f_{ 5/2}\rightarrow 2d_{ 3/2}$&$ 24.96$\\
   &$1f_{ 7/2}\rightarrow 1g_{ 9/2}$&$  7.17$&&$2p_{ 3/2}\rightarrow 3s_{ 1/2}$&$  5.76$&&$1f_{ 7/2}\rightarrow 1g_{ 9/2}$&$  5.32$\\
   &$2p_{ 1/2}\rightarrow 2d_{ 3/2}$&$  5.26$&&$1f_{ 5/2}\rightarrow 2d_{ 3/2}$&$  4.58$&&$2p_{ 1/2}\rightarrow 3s_{ 1/2}$&$ -3.36$\\
   &$2p_{ 3/2}\rightarrow 2d_{ 5/2}$&$  4.51$&&$1f_{ 7/2}\rightarrow 1g_{ 9/2}$&$  4.55$&&$2p_{ 3/2}\rightarrow 2d_{ 5/2}$&$  2.86$\\
   &$2p_{ 3/2}\rightarrow 3s_{ 1/2}$&$  2.56$&&$2p_{ 3/2}\rightarrow 2d_{ 5/2}$&$  3.46$&&$2p_{ 1/2}\rightarrow 2d_{ 3/2}$&$  1.76$\\
   &$1f_{ 5/2}\rightarrow 2d_{ 5/2}$&$ -2.46$&&$2p_{ 1/2}\rightarrow 3s_{ 1/2}$&$ -2.60$&&$1f_{ 5/2}\rightarrow 3d_{ 3/2}$&$  1.26$\\
   &$2p_{ 1/2}\rightarrow 3s_{ 1/2}$&$ -0.96$&&$1f_{ 5/2}\rightarrow 2d_{ 5/2}$&$  2.39$&&$1f_{ 5/2}\rightarrow 2d_{ 5/2}$&$ -1.05$\\
   &$1f_{ 7/2}\rightarrow 2d_{ 5/2}$&$ -0.56$&&$1f_{ 5/2}\rightarrow 3d_{ 3/2}$&$  1.40$&&$2p_{ 3/2}\rightarrow 3s_{ 1/2}$&$  1.00$\\
   &$1f_{ 5/2}\rightarrow 1g_{ 7/2}$&$  0.23$&&$2p_{ 3/2}\rightarrow 4s_{ 1/2}$&$  0.60$&&$1f_{ 7/2}\rightarrow 5g_{ 9/2}$&$  0.23$\\
   &$2p_{ 3/2}\rightarrow 5d_{ 5/2}$&$  0.22$&&$2p_{ 1/2}\rightarrow 4s_{ 1/2}$&$  0.57$&&$1f_{ 7/2}\rightarrow 4g_{ 9/2}$&$  0.22$\\
\hline
$p$&$1f_{ 7/2}\rightarrow 1g_{ 9/2}$&$  2.29$&&$1d_{ 3/2}\rightarrow 1f_{ 5/2}$&$  1.46$&&$1f_{ 7/2}\rightarrow 1g_{ 9/2}$&$  0.98$\\
   &$1f_{ 7/2}\rightarrow 2d_{ 5/2}$&$  0.96$&&$1f_{ 7/2}\rightarrow 1g_{ 9/2}$&$  1.44$&&$1f_{ 7/2}\rightarrow 2d_{ 5/2}$&$  0.77$\\
   &$1d_{ 3/2}\rightarrow 2p_{ 1/2}$&$  0.68$&&$1f_{ 7/2}\rightarrow 2d_{ 5/2}$&$  0.52$&&$1d_{ 3/2}\rightarrow 2p_{ 1/2}$&$  0.51$\\
   &$1d_{ 3/2}\rightarrow 1f_{ 5/2}$&$  0.50$&&$1d_{ 3/2}\rightarrow 2p_{ 1/2}$&$  0.33$&&$1d_{ 3/2}\rightarrow 1f_{ 5/2}$&$  0.46$\\
   &$1d_{ 3/2}\rightarrow 2p_{ 3/2}$&$  0.37$&&$1f_{ 7/2}\rightarrow 5g_{ 9/2}$&$  0.29$&&$1f_{ 7/2}\rightarrow 5g_{ 9/2}$&$  0.36$\\
   &$2s_{ 1/2}\rightarrow 2p_{ 3/2}$&$ -0.35$&&$1f_{ 7/2}\rightarrow 6g_{ 9/2}$&$  0.22$&&$1f_{ 7/2}\rightarrow 4g_{ 9/2}$&$  0.26$\\
   &$1f_{ 7/2}\rightarrow 5g_{ 9/2}$&$  0.34$&&$1f_{ 7/2}\rightarrow 4g_{ 9/2}$&$  0.19$&&$1f_{ 7/2}\rightarrow 6g_{ 9/2}$&$  0.22$\\
   &$1f_{ 7/2}\rightarrow 4g_{ 9/2}$&$  0.23$&&$1d_{ 3/2}\rightarrow 5f_{ 5/2}$&$  0.16$&&$1d_{ 3/2}\rightarrow 2p_{ 3/2}$&$  0.20$\\
   &$1f_{ 7/2}\rightarrow 6g_{ 9/2}$&$  0.22$&&$1d_{ 5/2}\rightarrow 4f_{ 7/2}$&$  0.15$&&$1d_{ 5/2}\rightarrow 4f_{ 7/2}$&$  0.18$\\
   &$1d_{ 5/2}\rightarrow 4f_{ 7/2}$&$  0.19$&&$1d_{ 3/2}\rightarrow 2p_{ 3/2}$&$  0.14$&&$1d_{ 3/2}\rightarrow 4f_{ 5/2}$&$  0.17$\\
\hline\hline
\end{tabular}
\end{center}
\end{table*}
%

% >>>>>>>>>>>>>>>>>>>>>>>>>>>>>>>>>>>>>>>>>>>>>>>>>>>>>>>>>>>>>>>>>>>>
% REFERENCES.

%
% >>>>>>>>>>>>>>>>>>>>>>>>>>>>>>>>>>>>>>>>>>>>>>>>>>>>>>>>>>>>>>>>>>>>
\end{document}